\documentclass[twocolumn]{aastex62}
\usepackage{natbib}
\graphicspath{{./}{figures/}}
\usepackage{color,subfigure,lineno} 

\newcommand{\hst}{{\it HST}}

\def\NaI{Na\,{\sc i}}
\def\CaII{Ca\,{\sc ii}}

\shorttitle{VODKA: HST Discovers Double Quasars}
\shortauthors{Chen et al.}

\begin{document}

\title{Varstrometry for Off-nucleus and Dual sub-Kpc AGN (VODKA): Hubble Space Telescope Discovers Double Quasars}


\author[0000-0002-9932-1298]{Yu-Ching Chen}
\affiliation{Department of Astronomy, University of Illinois at Urbana-Champaign, Urbana, IL 61801, USA}
\affiliation{National Center for Supercomputing Applications, University of Illinois at Urbana-Champaign, Urbana, IL 61801, USA}
\affiliation{Center for AstroPhysical Surveys, National Center for Supercomputing Applications, Urbana, IL, 61801, USA}

\author[0000-0003-4250-4437]{Hsiang-Chih Hwang}
\affiliation{Department of Physics and Astronomy, Johns Hopkins University, Baltimore, MD 21210, USA }

\author[0000-0003-1659-7035]{Yue Shen}
\affiliation{Department of Astronomy, University of Illinois at Urbana-Champaign, Urbana, IL 61801, USA}
\affiliation{National Center for Supercomputing Applications, University of Illinois at Urbana-Champaign, Urbana, IL 61801, USA}

\author[0000-0003-0049-5210]{Xin Liu}
\affiliation{Department of Astronomy, University of Illinois at Urbana-Champaign, Urbana, IL 61801, USA}
\affiliation{National Center for Supercomputing Applications, University of Illinois at Urbana-Champaign, Urbana, IL 61801, USA}

\author[0000-0001-6100-6869]{Nadia L. Zakamska}
\affiliation{Department of Physics and Astronomy, Johns Hopkins University, Baltimore, MD 21210, USA }

\author[0000-0001-6100-6869]{Qian Yang}
\affiliation{Department of Astronomy, University of Illinois at Urbana-Champaign, Urbana, IL 61801, USA}

\author[0000-0002-0311-2812]{Jennifer I. Li}
\affiliation{Department of Astronomy, University of Illinois at Urbana-Champaign, Urbana, IL 61801, USA}

\begin{abstract}

Dual supermassive black holes (SMBHs) at $\sim$\,kpc scales are the progenitor population of SMBH mergers and play an important role in understanding the pairing and dynamical evolution of massive black holes in galaxy mergers. Because of the stringent resolution requirement and the apparent rareness of these small-separation pairs, there are scarce observational constraints on this population, with few confirmed dual SMBHs at $<10\,$kpc separations at $z>1$. Here we present results from a pilot search for kpc-scale dual quasars selected with {\it Gaia} Data release 2 (DR2) astrometry and followed up with {\it Hubble Space Telescope} ({\it HST}) Wide Field Camera 3 dual-band (F475W and F814W) snapshot imaging. Our targets are quasars primarily selected with the varstrometry technique, i.e., light centroid jitter caused by asynchronous variability from both members in an unresolved quasar pair, supplemented by sub-arcsec pairs already resolved by {\it Gaia} DR2. We find an overall high fraction of {\it HST}-resolved pairs among the varstrometry-selected quasars (unresolved in {\it Gaia} DR2), $\sim 30-50\%$, increasing toward high redshift ($\sim 60-80\%$ at $z>1.5$). We discuss the nature of the 45 resolved sub-arcsec pairs based on {\it HST} and supplementary data. A substantial fraction ($\sim 40\%$) of these pairs are likely physical quasar pairs or gravitationally lensed quasars. We also discover a triple quasar candidate and a quadruply lensed quasar, which is among the smallest-separation quadruple lenses.
These results provide important guidelines to improve varstrometry selection and follow-up confirmation of $\sim$~kpc-scale dual SMBHs at high redshift.


\end{abstract}


\section{Introduction}\label{sec:introduction}

Since most massive galaxies harbor a central supermassive black hole \citep[SMBH;][]{kormendy95,KormendyHo2013}, galaxy mergers should result in the formation of dual SMBHs and eventually binary SMBHs \citep{begelman80}. Dual SMBHs are the precursors of binary SMBHs. A binary SMBH is on a compact ($\lesssim 10$~parsec) orbit in its own potential whereas a dual SMBH is on wider orbits and evolving in the potential of the (merged) host galaxy. Theory predicts strong gravitational waves (GWs) from the final coalescence of merging SMBHs -- a ``standard siren'' for cosmology and a direct testbed for strong-field general relativity \citep{hughes09,Centrella2010}. The presence of gas accretion during the merger process can dramatically change the expectations for the duration of the gravitational-wave-emitting phase \citep{Bogdanovic2021}. 

Before the binary SMBH stage, the two black holes spend several hundred million years in the dual SMBH stage as the galaxy merger proceeds. The observational search of close ($\sim$~kpc) dual quasars at $1<z<3$ is particularly important for constraining the merger process, because the effects of mergers are believed to be the most significant in the high-redshift, high-luminosity, and close-separation regime \citep{hopkins08}. The redshift range $1<z<3$ is also where we expect the horizon of the stochastic gravitational wave background from the ensemble of SMBH binaries \citep{Arzoumanian2018a}.

\begin{figure*}
  \centering
    \includegraphics[width=1.0\textwidth]{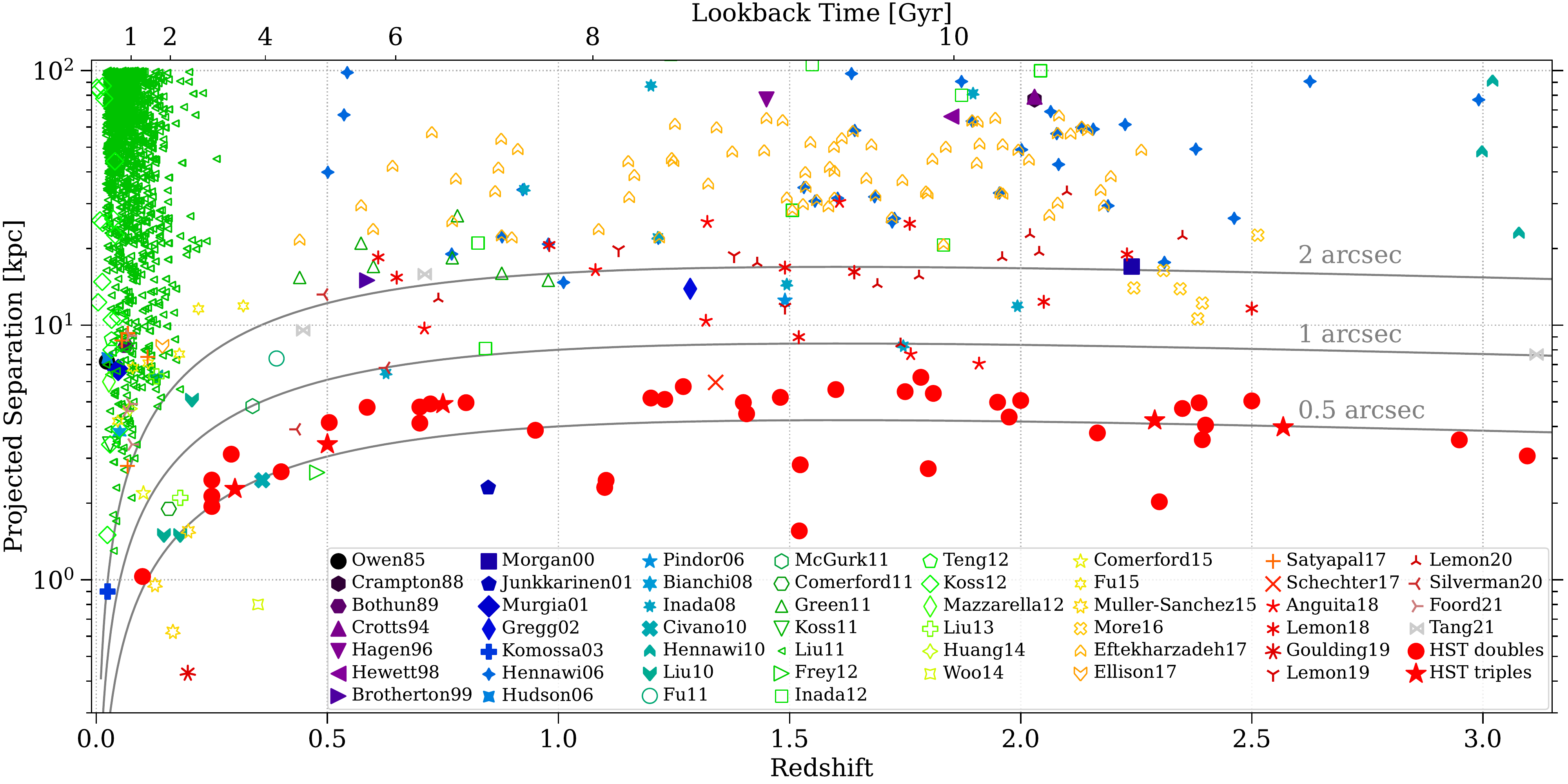}
    \caption{
    Quasars with double/triple sources discovered by {\it HST}/WFC3 imaging in comparison to ther reported dual quasars in the literature \citep{Owen85,Crampton88,Bothun89,Crotts94,Hagen96,Hewett98,Brotherton99,Morgan00,junkkarinen01,Murgia01,Gregg02,Komossa03,Hennawi06,Hudson06,Pindor06,Bianchi08,Inada08,Civano10,Hennawi10,Liu10,Fu11,McGurk11,Comerford11,Green11,Koss11,Liu11,Frey12,Inada12,Teng12,Koss12,Mazzarella12,Liu13,Huang14,Woo14,Comerford15,Fu15,Muller-Sanchez15,More16,Eftekharzadeh17,Ellison17,Satyapal17,Schechter17,Anguita18,Lemon18,Goulding19,Lemon19,Lemon20,Silverman20,Foord2021,Tang21}. A dual quasar is defined as a quasar pair with velocity differences $<2000\,{\rm km\,s^{-1}}$. Shown here is the projected physical separation as a function of redshift. No confirmed dual quasar is known at $r_p<6$ kpc at $z>1$. Our sample probes this new parameter space.
    }
    \label{fig:z_rp}
\end{figure*}

In the past decade, significant progress has been made toward finding concrete evidence for active merging SMBHs in kpc-scale dual active galactic nuclei (AGN) \citep{DeRosa2020}. 
The majority of known dual AGN are at low redshifts and/or at large physical separations ($>$20 kpc, \autoref{fig:z_rp}), and many have relatively low luminosities. Little is known about dual SMBHs at the cosmic ``high noon'' (i.e., $1<z<3$), when both the number density of luminous quasars and the global star formation rate density peak \citep{richards06,Madau2014}.  Finding small-separation dual quasars at high redshifts is extremely difficult observationally due to the typical sizes of the point-spread functions (PSFs) of ground-based optical surveys.
There is only one 2.3 kpc dual quasar known at $z\sim 0.8$ from a serendipitous discovery \citep{junkkarinen01,Shields12} and five dual quasars with 6 $<r_p<$ 10 kpc at $0.5<z\lesssim3$ from systematic searches \citep[][see \autoref{fig:z_rp} for details]{Inada12,More16,Lemon18,Silverman20,Tang21}. From a theoretical perspective, the dynamical evolution of dual SMBHs in merged galaxies can be well described by dynamical friction before entering the GW-dominated regime \citep{Yu2002,Merritt2013,ChenYF2020}, which results in shorter time spans on smaller separations. Therefore, the frequency of $\lesssim$\,kpc dual SMBHs is expected to be substantially lower than those on $\sim$~tens of kpc separations, requiring an efficient targeting scheme to identify candidates for follow-up confirmation.

Although not the focus of this work, searches for binary SMBHs ($\lesssim 10\,$parsec) are even more challenging. There are only two reported cases of (candidate) binary SMBHs resolved with the Very Long Baseline Array (VLBA) imaging at 7 parsec \citep{Rodriguez2006} and 0.35 parsec \citep{Kharb2017} from serendipitous searches. While searches using indirect methods \citep[from periodic light curves or radial velocity shifts of broad emission lines,][]{eracleous11,Shen2013,Liu2014,Graham2015,Charisi2016,Runnoe2017,Wang2017,Guo2019,Liutt2019,ChenYC2020,Liao2021} have yielded candidates of sub-pc binary SMBHs, alternative scenarios involving single SMBHs remain possible. Direct imaging confirmation for most of these sub-pc binary candidates is not feasible even with Very Long Baseline Interferometry (VLBI) in radio due to strict angular resolution requirement \citep{burke11}.


\citet{ShenHwang2019} and \citet{HwangShen2020} have proposed a new astrometric technique, {\tt varstrometry}, with {\it Gaia} data \citep{GaiaDR2} to discover close (sub-arcsecond) dual/lensed quasars \citep[also see][]{williams1995,LiuY_2015,Springer2021}. It takes advantage of the ubiquitous quasar variability, which in dual or lensed quasars results in an astrometric noise signature when the asynchronous variability from the two quasars causes centroid jitters in the unresolved system. {\it Gaia}'s full-sky coverage and depth to $G{\sim}21$ mag (reaching ${\sim}7\times10^5$ extragalactic objects) enable a large-scale systematic search in the poorly explored sub-arcsecond regime at high redshift \citep{HwangShen2020,Shen2021}. 

In this work, we report the first results from a pilot \hst\ dual-band imaging program to follow up the {\it Gaia}-astrometry-selected dual/lensed quasar candidates. It reveals compact double or triple sources in the 45 targets, 40\% of which are possible dual/lensed quasars based on the \hst\ two-band colors. The targets that are not quasar-star superpositions are more likely to be dual quasars instead of lenses, which require very massive lenses at high redshifts given the non-detection of the lens galaxy in the \hst\ image \citep{Shen2021}.

The paper is organized as follows. In \S\ref{sec:data}, we describe the methodology and the data we use for this study. We present our main results in \S\ref{sec:results} and discuss their implications and future prospects in \S\ref{sec:discussions}. We summarize our findings in \S\ref{sec:conclusions}. Throughout this paper we focus on optically unobscured, broad-line quasars, for which we can apply this {\tt varstrometry} technique, and all physical separations are the projected separation. A flat $\Lambda$CDM cosmology is adopted throughout with $\Omega_\Lambda=0.7$, $\Omega_m$=0.3, and $H_0=70\,{\rm km\,s^{-1}Mpc^{-1}}$.


\section{Observations and Data Analysis}\label{sec:data}

\subsection{Target Selection}\label{sec:target_selection}

Our parent quasar sample includes both spectroscopically confirmed SDSS DR7 quasars \citep{Shen2011b} and DR14 quasars \citep{Paris2018}, and photometrically selected quasar candidates based on WISE \citep{WISE,Mateos2012,Secrest2015}. The target selection criteria are mainly based on {\it Gaia} DR2 \citep{GaiaDR2}. Our targets are selected as {\it Gaia}-unresolved quasars with excess astrometric noise (i.e., {\tt varstrometry} selection) and quasars with a sub-arcsec companion resolved in {\it Gaia} data \citep{HwangShen2020}. For {\it Gaia}-unresolved sources, we use {\tt astrometric\_excess\_noise} \citep{Lindegren2012,Lindegren2018}  as a surrogate for the astrometric jitters caused by {\tt varstrometry} \citep{HwangShen2020,ShenHwang2019}. For {\it Gaia}-resolved sources, we  prioritize pairs with small angular separations (0\farcs4-0\farcs7) and use other information (see below) to reduce chance superposition stars.  The {\it Gaia}-resolved sources are intended to increase the statistics on sub-arcsec pairs in quasars, since \hst\ is able to resolve these pairs and derive color/morphology information, though they are at wider separation due to the minimal separation limit of 0\farcs4 in {\it Gaia} DR2 \citep{Arenou2018}


The detailed target selection and target categories are:
\begin{enumerate}
    \item SDSS quasars with a single {\it Gaia} match within 3\arcsec\ (40 targets selected). We use both the SDSS DR7 quasar catalog \citep{Shen2011b} and the DR14 quasar catalog \citep{Paris2018}. Because the extended structure of low-redshift host galaxies may also result in a high {\tt astrometric\_excess\_noise} \citep{HwangShen2020}, we apply a redshift cut of $z>0.5$ and require that the source is flagged as non-extended in all filters of Pan-STARRS1 \citep{Chambers2016}. To reduce contamination from foreground stars, we exclude sources where the SDSS fiber (2\arcsec\ or 3\arcsec) spectra have obvious stellar features. Then, we select 8 targets that have {\tt astrometric\_excess\_noise}$>3.39$\,mas and {\it Gaia} $G$-band magnitude $<19.5$\,mag, and additional 32 targets that have {\tt astrometric\_excess\_noise}$>1.44$\,mas and {\it Gaia} $G$-band magnitude $<19.1$\,mag. These magnitude cuts restrict the selection to reliable {\it Gaia} astrometry. 
    
    
    \item WISE+Pan-STARRS1 quasars with a single {\it Gaia} match within 3\arcsec\ (40 targets selected). We use an all-sky photometric quasar sample selected with WISE data \citep{Mateos2012,Secrest2015} and the cross-match catalog provided by {\it Gaia} DR2 \citep{Marrese2019}. We further require that the target is flagged as non-extended in all filters in Pan-STARRS1 \citep{Chambers2016}. To reduce contamination from stars misclassified as quasars in the photometric quasar catalog, we exclude targets that have non-zero total proper motions at $>3\sigma$. Then, we select 40 targets that have the largest \texttt{astrometric\_excess\_noise} ($>1.52$ mas) with {\it Gaia} $G$-band magnitude $<19.15$\,mag.  
    
    \item WISE-only quasars without SDSS and Pan-STARRS1 information and with a single {\it Gaia} match within 3\arcsec\ (20 targets selected). We select WISE quasars \citep{Mateos2012,Secrest2015,Marrese2019} that are not covered by Pan-STARRS1 \citep{Chambers2016}. Without the optical imaging information, this sample has a higher risk of being affected by extended host galaxies at low redshifts. Specifically, we find that if we select the WISE quasars with highest \texttt{astrometric\_excess\_noise} directly, our sample is dominated by low-redshift, extended galaxies upon inspecting their optical images from the DECam Legacy Survey \citep[DECaLS,][]{DECaLS}. 
    
    To improve the selection, we impose an additional criterion on {\tt phot\_bp\_rp\_excess\_factor}, which is a quality indicator for {\it Gaia} photometry \citep{Riello2018}. In {\it Gaia} DR2, G-band photometry is measured by PSF fitting, while the BP and RP photometry is the sum of fluxes in a $3\farcs5\times2\farcs1$ window \citep[no de-blending treatment,][]{Riello2018}. Therefore, the ratio between the G-band flux and the sum of BP- and RP-fluxes ({\tt phot\_bp\_rp\_excess\_factor}) can be used as a measure of how PSF-like a target is. Specifically, here we select WISE-only quasars with {\tt phot\_bp\_rp\_excess\_factor} $<1.5$. Then we select the targets with \texttt{astrometric\_excess\_noise} between 1.5 and 2.5 \,mas, {\it Gaia} proper motions consistent with 0 within 3$\sigma$, and {\it Gaia} $G$-band magnitude $<19.3$\,mag. We avoid \texttt{astrometric\_excess\_noise}$>2.5$\,mas because the images from DECaLS (if available) suggest that many of them may still be affected by extended host galaxies. A color cut of BP-RP$<2.5$ is also imposed because we find it particularly useful to avoid low-redshift ($z<0.5$) galaxies. Despite precautions, this WISE-only sample may have a high rate of contamination by stars and/or by low-redshift host galaxies. We thus prioritize the targets with larger \texttt{astrometric\_excess\_noise} and limit the number of the proposed \hst\ targets to 20. 
    
    \item SDSS quasars with multiple {\it Gaia} matches (20 targets selected). The selection is similar to the single-matched SDSS quasars, but here the targets have multiple {\it Gaia} matches within 3\arcsec. To reduce chance projection of foreground stars, we exclude sources that have non-zero parallaxes or non-zero proper motions with significance $>3\sigma$, and exclude those having obvious stellar features in the SDSS spectra. Two targets have three {\it Gaia} matches (J121135.93+354417.6 and J133039.82$-$001035.7), and the other targets have two {\it Gaia} matches. We prioritize targets with small angular separations (0\farcs52--0\farcs73), which are usually unresolved by ground-based surveys. Since this selection does not involve {\tt astrometric\_excess\_noise}, we do not impose any cuts on spectroscopic redshift and G-band magnitude.
    
    \item WISE+Pan-STARRS1 quasars with multiple {\it Gaia} matches (20 targets selected). The selection is similar to the single-matched WISE+Pan-STARRS1 quasars, but here the targets have multiple {\it Gaia} matches within 3\arcsec. We exclude sources that have non-zero parallaxes or proper motions with significance $>3\sigma$. We prioritize targets with small angular separations (0\farcs48--0\farcs66).
    
    \item  WISE-only quasars with multiple {\it Gaia} matches (10 targets selected). We exclude sources that have non-zero parallaxes or proper motions with significance $>3\sigma$, and we prioritize targets with small angular separations (0\farcs46--0\farcs60).
\end{enumerate}

\begin{table}
 \caption{Summary of \hst\ sample statistics. Listed are the number of dual/lensed quasar candidates selected using {\it Gaia} DR2, the number of candidates observed by {\it HST}, and the number of candidates containing multiple sources.}
 \label{tab:number_stat}
\addtolength{\tabcolsep}{-3pt}
 \begin{tabular}{lccc}
  \hline\hline
 Target Category  & \# targeted & \# observed & \# multiple \\
  \hline
  1. SDSS Single & 40 & 17 & 9\\
  2. WISE+PS1 Single & 40 & 26 & 7\\
  3. WISE-only Single & 20 & 13 & 2 \\
  4. SDSS Multiple & 20 & 7 & 7 \\    
  5. WISE+PS1 Multiple & 20 & 15 & 15 \\
  6. WISE-only Multiple & 10 & 6 & 5 \\
  \hline
  Total & 150 & 84 & 45 \\
  \hline
  \end{tabular}

\end{table}

\begin{figure}
  \centering
    \includegraphics[width=\columnwidth]{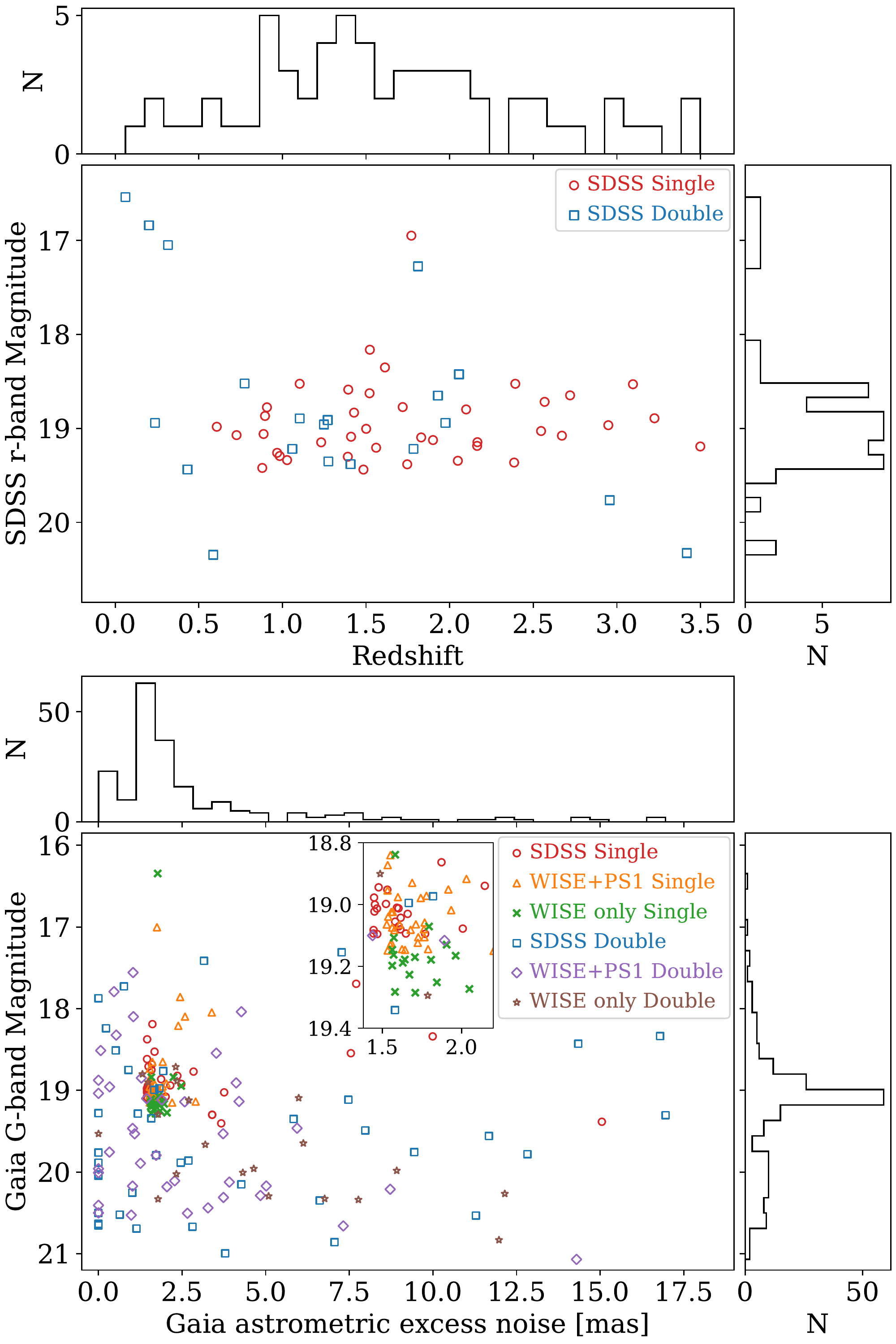}
    \caption{{\it Top:} Distribution of redshift and SDSS r-band PSF magnitude for the 60 targets selected from SDSS. {\it Bottom:} Distribution of the {\it Gaia} {\tt astrometric\_excess\_noise} and G-band magnitude for the final target sample of 150. 
    }
    \label{fig:mag_z_noise}
\end{figure}

Our final target sample for the \hst\ follow-up observations consists of 150 quasars. \autoref{tab:number_stat} summarizes the numbers of the proposed targets, the observed targets and the pairs discovered by the \hst\ in each category. The top panel of \autoref{fig:mag_z_noise} shows the distributions of redshift and SDSS $r$-band PSF magnitude for the 60 targets selected from SDSS. The bottom panel of \autoref{fig:mag_z_noise} shows the distributions of {\it Gaia} {\tt astrometric\_excess\_noise} and $G$-band  magnitude for the final target sample.

\subsection{{\it HST}/WFC3 F475W and F814W Imaging}

We observed 84 out of the 150 targets with the Wide Field Camera 3 (WFC3) on board the \hst\ in Cycle 27 and 28 from 2019 October 9 UT to 2021 September 23 UT (Program ID: SNAP-15900; PI: Hwang). Each target was imaged in the UVIS/F475W (with pivot wavelength $\lambda_{p}=$ 4773{\rm \AA\ } and effective width of 1344{\rm \AA\ }) and UVIS/F814W (with $\lambda_{p}=$ 8024{\rm \AA\ } and effective width of 1536{\rm \AA\ }) filters within a single \hst\ SNAP orbit to help distinguish stars and quasars. The typical net exposure times are 360s in the F475W filter and 400s in the F814W filter for each target and are scaled for fainter targets to reach similar signal-to-noise ratios. \autoref{tab:HST_obs_info} lists the spectroscopic and photometric properties and the observation details for all the observed targets.  Target names are in the form of ``hhmmss.ss$\pm$ddmmss.s'' based on {\it Gaia} DR2; the coordinates of the brightest source are used if multiple sources are detected in {\it Gaia}. Magnitudes are reported in ST magnitude ($m_{\rm ST}$ = $-2.5{\rm log}(F_\lambda)-$ 21.1), where $F_{\lambda}$ is in units of erg s$^{-1}$ cm$^{-2}$ \AA$^{-1}$.

The F475W filter covers rest-frame 1600--2200 {\rm \AA\ } and the F814W filter covers rest-frame 2800-3800 {\rm \AA\ } at the median redshift $z\sim 1.5$ of our targets. They sample the ultraviolet part of the quasar spectrum. We use these two bands to derive the F475W$-$F814W color to facilitate the interpretation of resolved pairs.

The WFC3-UVIS charge-coupled device (CCD) has a sampling of 0\farcs04 pixel$^{-1}$. The observations were dithered at 4 positions to properly sample the PSF and to reject cosmic rays and bad pixels. A 512$\times$512 sub-array was employed, yielding a field of view (FOV) of 20.5$''\times$20.5$''$. 

We reduced the \hst\ images following standard procedures. We used the \textsf{calwf3} and \textsf{MultiDrizzle} tasks from the STSDAS package in PyRAF. We processed the images with \textsf{MultiDrizzle} to correct for geometric distortion and pixel area effects. We combined the dithered frames and rejected cosmic rays and hot pixels. The resulting images are photometrically and astrometrically calibrated. \textsf{MultiDrizzle} relies on the positions of guide stars for absolute astrometric calibration. The absolute astrometric accuracy of the reduced images is limited by the positional uncertainty of guide stars ($\gtrsim$0\farcs2) and the calibration uncertainty of the fine guidance sensor to the instrument aperture ($\sim$0\farcs015). The image relative astrometric accuracy is limited by the uncertainty in the geometric distortion correction of the camera. The typical relative astrometric accuracy is 0\farcs004.

\subsection{Surface Brightness Decomposition}

We perform 2-dimensional (2-D) surface brightness decomposition with GALFIT \citep{Peng2010}. GALFIT fits the image with 2-D functions including PSF, S${\rm \acute{e}}$rsic profiles and various structures such as rings and spiral arms for galaxies and point sources.

The PSF model is constructed using a nearby isolated star in the same field for each target. For those targets without nearby field stars or when the field stars are too faint, we use a general model made from a bright isolated star in one of the visits. 

Since our SNAP observations are generally too shallow to detect faint features, a PSF model is often sufficient to fit each component of the resolved pair. A S${\rm \acute{e}}$rsic component with S${\rm \acute{e}}$rsic index $n=4$ is added in the fit if the residuals show significant extended features.

\subsection{Photometric Redshift }

For the 60 observed targets without a spectroscopic redshift, we estimate photometric redshifts (photo-$z$s) with Skewt-QSO\footnote{https://github.com/qian-yang/Skewt-QSO}. Skewt-QSO can achieve a photo-$z$ accuracy, defined as the fraction of quasars with the difference between the photo-$z$ and the spectroscopic redshift within 0.1, of 72\%--79\% using optical photometry from Pan-STARRS1 or DECaLS with {\it WISE} {\it W}1 and {\it W}2 data \citep{Yang2017}. We use optical photometric data from Pan-STARRS1, or {\it Gaia} if Pan-STARRS1 is not available, in combination with {\it WISE} {\it W}1 and {\it W}2 mid-infrared photometry. The photo-$z$s are labeled with dagger symbols in \autoref{tab:HST_obs_info} and \autoref{tab:double_info}. 

\startlongtable
\begin{deluxetable*}{lcccccccc}
 \tablecaption{Target properties and log of the \hst\ observations.
 \label{tab:HST_obs_info}}
 \tablehead{ \colhead{Name} & \colhead{Redshift} & \colhead{$r$} & \colhead{$G$} & \colhead{Ast. excess noise} & \colhead{Std.(Flux)}  & \colhead{Obs. date} & \colhead{Exp. time} & \colhead{Cat.}  \\ 
  &  & \colhead{[mag]} & \colhead{[mag]} & \colhead{[mas]} & \colhead{[e$^-$ s$^{-1}$]} & \colhead{[UT]} & \colhead{[s]} &   \\ 
 \colhead{(1)} & \colhead{(2)} & \colhead{(3)} & \colhead{(4)} & \colhead{(5)} & \colhead{(6)} & \colhead{(7)} & \colhead{(8)} & \colhead{(9)}
 }
 \startdata
 J000514.20+702249.2$^{\star}$ & 0.7$^{\dagger}$ & \nodata & 18.54/18.85 & 3.522/1.283 & 72.1/21.6 & 2020-01-24 & 360/320 & 5 \\
 J004127.74$-$734706.4 & 1.3$^{\dagger}$ & \nodata & 16.35 & 1.771 & 225.1 & 2019-11-14 & 40/40 & 3 \\
 J014859.18$-$553037.5$^{\star}$ & 0.1$^{\dagger}$ & \nodata &  18.71/18.90 & 2.309/1.486 & 33.7/46.7 & 2020-08-24 & 320/320 & 6 \\
 J023959.20$-$190507.5 & 0.75$^{\dagger}$ & \nodata & 19.11 & 1.766 & 35.1 & 2020-07-15 & 360/600 & 2 \\
 J024134.91+780107.0$^{\star}$ & 2.35$^{\dagger}$ & \nodata & 18.95 & 1.917 & 23.6 & 2020-01-31 & 400/320 & 2 \\
 J024205.24$-$453805.2 & 1.3$^{\dagger}$ & \nodata & 19.16 & 1.571 & 40.5 & 2020-08-28 & 360/400 & 3 \\
 J024628.63+692234.0$^{\star}$ & 0.95$^{\dagger}$ & \nodata & 18.04/19.12 & 4.272/1.892 & 12.6/16.0 & 2019-10-13 & 1080/80 & 5 \\ 
 J024804.31$-$281449.9& 1.9$^{\dagger}$ & \nodata & 18.98 & 1.598 & 34.4 & 2019-10-20 & 320/400 & 2 \\
 J024929.19$-$002104.3 & 1.428 & 18.84 & 18.71 & 1.497 & 48.4 & 2019-10-19 & 360/320 & 1 \\
 J025359.01$-$282653.6 & 2.0$^{\dagger}$ & \nodata & 19.02 & 1.566 & 48.0 & 2021-08-25 & 520/440 & 2 \\
 J025843.71+690543.4 & 0.1$^{\dagger}$ & \nodata & 18.92 & 2.031 & 37.9 & 2019-10-12 & 1320/240 & 2 \\
 J025956.93$-$735316.8 & 0.85$^{\dagger}$ & \nodata & 19.11 & 1.573 & 39.8 & 2019-11-09 & 360/480 & 3 \\
 J032233.97$-$551532.9 & 0.1$^{\dagger}$ & \nodata & 18.80/19.09\tablenotemark{a} & 1.314/5.986\tablenotemark{a} & 98.3/24.6\tablenotemark{a} & 2020-10-04 & 320/320 & 6 \\
 J034828.66$-$401513.1$^{\star}$ & 2.4$^{\dagger}$ & \nodata & 19.29 & 1.708 & 42.0 & 2020-09-30 & 400/440 & 3 \\
 J045528.99$-$445637.6$^{\star}$ & 0.5$^{\dagger}$ & \nodata & 19.98/20.83 & 8.917/11.975 & 27.7/15.0 & 2020-08-09 & 360/400 & 6 \\
 J045905.23$-$071407.1$^{\star}$ & 0.25$^{\dagger}$ & \nodata & 21.07/20.21 & 14.294/8.723 & 13.5/23.0 & 2020-02-05 & 360/400 & 5 \\
 J053620.23+503826.2$^{\star}$ & 1.8$^{\dagger}$ & \nodata & 17.86 & 2.442 & 94.2 & 2020-08-07 & 320/320 & 2 \\
 J054721.54$-$321309.5  & 0.3$^{\dagger}$ & \nodata & 19.04 & 1.537 & 50.0 & 2019-11-05 & 360/320 & 2 \\
 J055321.99+374004.5 & 0.35$^{\dagger}$ & \nodata & 19.12 & 1.556 & 68.5 & 2019-12-14 & 480/400 & 2 \\
 J055455.69+511252.8$^{\star}$ & 0.7$^{\dagger}$ & \nodata & 18.51/20.50 & 0.066/2.661 & 51.8/12.8 & 2020-02-23 & 320/320 & 5 \\
 J060155.15$-$071007.3  & 1.95$^{\dagger}$ & \nodata & 19.08 & 1.576 & 33.4 & 2019-12-09 & 480/400 & 2 \\
 J060734.57$-$064838.5$^{\star}$ & 2.3$^{\dagger}$ & \nodata & 18.10 & 2.587 & 91.1 & 2020-04-23 & 1320/80 & 2 \\
 J062026.53$-$071144.2 & 0.75$^{\dagger}$ & \nodata & 19.02 & 1.935 & 33.0 & 2020-01-30 & 360/400 & 2 \\
 J062903.08$-$753642.8$^{\star}$ & 0.3$^{\dagger}$ & \nodata & 19.65/20.02 & 6.126/2.329 & 72.0/25.6 & 2020-05-09 & 360/400 & 6 \\
 J074800.55+314647.4$^{\star}$ & 1.407 & 19.32 & 20.25/20.05 & 0.0/1.008 & 27.3/17.9 & 2020-04-18 & 360/400 & 4 \\
 J074817.13+191003.0$^{\star}$ & 3.096 & 18.49 & 18.94 & 2.147 & 78.8 & 2021-05-06 & 320/320 & 1 \\
 J074922.96+225511.7$^{\star}$ & 2.166 & 19.11 & 18.62 & 1.452 & 95.4 & 2020-01-05 & 320/400 & 1 \\
 J074930.93+505859.7 & 0.907 & 18.77 & 19.06 & 1.581 & 24.7 & 2019-12-01 & 360/480 & 1 \\
 J075350.57+424743.9$^{\star}$ & 1.523 & 18.14 & 18.19 & 1.613 & 64.5 & 2020-02-05 & 320/320 & 1 \\
 J075824.26+145752.4$^{\star}$ & 2.568 & 18.75 & 19.30 & 3.400 & 46.7 & 2020-09-28 & 360/400 & 1 \\
 J080414.01$-$443515.4 & 1.85$^{\dagger}$ & \nodata & 19.20 & 1.562 & 25.1 & 2020-07-31 & 360/400 & 3 \\
 J081603.79+430521.1$^{\star}$ & 0.586 & 20.08 & 20.86/20.67 & 7.062/2.810 & 32.5/20.2 & 2019-12-08 & 360/400 & 4 \\
 J082341.08+241805.6$^{\star}$ & 1.811 & 17.17 & 18.24/17.87 & 0.227/0.0 & 84.1/120.5 & 2020-01-16 & 320/320 & 4 \\
 J084129.77+482548.4$^{\star}$ & 2.949 & 18.87 & 19.30 & 3.399 & 37.4 & 2019-11-30 & 360/400 & 1 \\
 J090408.66+333205.2$^{\star}$ & 1.103 & 18.50 & 18.55 & 7.546 & 93.7 & 2020-01-24 & 320/320 & 1 \\
 J090501.12+585902.5$^{\star}$ & 2.386 & 19.33 & 18.53 & 1.676 & 63.6 & 2020-04-29 & 320/320 & 1 \\
 J091826.08+351243.1$^{\star}$ & 1.521 & 18.64 & 18.77 & 2.845 & 45.7 & 2021-01-13 & 360/320 & 1 \\
 J091938.93+621951.1$^{\star}$ & 1.270 & 18.81 &  20.52/18.75 & 0.633/0.898 & 12.8/34.5 & 2020-12-13 & 320/320 & 4 \\
 J094007.41+334609.5$^{\star}$ & 1.784 & 19.04 & 19.49/19.76 & 7.978/9.445 & 89.8/17.3 & 2020-01-07 & 360/400 & 4 \\
 J095611.62+425459.7 & 0.968 & 19.26 & 19.01 & 1.602 & 78.5 & 2019-10-09 & 520/600 & 1 \\
 J102531.00+414024.0 & 1.746 & 19.37 & 19.08 & 2.008 & 32.2 & 2020-06-16 & 360/400 & 1 \\
 J105929.04$-$503739.5$^{\star}$ & 0.25$^{\dagger}$ & \nodata & 18.84 & 2.243 & 32.5 & 2021-08-25 & 320/320 & 3 \\
 J111841.61+220015.9 & 1.394 & 18.59 & 18.98 & 1.448 & 38.3 & 2019-11-02 & 320/400 & 1 \\
 J113302.56+475623.0 & 0.984 & 19.29 & 18.95 & 1.531 & 33.8 & 2020-01-25 & 360/400 & 1 \\
 J122734.48$-$491202.5$^{\star}$ & 0.25$^{\dagger}$ & \nodata & 20.30/20.01 & 5.086/4.316 & 28.5/37.8 & 2020-07-21 & 360/400 & 6 \\
 J131416.01$-$491218.2$^{\star}$ & 2.29 & \nodata & 19.29/19.53 & 1.787/0.0 & 27.617.6 & 2019-12-26 & 400/440 & 6 \\
 J132614.08+063909.9 & 2.165 & 19.19 & 19.07 & 1.604 & 23.2 & 2020-01-05 & 360/400 & 1 \\
 J133040.00$-$125135.7 & 2.4$^{\dagger}$ & \nodata & 19.06 & 1.609 & 46.2 & 2019-12-18 & 360/480 & 2 \\
 J133952.42$-$074828.6 & 0.6$^{\dagger}$ & \nodata & 19.07 & 1.562 & 30.7 & 2020-02-24 & 360/400 & 2 \\
 J135531.37$-$084212.1 & 1.05$^{\dagger}$ & 18.02 & 18.84 & 1.551 & 35.5 & 2020-04-10 & 320/320 & 2 \\
 J140518.25$-$050607.7 & 1.2$^{\dagger}$ & \nodata & 19.11 & 1.727 & 50.8 & 2020-02-04 & 520/600 & 2 \\
 J142346.38$-$233844.7 & 0.15$^{\dagger}$ & \nodata & 18.98 & 1.745 & 31.0 & 2021-04-27 & 360/320 & 2 \\
 J151041.67+022134.3 & 1.232 & 19.15 & 18.94 & 1.477 & 28.3 & 2020-05-19 & 320/400 & 1 \\
 J155542.38$-$082826.9 & 0.3$^{\dagger}$ & \nodata & 19.02 & 1.556 & 30.7 & 2020-08-25 & 360/320 & 2\\
 J161349.51$-$264432.5$^{\star}$ & 1.1$^{\dagger}$ & \nodata & 19.15 & 2.203 & 84.7 & 2020-01-31 & 480/400 & 2 \\
 J162319.31$-$044703.2 & 1.55$^{\dagger}$ & \nodata & 18.95 & 1.533 & 50.8 & 2020-06-08 & 360/400 & 2 \\
 J164818.07+415550.1$^{\star}$ & 2.393 & 18.50 & 18.38 & 1.456 & 50.3 & 2021-06-03 & 320/320 & 1 \\
 J164941.29+081233.5$^{\star}$ & 1.4$^{\dagger}$ & \nodata & 18.32/19.46 & 0.530/1.020 & 76.2/17.4 & 2020-09-10 & 320/320 & 5 \\
 J171139.97$-$161147.9$^{\star}$ & 0.75$^{\dagger}$ & \nodata & 20.29/19.46 & 4.842/5.938 & 14.9/46.5 & 2020-05-06 & 360/400 & 5 \\
 J173222.88$-$133535.2$^{\star}$ & 0.292 &\nodata  & 19.08 & 1.762 & 49.1 & 2019-10-22 & 520/400 & 2 \\
 J175543.18+422924.2$^{\star}$ & 1.95$^{\dagger}$ & 17.66 & 20.44/17.79 & 3.275/0.464 & 16.2/80.9 & 2020-02-17 & 320/320 & 5 \\
 J180409.55+323029.5$^{\star}$ & 0.504 & \nodata & 19.14/19.76 & 4.203/0.327 & 63.8/17.6 & 2020-01-06 & 360/320 & 5 \\
 J183353.81$-$475524.9 & 0.65$^{\dagger}$ & \nodata & 19.23 & 1.671 & 30.0 & 2020-03-24 & 400/440 & 3 \\
 J184520.85$-$232227.6$^{\star}$ & 0.8$^{\dagger}$ & \nodata & 19.03/19.53 & 0.0/1.081 & 30.9/15.3 & 2021-09-23 & 320/320 & 5 \\
 J185226.10+483315.0$^{\star}$ & 1.480 & \nodata & 19.14 & 2.906 & 30.9 & 2020-02-06 & 360/400 & 2 \\
 J185728.65+704811.3$^{\star}$ & 1.230 & \nodata & 19.53/20.53 & 3.727/0.980 & 23.4/14.9 & 2019-10-20 & 360/400 & 5 \\
 J192843.97+643244.1 & 0.35$^{\dagger}$ & \nodata & 19.14 & 1.789 & 36.2 & 2019-11-02 & 520/400 & 2 \\
 J193547.95$-$554946.4 & 1.55$^{\dagger}$ & \nodata & 19.15 & 1.563 & 21.0 & 2020-03-17 & 360/400 & 3 \\
 J193711.89$-$163254.6 & 0.25$^{\dagger}$ & \nodata & 18.97 & 1.779 & 32.8 & 2020-03-15 & 360/400 & 2 \\
 J193718.81$-$182132.2$^{\star}$ & 1.2$^{\dagger}$ & \nodata & 20.31/20.11 & 3.736/2.279 & 14.0/15.5 & 2020-07-02 & 360/400 & 5 \\
 J194200.49$-$514724.2 & 0.65$^{\dagger}$ & \nodata & 19.18 & 1.807 & 51.7 & 2020-03-14 & 360/480 & 3 \\
 J194349.74$-$023819.1$^{\star}$ & 1.6$^{\dagger}$ &\nodata & 18.96/20.17 & 0.333/1.018 & 24.4/18.1 & 2020-06-27 & 320/320 & 5 \\ 
 J194859.87$-$341200.2 & 0.3$^{\dagger}$ & \nodata & 19.15 & 1.643 & 50.3 & 2020-03-18 & 360/240 & 2 \\
 J204848.00+625858.3$^{\star}$ & 2.5$^{\dagger}$ & \nodata & 20.41/20.18 & 0.0/2.047 & 17.8/16.5 & 2019-10-27 & 360/400 & 5 \\
 J205000.01$-$294721.7$^{\star}$ & 1.75$^{\dagger}$ & \nodata & 20.12/18.91 & 3.913/4.117 & 17.6/51.0 & 2020-03-21 & 360/320 & 5 \\
 J212243.01$-$002653.8 $^{\star}$ & 1.975 & 18.71 & 19.89/19.34 & 0.0/1.580 & 15.0/40.6 & 2019-12-04 & 360/400 & 4 \\
 J215444.04+285635.3$^{\star}$ & 0.723 & 19.62 & 20.35/19.35 & 6.615/5.831 & 17.4/56.1 & 2019-12-02 & 360/400 & 4 \\
 J215732.70$-$510730.4 & 1.15$^{\dagger}$ & \nodata & 19.25 & 1.844 & 24.0 & 2020-05-10 & 400/440 & 3 \\
 J221849.86$-$332243.6$^{\star}$ & 0.35$^{\dagger}$ & \nodata & 20.66/20.17 & 7.320/5.016 & 19.8/27.6 & 2020-08-21 & 360/400 & 5\\
 J230223.02$-$423647.8 & 1.0$^{\dagger}$ & \nodata & 19.13 & 1.906 & 26.9 & 2020-08-22 & 400/400 & 3 \\
 J232157.47$-$113055.2 & 0.75$^{\dagger}$ & \nodata & 19.12 & 1.724 & 29.2 & 2020-08-24 & 360/480 & 2 \\
 J232412.70+791752.3$^{\star}$ & 0.4$^{\dagger}$ & \nodata & 17.00 & 1.749 & 103.3 & 2020-01-30 & 320/320 & 2 \\
 J233643.38$-$480152.3 & 0.8$^{\dagger}$ & \nodata & 19.18 & 1.642 & 57.5 & 2020-08-28 & 360/400 & 3 \\
 J235128.30$-$503418.5 & 0.6$^{\dagger}$ & \nodata & 19.28 & 1.580 & 29.5 & 2020-07-31 & 400/440 & 3 \\
 \enddata
 \tablecomments{Col. 1: Coordinate names in the form of ``hhmmss.ss$\pm$ddmmss.s'' based on {\it Gaia} DR2; the coordinates of the brightest source are used when multiple sources are detected in {\it Gaia}. Targets with multiple sources in \hst\ are marked with a star.} Col. 2: Spectroscopic redshift or photometric redshift when denoted with ``$\dagger$''. Col. 3: SDSS $r$-band PSF magnitude. Col. 4: {\it Gaia} $G$-band mean magnitude. Col. 5: {\it Gaia} {\tt astrometric\_excess\_noise}. Col. 6: Standard deviation of {\it Gaia} $G$-band flux (={\tt phot\_g\_mean\_flux\_error}$\times$({\tt phot\_g\_n\_obs})$^{\frac{1}{2}}$). Col. 7: \hst\ Observation date. Col. 8: \hst\ Exposure time in F475W and F814W filters, respectively. Col. 9: Target category (see Table~\ref{tab:number_stat}). \tablenotetext{a}{Though {\it Gaia} DR2 detects two sources in J0322$-$5515, the \hst\ observations only reveal one source. Recent {\it Gaia} EDR3 \citep{GaiaEDR3} only shows one source in J0322$-$5515, which suggests that the second source in {\it Gaia} DR2 is an artifact.} 
\end{deluxetable*}
\clearpage
\startlongtable  
\begin{deluxetable*}{lccccccccc}
\addtolength{\tabcolsep}{-2.5pt}
 \tablecaption{Properties of the 45 \hst\ targets with multiple cores.
 \label{tab:double_info}}
 \tablehead{ \colhead{Name} & \colhead{Redshift} & \colhead{$\Delta\theta$} & \colhead{$r_p$} & \colhead{P.A.} & \colhead{F475W}  & \colhead{F814W} & \colhead{Cat.} & \colhead{Classification}\\ 
  &  & \colhead{[arcsec]} & \colhead{[kpc]} & \colhead{[deg]} & \colhead{[mag]} & \colhead{[mag]}  \\ 
 \colhead{(1)} & \colhead{(2)} & \colhead{(3)} & \colhead{(4)} & \colhead{(5)} & \colhead{(6)} & \colhead{(7)} & \colhead{(8)} & \colhead{(9)}
 }
 \startdata
J000514.20+702249.2 & 0.7$^{\dagger}$ & 0.58 & 4.1 & 279.8 & 19.81/20.08 & 19.02/19.08 & 5 & dual/lensed quasar \\
J014859.18$-$553037.5 & 0.1$^{\dagger}$ & 0.56 & 1.0 & 181.6 & 18.51/20.16 & 19.82/18.88 & 6 & quasar+star\\
J024134.91+780107.0 & 2.35$^{\dagger}$ & 0.58 & 4.7 & 126.8 & 19.44/23.19 & 19.56/21.36 & 2 & quasar+star\\
J024628.63+692234.0 & 0.95$^{\dagger}$ & 0.49 & 3.9 &  49.2 & 20.90/22.59 & 17.93/18.89  & 5 &  quasar+star\\ 
J034828.66$-$401513.1 & 2.4$^{\dagger}$ & 0.50 & 4.1 & 186.7 & 19.26/21.07 & 20.20/21.96 & 3 & dual/lensed quasar\\
J045528.99$-$445637.6 & 0.5$^{\dagger}$ & 0.56 & 3.4 & 93.1 & 20.99/21.10/22.87 & 21.99/22.44/24.00 & 6 & triple quasar\tablenotemark{d}\\
J045905.23$-$071407.1 & 0.25$^{\dagger}$ & 0.63 & 2.5 & 154.6 & 21.21/21.98 & 22.34/22.39 & 5 & dual quasar\tablenotemark{d} \\
J053620.23+503826.2 & 1.8$^{\dagger}$ & 0.32 & 2.7 & 162.3 & 18.68/20.01 & 18.52/19.90 & 2 & dual/lensed quasar\\
J055455.69+511252.8 & 0.7$^{\dagger}$ & 0.67 & 4.8 & 77.4 & 18.56/21.36 & 19.33/20.94  & 5 & quasar+star\\
J060734.57$-$064838.5 & 2.3$^{\dagger}$ & 0.25 & 2.0 & 196.3 & 21.29/23.87 & 18.11/19.69 & 2 & quasar+star\\
J062903.08$-$753642.8 & 0.3$^{\dagger}$ & 0.51 & 2.3 & 61.1 & 19.96/22.04/24.10 & 21.12/20.61/22.28 & 6 & quasar+stars \\
J074800.55+314647.4 & 1.407 & 0.53 & 4.5 & 171.1 & 20.13/20.54 & 21.04/20.58 & 4 & dual/lensed quasar \\
J074817.13+191003.0 & 3.096 & 0.40 & 3.1 & 52.1 & 18.93/20.92 & 19.78/19.83 & 1 & quasar+star \\
J074922.96+225511.7 & 2.166 & 0.46 & 3.8 & 56.1 & 19.18/20.78 & 20.02/21.39 & 1 & \textbf{dual quasar\tablenotemark{a}} \\
J075350.57+424743.9 & 1.523 & 0.33 & 2.8 & 306.9 & 18.06/21.49 & 18.93/20.43 & 1 & quasar+star\\
J075824.26+145752.4 & 2.568 & 0.50 & 4.0 & 69.0 & 19.58/21.90/23.57 & 20.07/21.54/21.93  & 1 & quasar+stars\\
J081603.79+430521.1 & 0.586 & 0.72 & 4.8 & 275.3 & 20.61/22.34 & 21.75/20.65 & 4 &  \textbf{quasar+star\tablenotemark{c}} \\
J082341.08+241805.6 & 1.811 & 0.64 & 5.4 & 183.2 & 18.19/18.58 & 18.54/18.86 & 4 & dual/lensed quasar \\
J084129.77+482548.4 & 2.949 & 0.46 & 3.5 & 132.0 & 19.48/19.81 & 20.04/20.61 & 1 & \textbf{dual quasar\tablenotemark{a}} \\
J090408.66+333205.2 & 1.103 & 0.30 & 2.5 & 104.2 & 18.65/20.69 & 19.75/20.08  & 1 &  quasar+star\\
J090501.12+585902.5 & 2.386 & 0.61 & 5.0 & 228.0 & 18.62/22.87 & 19.72/21.36 & 1 & quasar+star \\
J091826.08+351243.1 & 1.521 & 0.18 & 1.6 & 260.3 & 19.13/21.92 & 19.75/20.10  & 1 & quasar+star \\
J091938.93+621951.1 & 1.270 & 0.69 & 5.7 & 333.3 & 19.11/22.61 & 19.65/20.39 & 4 & quasar+star \\
J094007.41+334609.5 & 1.784 & 0.74 & 6.2 & 239.2 & 19.81/20.89 & 20.31/20.07 & 4 & quasar+star\\
J105929.04$-$503739.5 & 0.25$^{\dagger}$ & 0.54 & 2.1 & 223.2 & 19.00/22.62 & 19.45/21.20 & 3 & quasar+star \\
J122734.48$-$491202.5 & 0.25$^{\dagger}$ & 0.50 & 1.9 & 301.8 & 20.71/21.28 & 20.74/20.89 & 6 & quasar+star\\
J131416.01$-$491218.2 & 2.29 & 0.52 & 4.2 & 160.5 & 19.38/20.93/24.26 & 20.13/19.65/22.27 & 6 & \textbf{quasar+stars\tablenotemark{b}} \\
J161349.51$-$264432.5 & 1.1$^{\dagger}$ & 0.28 & 2.3 & 250.5 & 20.02/23.19 & 20.78/21.67 & 2 & quasar+star \\
J164818.07+415550.1 & 2.393 & 0.44 & 3.5 & 39.9 & 18.40/21.02 & 19.44/21.79 & 1 & dual/lensed quasar \\
J164941.29+081233.5 & 1.4$^{\dagger}$ & 0.59 & 5.0 & 1.9 & 19.07/19.73 & 19.55/20.26 & 5 & dual/lensed quasar\\
J171139.97$-$161147.9 & 0.75$^{\dagger}$ & 0.67 & 4.9 & 128.6 & 20.69/20.99/24.87 & 20.34/20.76/22.66 & 5 & dual/lensed quasar+star\\
J173222.88$-$133535.2 & 0.292& 0.72 & 3.2 & 329.0 & 19.98/21.37 & 19.43/21.08 & 2 & \textbf{quasar+star\tablenotemark{b}} \\
J175543.18+422924.2 & 1.95$^{\dagger}$ & 0.59 & 5.0 & 12.0 & 17.72/21.85 & 18.57/20.86 & 5 & quasar+star\\
J180409.55+323029.5 & 0.504 & 0.68 & 3.6 & 181.2 & 19.09/20.68 & 20.23/20.15 & 5 & \textbf{quasar+star\tablenotemark{b}} \\
J184520.85$-$232227.6 & 0.8$^{\dagger}$ & 0.66 & 5.0 & 170.4 & 19.51/20.04 & 19.88/20.02 & 5 & dual/lensed quasar \\
J185226.10+483315.0 & 1.480 & 0.62 & 5.2 & 190.6 & 18.94/22.49 & 19.80/21.65 & 2 & \textbf{quasar+star\tablenotemark{b}}\\
J185728.65+704811.3 & 1.230 & 0.61 & 5.0 & 160.4 & 19.87/23.15 & 20.47/20.27 & 5 &  \textbf{quasar+star\tablenotemark{b}} \\
J193718.81$-$182132.2 & 1.2$^{\dagger}$ & 0.62 & 5.2 & 188.4 & 20.25/20.59 & 21.09/20.98 & 5 & dual/lensed quasar \\
J194349.74$-$023819.1 & 1.6$^{\dagger}$ & 0.66 & 5.6 & 174.7 & 19.25/21.04 & 19.79/20.57 & 5 & quasar+star \\ 
J204848.00+625858.3 & 2.5$^{\dagger}$ & 0.62 & 5.0 & 297.1 & 20.33/20.87 & 20.58/20.94 & 5 & dual/lensed quasar\\
J205000.01$-$294721.7 & 1.75$^{\dagger}$ & 0.65 & 5.5 & 280.4 & 19.18/20.64 & 19.59/20.67 & 5 &  dual/lensed quasar\\
J212243.01$-$002653.8 & 1.975 & 0.52 & 4.4 & 313.8 & 19.20/20.38 & 19.96/20.48 & 4 & dual/lensed quasar\\
J215444.04+285635.3 & 0.723 & 0.68 & 4.9 & 193.0 & 19.53/20.89 & 20.90/20.98  & 4 & quasar+star \\
J221849.86$-$332243.6 & 2.0\tablenotemark{e} & \nodata & \nodata & \nodata & 21.11/21.25/21.89/22.73  & 21.10/21.30/21.77/22.68 & 5 & quad lens\\
J232412.70+791752.3 & 0.4$^{\dagger}$ & 0.49 & 2.7 & 310.5 & 17.58/18.93 & 17.55/19.57 & 2 & dual/lensed quasar\\
 \enddata
 \tablecomments{Col. 1: J2000 coordinates in the form of ``hhmmss.ss$\pm$ddmmss.s''. Col. 2: Spectroscopic redshift or photometric redshift when denoted with ``$\dagger$''. Col. 3: Angular separation. Col. 4: Projected physical separation based on the redshift. Col. 5: Position Angle between the two brightest sources  in degree east of north. Col. 6: \hst\ F475W ST magnitude of each source. Col. 7: \hst\ F814W ST magnitude of each source. Col. 8: Target Category. Col. 9: Best-effort classification. The term ``stars" in classification means more than one PSF components with star-like color, instead of an extended source. Targets which are confirmed spectroscopically are marked in bold.}
 \tablenotetext{a}{See detailed discussion in \citet{Shen2021}.}
 \tablenotetext{b}{The classifications are based on the spatially resolved optical spectra from Gemini.}
 \tablenotetext{c}{The classification is based on the unresolved SDSS spectra.}
 \tablenotetext{d}{The targets reveal irregular merger features.}
 \tablenotetext{e}{Using a fiducial redshift of 2.0 for the quadruply lensed quasar. The photo-z is biased due to the contamination of the foreground galaxy}
\end{deluxetable*}

\subsection{Initial Optical Spectroscopic Follow-up Observations}

We are conducting follow-up observations of \hst-resolved pairs, and will present the final results of our follow-up observations in a future paper. Here we present confirmation and redshift measurements from ground-based Gemini optical spectroscopy for the subset of targets that have been followed-up so far.

The Gemini GMOS spectroscopic follow-ups (GN-2020A-DD-106 and GS-2020A-DD-106; PI: Liu, GN-2020A-Q-232; PI: Chen) were conducted between February 2020 and August 2020 for seven targets. The spectra cover the observed wavelength of 4000\AA--10040\AA~ with a resolving power $R$ of 421. We are able to decompose the (marginally) spatially resolved sources and obtain the spectra of each source for six targets, except for J0904+3332 due to its small pair separation of 0\farcs3. Out of the six targets, one is the a dual quasar reported in \citet{Shen2021} and five are star-quasar superpositions, whose spectra are shown in \autoref{fig:spectra_star}.




\section{Results}\label{sec:results}

\begin{figure*}
\centering
 \includegraphics[width=\textwidth]{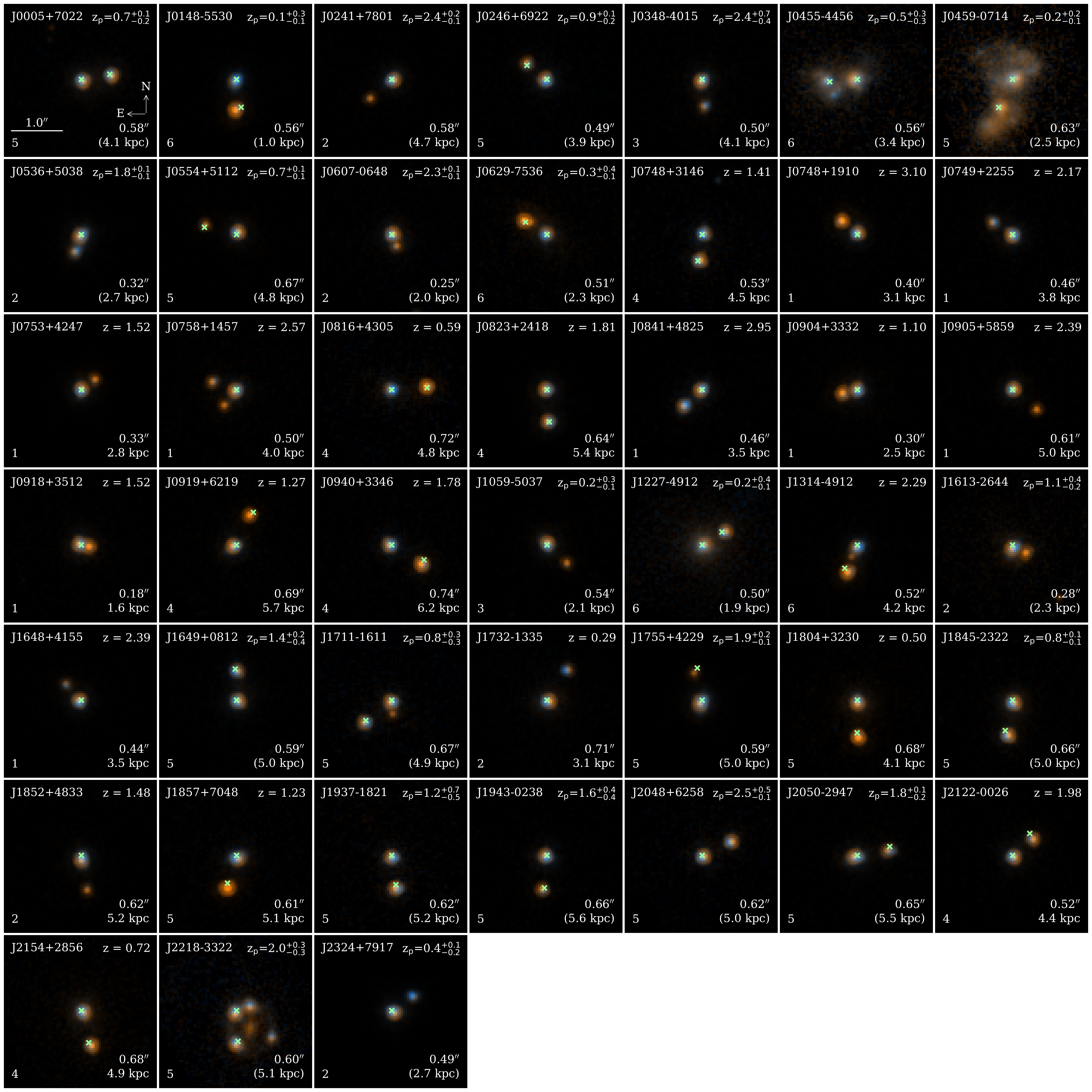}
 \caption{
\hst/WFC3 color composite images (F475W in blue, F814W in red, and the average of F475W and F814W in green) of the 45 targets showing multiple sources in the \hst\ images. North is up and east is to the left. The {\it Gaia} DR2 source positions are marked by the green crosses. Labeled in each panel are the angular and physical separations between the double sources (physical separations calculated using photo-zs are in parenthesises). The target category of each source is labeled in the bottom-left corner. 
}
 \label{fig:hst_double}
\end{figure*}

\begin{figure*}
\centering
 \includegraphics[width=1.0\textwidth]{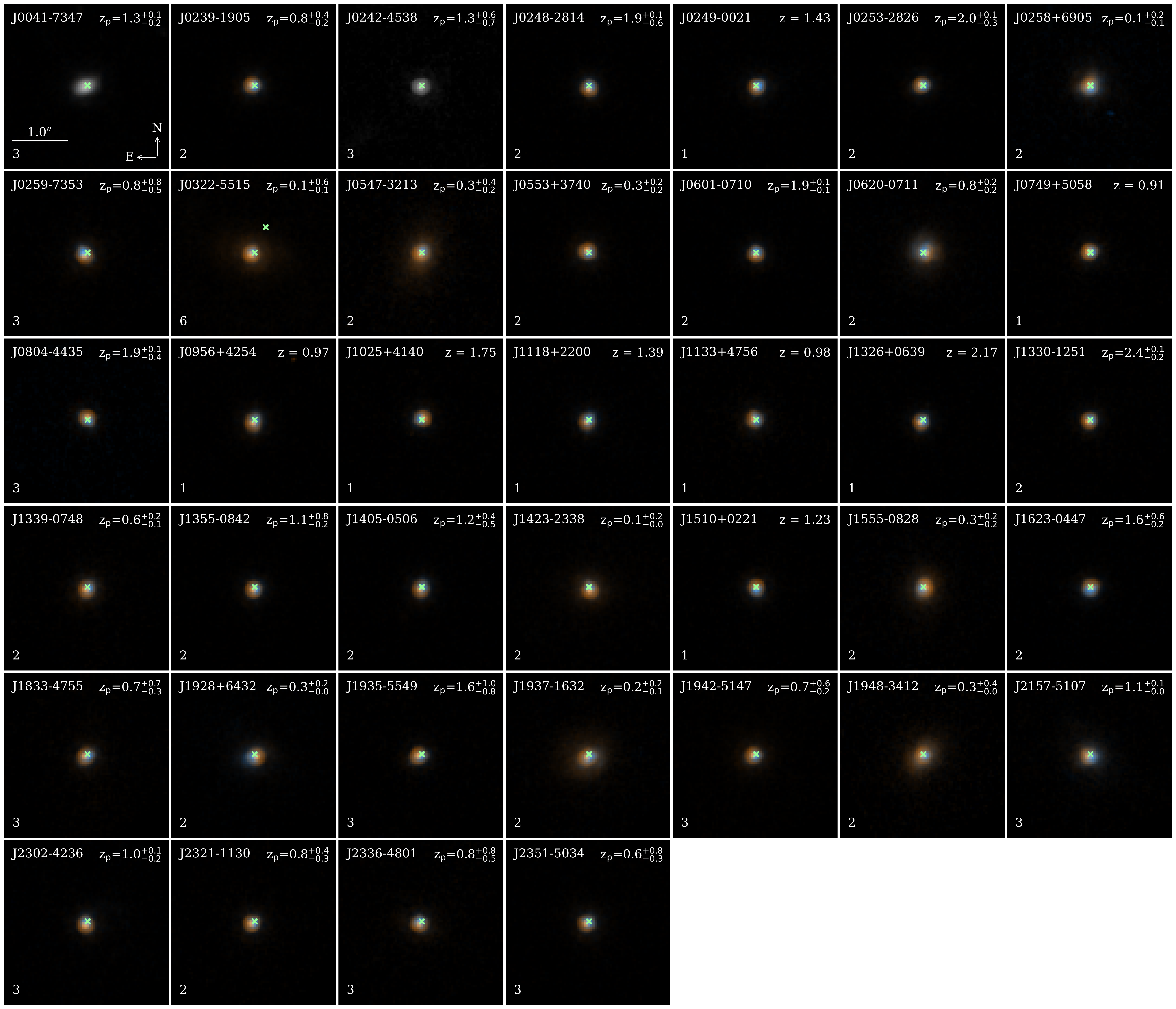}
 \caption{
Same as Fig.~\ref{fig:hst_double}, but for the 39 targets that appear single in \hst\ imaging.}
 \label{fig:hst_single}
\end{figure*}

We present our \hst\ follow-up observations of the 84 observed targets. \autoref{fig:hst_double} shows the \hst\ color-composite images of the 45 targets that have sub-arcsec multiple cores resolved at the \hst\ resolution. \autoref{fig:hst_single} shows the \hst\ images of the remaining 39 targets which have a single unresolved core with \hst. Among the 45 quasars with multiple cores in \hst, 26 have multiple {\it Gaia} source matches and 19 have a single matched source in {\it Gaia}. 

We perform GALFIT decomposition of the \hst-resolved targets, and compile the decomposed PSF fluxes in \autoref{tab:double_info}. We classify each target based on the morphology, color information as well as follow-up spectroscopy. For the 39 single \hst-unresolved sources, the target is classified as ``unresolved single'' if a single PSF can well fit the surface brightness ($\chi^2_{\nu}$ $<$ 1.3 in the 6{\arcsec}$\times$6{\arcsec} F814W GALFIT region); otherwise the target is classified as ``extended host'' when additional S${\rm \acute{e}}$rsic components are needed to reduce $\chi^2_{\nu}$.  For the 45 quasars showing multiple sources in \hst, we use color information of each core to distinguish star-quasar superposition (see \S\ref{subsec:star-quasar} for details) along with follow-up spectroscopy for several targets. If the system has a red color in either core or has a large color difference, it is classified as ``quasar+star'' and as ``dual/lensed quasar'' if the colors of both cores are quasar-like.

In addition to the classifications defined above, we also add ``dual quasar'', which include the two confirmed dual quasars (J0749+2255 and J0841+4825) from \citet{Shen2021} and a dual quasar candidate (J0459$-$0714) showing a tidal feature, ``triple quasar'', which includes a triple quasar candidate (J0455$-$4456) showing a tidal feature, and ``quad lens", which includes a quadruply lensed quasar (J2218$-$3322) with the Einstein ring and the central lens galaxy. The classification for each of those targets is discussed in detail in \S\ref{sec:dualvslens}.


\begin{figure}
  \centering
    \includegraphics[width=\columnwidth]{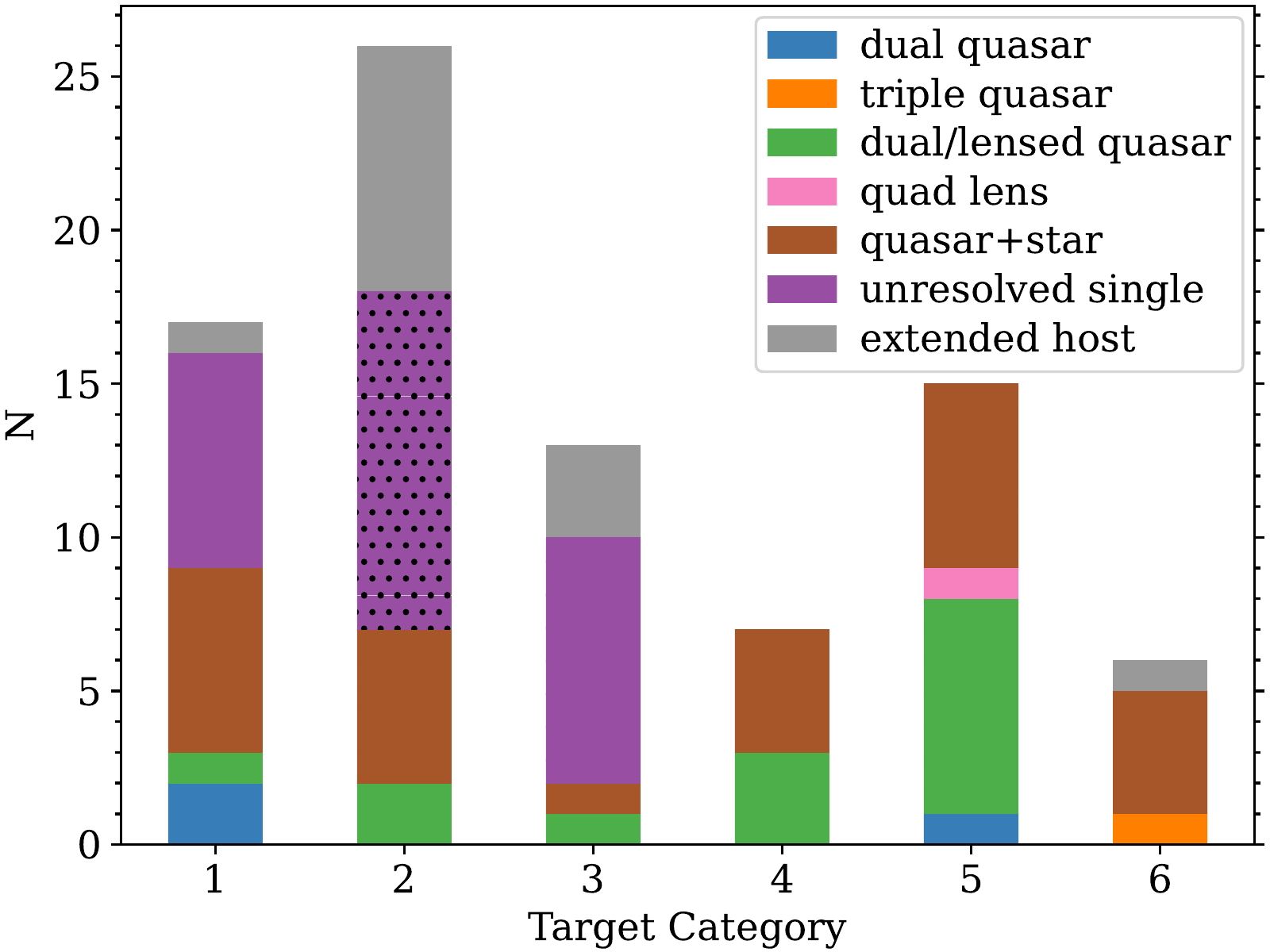}
    \caption{Numbers of targets for each classification in the six target categories. The six target categories are: 1. SDSS Single, 2. WISE+PS1 Single, 3. WISE-only Single, 4. SDSS Multiple, 5. WISE+PS1 Multiple, and 6. WISE-only Multiple. The details of each target category are described in \autoref{sec:target_selection}.
    }
    \label{fig:N_class}
\end{figure}



Our \hst\ program has revealed a large fraction of resolved sub-arcsec pairs in our {\it Gaia}-selected quasar targets; additional follow-up observations are required to confirm the nature of these resolved pairs. We summarize our source classifications for the six target categories in \autoref{fig:N_class}. Below we provide our best effort in classifying these resolved pairs based on the existing data. 

\subsection{Chance Superposition with Stars}
\label{subsec:star-quasar}

\begin{figure}
  \centering
    \includegraphics[width=\columnwidth]{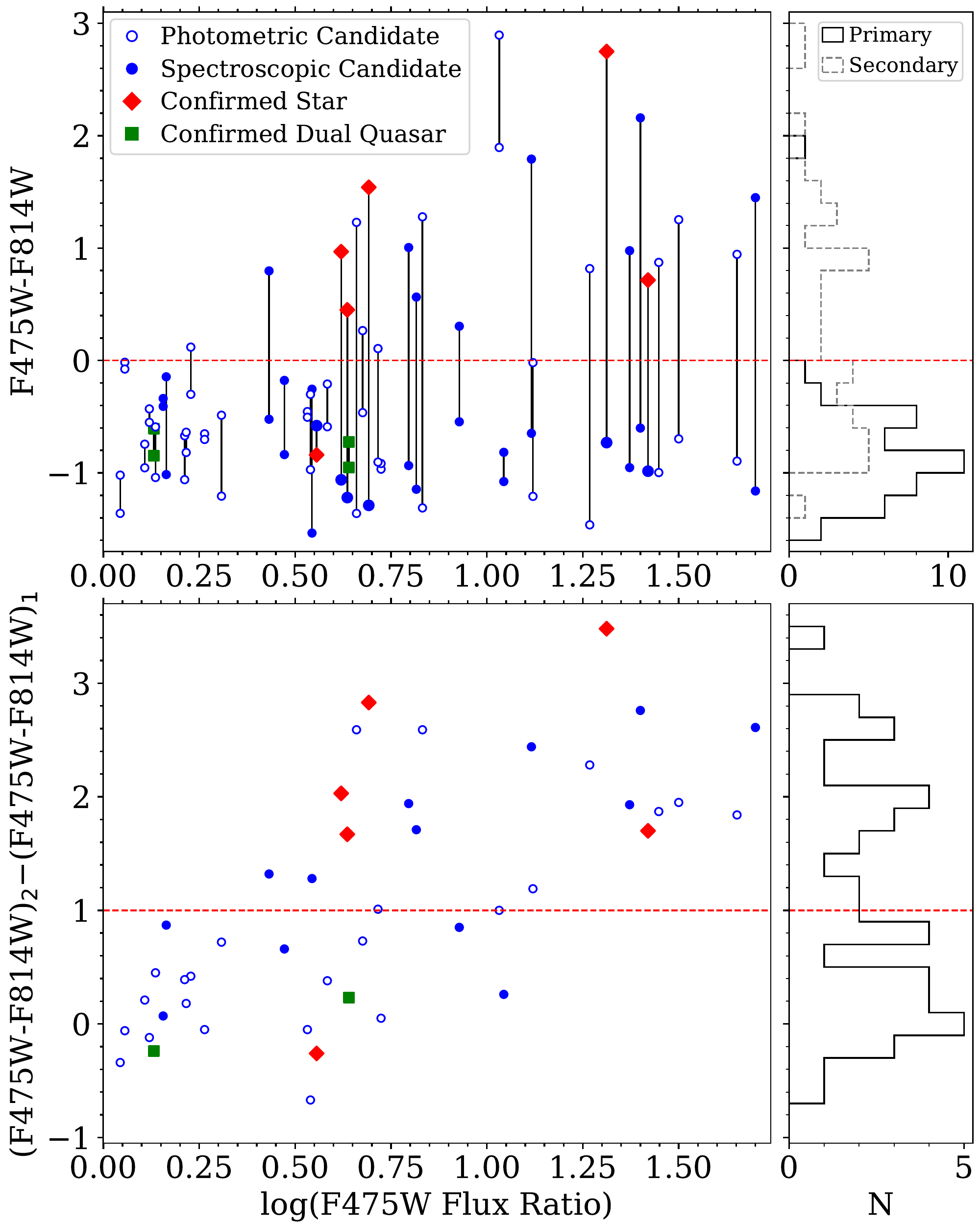}
    \caption{{\it Top:} F475W$-$F814W color vs. F475W flux ratio of two sources for the 45 targets with \hst-resolved multiple sources. The black vertical lines connect two sources in the same system. {\it Bottom:} F475W$-$F814W color difference vs. F475W flux ratio of the two sources. For targets with more than two sources, we only plot the two brightest sources. The red diamonds represent the spectroscopically confirmed star-quasar superpositions, the green squares represent the two confirmed dual quasars in \citet{Shen2021}, and the filled (open) circles represent the systems with spectroscopic (photometric) redshift. The red dashed lines represent the two color selection criteria for identifying stars.
    }
    \label{fig:cands_color}
\end{figure}

The dual-band \hst\ colors can be used to assess the likelihood of star-quasar superposition. If the resolved pairs have similar optical colors, they are more likely to be both quasars (either a dual quasar or a lensed quasar). However, there are exceptions where the physical quasar pairs can have different colors such as the $z$=1.285 dual quasar (with a separation of 1\farcs67) SDSS J233646.2$-$010732.6, which consists of a standard quasar with a blue continuum and a broad absorption line (BAL) quasar \citep{Gregg02}. 

We use the F475W$-$F814W color to identify possible star-quasar superpositions. Given some of the targets are at low Galactic latitudes, we correct for the Galactic extinction with {\sc dustmaps} \citep{Green2018} using the Schlegel, Finkbeiner \& Davis (SFD) map of interstellar dust \citep{Schlegel1998} and convert the $E(B-V)$ values in SFD to extinctions in the \hst\ F475W and F814W filers using Table 6 in \citet{Schlafly2011} for $R_V=3.1$. The de-reddened F475W$-$F814W colors of each source for the 45 \hst-resolved pairs are shown in the top panel of \autoref{fig:cands_color}. The color differences (color of the fainter source minus color of the brighter source) within each pair are presented in the bottom panel of \autoref{fig:cands_color}.

We use two criteria to identify possible star-quasar superpositions:
\begin{enumerate}
    \item F475W$-$F814W $>$ 0 mag, OR
    \item (F475W$-$F814W)$_2-$(F475W$-$F814W)$_1 >$ 1 mag,
\end{enumerate}
where 1 is the brightest source in F475W and 2 is the second brightest (or faintest) source in F475W. The system is classified as a star-quasar pair if it meets either criterion. For the few targets with three sources, we use the F475W$-$F814W color of each source and the color difference between the brightest source and the second brightest (or faintest) source to identify possible star contamination.

We convolve quasar's SED templates from \citet{VandenBerk2001,Glikman2006} and stellar templates from \citet{Pickles1998} with \hst\ bandpasses using {\sc pysynphot} to produce the synthetic F475W$-$F814W colors of canonical quasars and different types of stars. The colors of quasars at $z=0-3$ are between $-0.3$ and $-1.0$ mag. The first criterion separates quasars from foreground K and M type stars. Though we cannot separate F and G type stars from quasars due to similar continuum slope, they should be much rarer than K and M type stars \citep{Kroupa1993,Kirkpatrick2012,Bovy2017}. The second criterion removes those with large color discrepancy, which are likely to be star-quasar superpositions. The caveat of the color selections is that it will remove real quasar pairs with different colors such as normal+BAL quasars. It is also noted that some stars might be unresolved compact stellar clusters in a companion galaxy, whose two-band HST colors may be similar to those of single stars. 

\begin{figure*}
  \centering
    \includegraphics[width=0.325\textwidth]{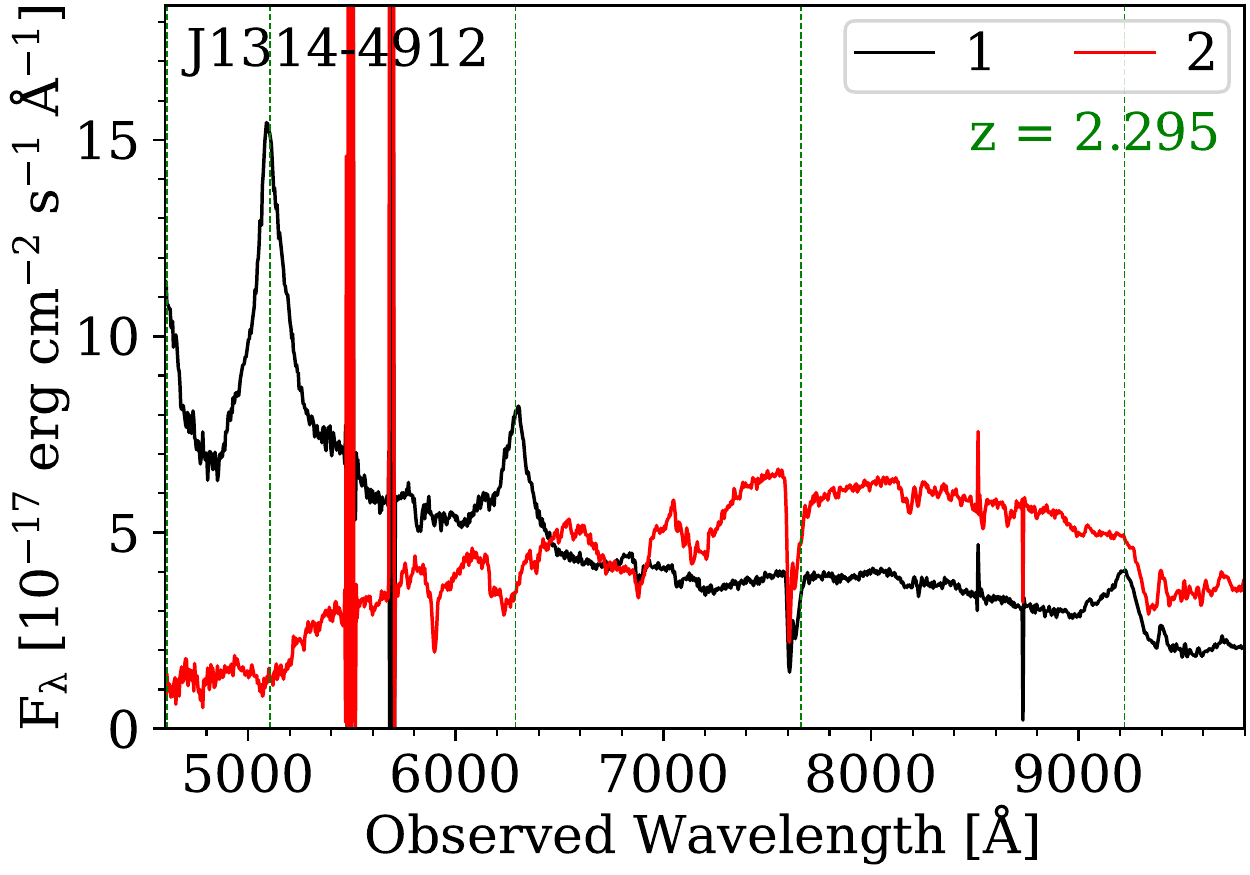}
    \includegraphics[width=0.325\textwidth]{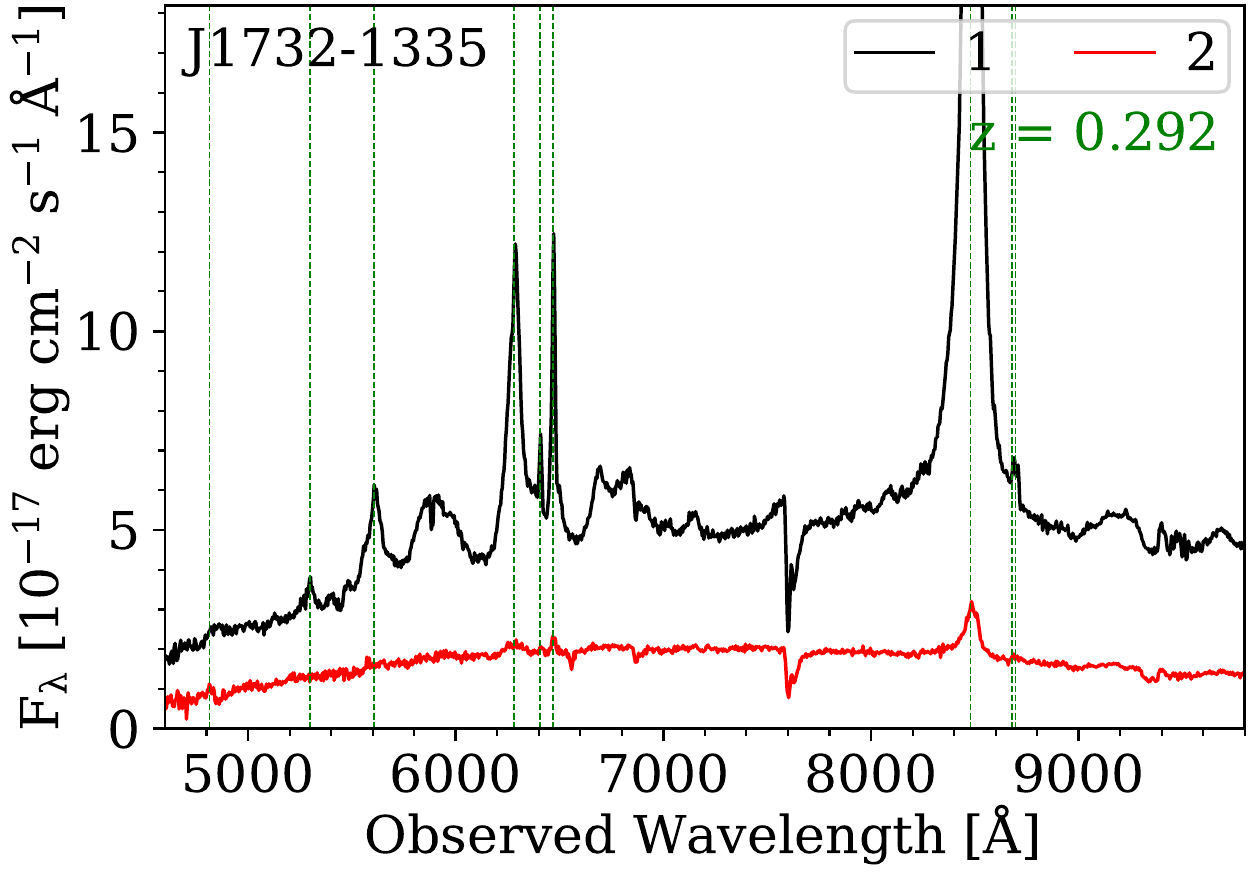}
    \includegraphics[width=0.325\textwidth]{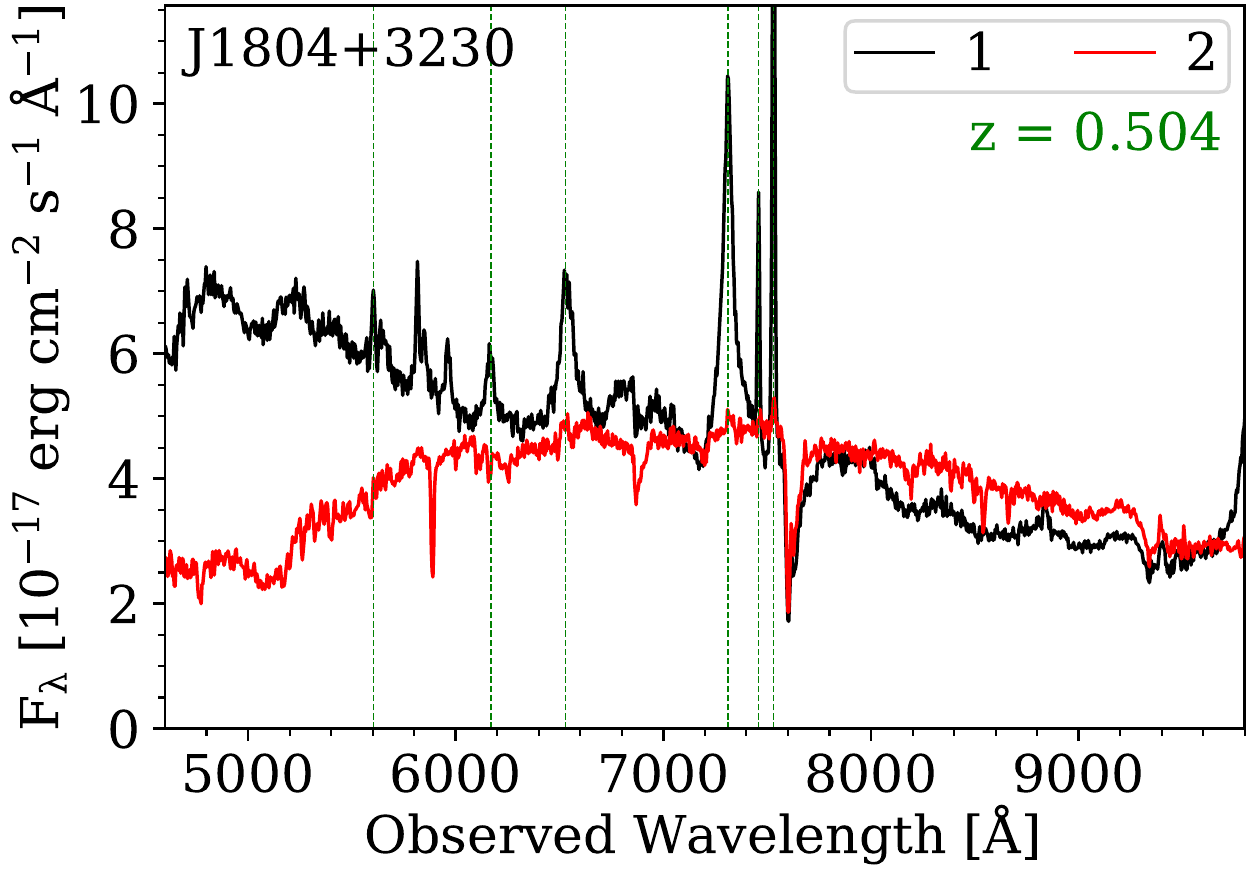}
    \includegraphics[width=0.325\textwidth]{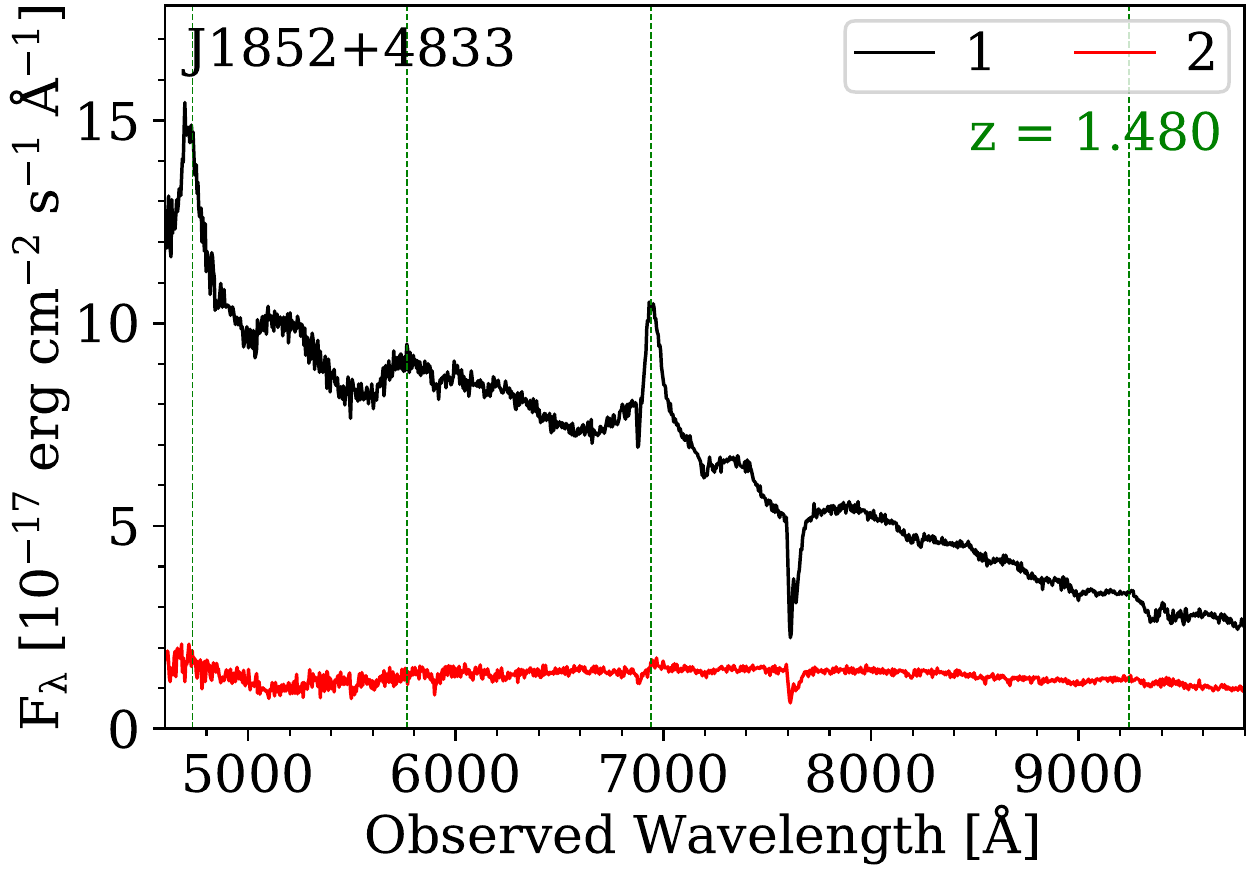}
    \includegraphics[width=0.325\textwidth]{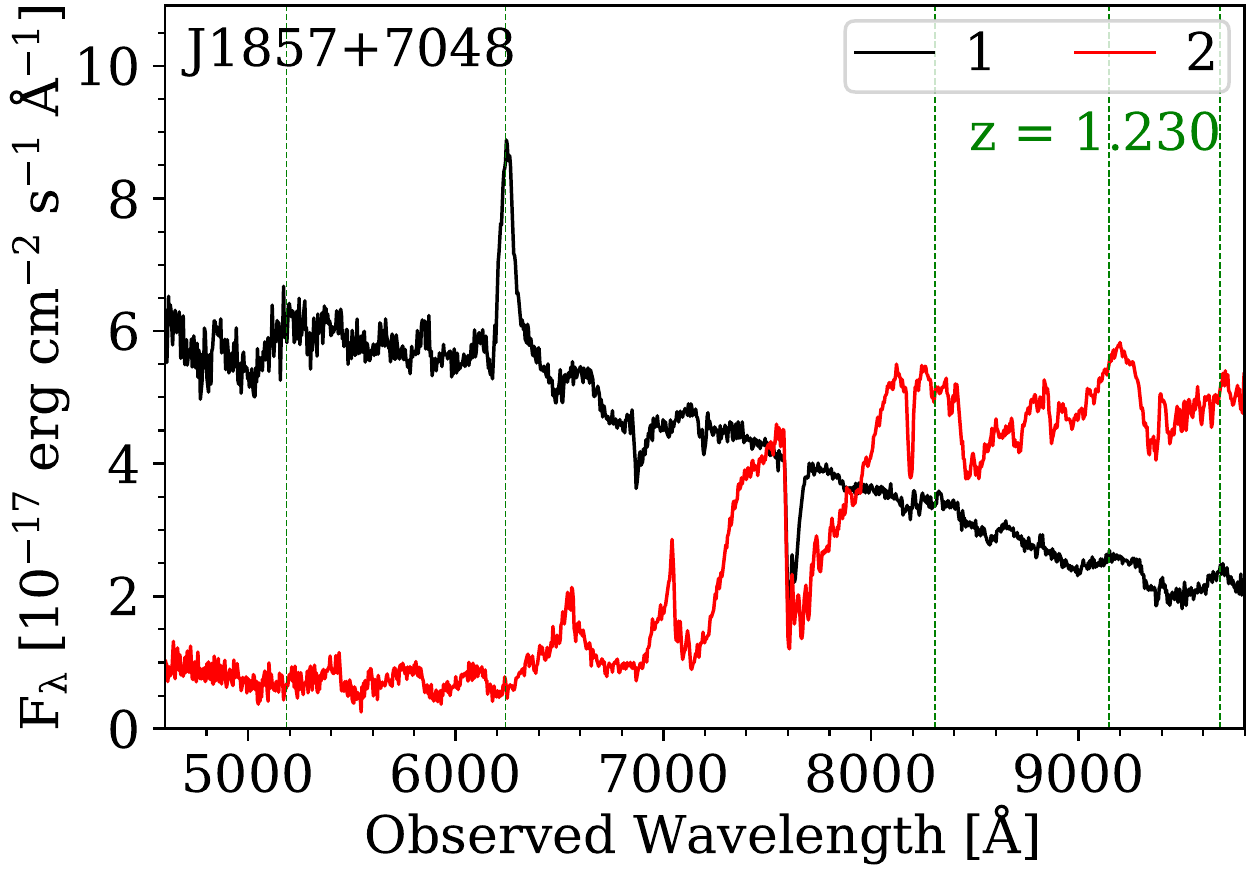}
    \caption{ Gemini GMOS optical spectra of the 5 confirmed star-quasar superpositions. Quasar's spectra are in black and star's spectra are in red. The green vertical lines mark the quasar's emission lines at the given redshift shown in the top-right corner.
    }
    \label{fig:spectra_star}
\end{figure*}

There are five star-quasar superpositions confirmed by Gemini long-slit spectroscopy, shown in \autoref{fig:spectra_star}. The stellar absorption features are seen in the spectra of all the companions. J1857+7048 shows strong metal absorption lines which indicates that the companion is an M type star. The \NaI\ $\lambda\lambda$ 5890,5896 \AA\ absorption lines as well as the \CaII\ $\lambda\lambda$ 8498,8542,8662 \AA\ lines are seen in J1314+4912, J1804+3230 and J1852+4833, suggesting that their companions are K type stars. Though the interstellar medium in the Milky Way can also produce \NaI\ absorption lines \citep{Murga2015}, we only see \NaI\ lines in the companion's spectrum but not in the quasars spectrum, which disfavors this scenario. As for J1732$-$1335, the narrow H$\alpha$ absorption line at 6563{\AA} indicates that the companion is a G type star; the weak emission at 8479{\AA} is the flux leakage from the quasar due to the high flux contrast between the two cores and the marginally resolved spectra. With one additional confirmed case (J0816+4305) from its SDSS unresolved spectrum (missed during our initial \hst\ target selection), six targets have been spectroscopically confirmed as star-quasar superpositions. They are shown as red filled diamonds in \autoref{fig:cands_color}.

All the spectroscopically confirmed star-quasar superpositions are correctly identified using our color cuts except for J1732$-$1335, whose companion is likely a G type star. This test suggests our two color criteria are reasonable in identifying star-quasar pairs with an estimated success rate of ~80\% (i.e., 5/6). Based on these two color criteria and the follow-up spectroscopic observations, we classify 26 of the 45 HST multiples as (potential) star-quasar superpositions.

\subsection{Star-quasar Superposition Probability}

We estimate the probability of star-quasar superposition for the SDSS quasars (Category 1) by considering their surrounding stellar density. We start with the parent quasar sample which the \hst\ SDSS targets are selected from \citep{HwangShen2020}. Specifically, for this parent quasar sample, we require that there is only a single {\it Gaia} match within 3\,arcsec with a G-band magnitude $<19.5$, \texttt{visibility\_periods\_used}$\ge9$, and $z>0.5$, resulting in a sample of $\sim$79,000 quasars. Then, we offset their declination by 1 arcminute and search for nearby Gaia DR2 sources at the offset positions. 
All nearby sources around the offset coordinates are considered as star-quasar pairs. Because the star-quasar superposition rate depends on the surrounding stellar density, which further depends on the Galactic coordinates, our test ensures that the offset coordinates have similar Galactic coordinates (differ by only 1 arcmin) and therefore have similar star-quasar superposition rate as the original parent sample.

Our star-quasar superposition test finds that there are 11 offset coordinates that have another fainter source within 0\farcs5 (given the mean angular separation of 0\farcs406 for the single-{\it Gaia}-match quasars with \hst\ double cores). We only consider the pairs where the nearby sources are fainter than the quasars in {\it Gaia} G-band because a superposition of a quasar with a brighter stellar object would be removed from our sample by their stellar features in SDSS spectra, or is not included in the SDSS quasar catalog in the first place. Since {\it Gaia} DR2's completeness of nearby source identification drops significantly below 0\farcs7 \citep{Arenou2018}, we expect that most of these 11 pairs would remain unresolved in {\it Gaia} DR2 and appear as single-{\it Gaia}-match quasars. If these 11 mock pairs all have sufficiently high astrometric excess noise and become our \hst\ targets, then the stellar contamination fraction among the 40 single-{\it Gaia}-matched \hst\ targets (i.e., Category 1) is at most 30$\pm$10\% (assuming Poisson noise). This is a hard upper limit because some of them can be excluded by the stellar spectral features when the star is not much fainter than the quasar, and more importantly, star-quasar pairs may not necessarily have high astrometric excess noise that fall into our selection. For comparison, our color criteria identify 6 star-quasar pairs out of the 17 observed targets in Category 1, which results in a contamination rate of 6/17 ($35\%$), in broad agreement with this shifted position test. 

\subsection{Dual Quasars versus Lensed Quasars}\label{sec:dualvslens}

The primary goal of our program is to discover $\sim$\,kpc-scale dual quasars, especially at $z>1$. As illustrated in \autoref{fig:z_rp}, our targeted search has the potential to fill a redshift-separation regime largely uncharted by previous searches for dual quasars. We have followed up some promising candidate high-redshift kpc-scale dual quasars from the resolved 45 systems. J0749+2255 and J0841+4825 have been reported in \citet{Shen2021}, and favor the dual quasar scenario (although the lensed quasar scenario cannot be entirely ruled out). Based on statistical arguments, \citet{Shen2021} argued that the abundance of high-redshift sub-arcsec gravitational lens is insufficient to account for most of the resolved pairs in our systematic search. 

\begin{figure*}
  \centering
    \includegraphics[width=0.7\textwidth]{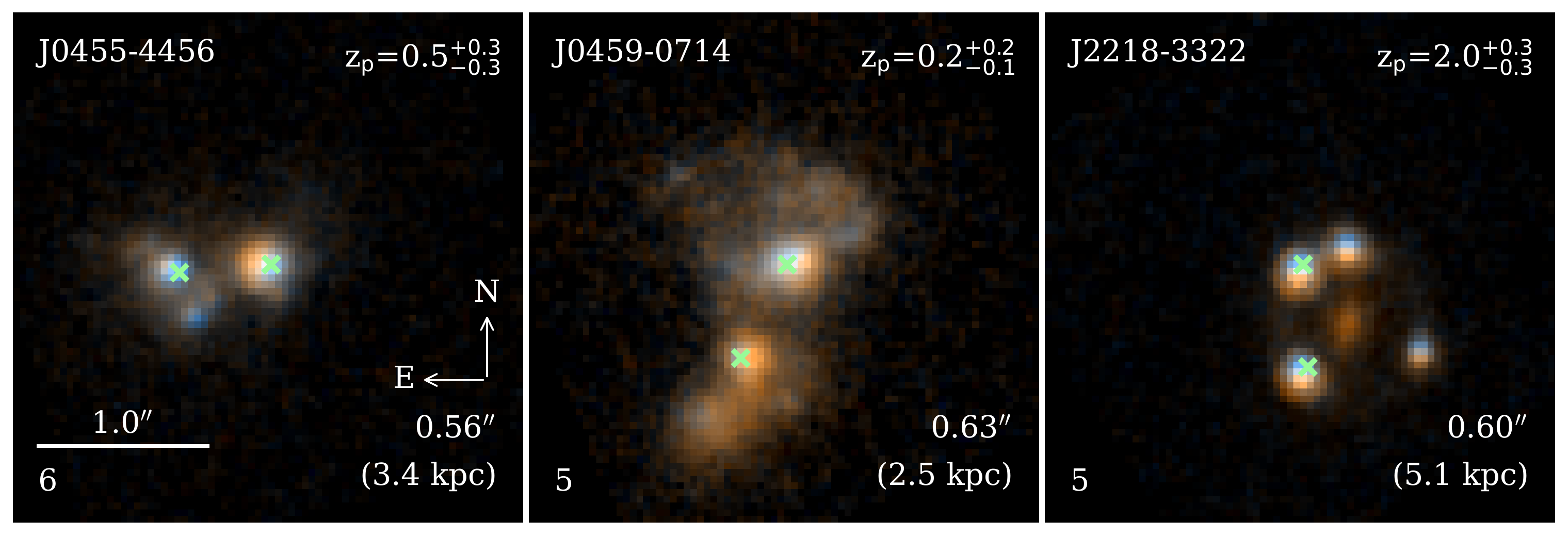}
    \caption{\hst/WFC3 color composite images of three highlighted targets. {\it Left}: A triple quasar candidate with irregular tidal feature. {\it Middle}: A dual quasar candidate with irregular tidal feature. {\it Right}: A quadruply lensed quasar.
    }
    \label{fig:double_highlight}
\end{figure*}

Besides the spectroscopically confirmed dual quasars and star-quasar superpositions, the \hst\ images revealed diagnostic morphology such as tidal features or lens galaxies in several targets (\autoref{fig:double_highlight}). J0455$-$4456 is classified as a triple quasar candidate, which shows an irregular tidal feature in the images and the F475W$-$F814W colors of its three sources are consistent with quasar color. Similar tidal features are seen in J0459$-$0714, which is likely a dual quasar. On the other hand, a lens galaxy and a weak Einstein ring in J2218$-$3322 confirm the lensing scenario. J2218$-$3322 is among the smallest-separation quadruply lensed quasars \citep{Blackburne2008}. Besides J0455$-$4456, there are also four targets (J0629$-$7536, J0758+1457, J1314$-$4912 and 1711$-$1611) consisting of three components, although not all components are classified as quasars based on their colors in each target. Those four targets are likely a dual/lensed quasar with a star or a quasar with two stars.

For other pairs without distinctive features, the hypothesis of lensed quasars cannot be easily ruled out. Our \hst\ optical imaging is too shallow and inefficient to rule in/out high-redshift lens in individual systems, except in rare cases where a potential lens is detected (e.g. J2218$-$3322) or a merger feature is seen (e.g. J0455$-$4456 and J0459$-$0714), supporting the dual quasar classification. Besides, given the general similarity of quasar spectra, nearly identical spectra do not necessarily imply a lensed quasar. For example, SDSS J1124+5710 and SDSS J1309+5617 in \citet{More16} have nearly identical spectra, but the slight differences in the emission line shapes and redshifts suggest that they are probably physical pairs. Future high-resolution IR imaging and spatially resolved spectroscopy are essential to distinguish dual quasars from lensed quasars.

\section{Discussion}\label{sec:discussions}

\subsection{Varstrometry Selection Efficiency Increases with Redshift}

\begin{figure}
  \centering
    \includegraphics[width=\columnwidth]{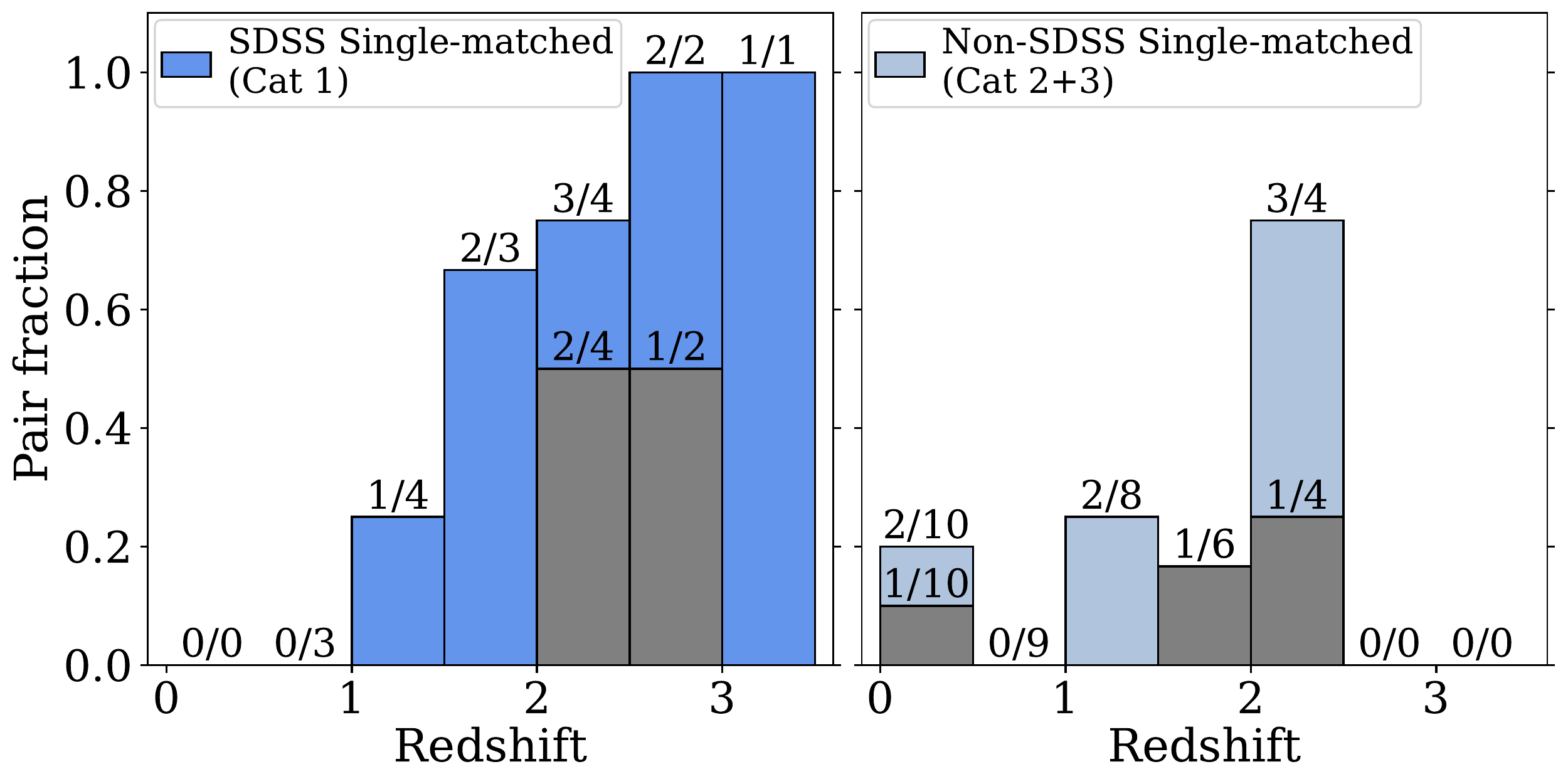}
    \caption{Pair fraction of the {\it Gaia} single-matched targets (i.e., {\tt varstrometry} selection) as a function of redshift. The pair fraction is defined as the number of doubles/multiples regardless of their physical nature divided by the number of \hst-observed targets. The values is also denoted on top of each bin. In addition to the pair fraction, the dual/lensed quasar fraction is shown in the grey histogram. {\it Left:} The SDSS single-matched targets only. {\it Right:} The non-SDSS single-matched targets. The redshifts are from photo-$z$s if spectroscopic redshifts are unavailable. 
    }
    \label{fig:pair_fraction}
\end{figure}

Among the {\it Gaia} single-matched targets observed by \hst, those with spectroscopic redshifts from SDSS (Category 1) have the highest pair fraction of 53\% and those from PS1+WISE (Category 2) have a moderate pair fraction of 27\%; in contrast, those with WISE-only selection (Category 3) have the lowest pair fraction of 15\% (\autoref{tab:number_stat}). The high success rate of the SDSS targets is expected given that we were able to exclude the low-redshift ($z<0.5$) targets. As for the PS1+WISE targets, without spectroscopic redshifts, the sample might still contain low-$z$ host galaxies that are unresolved in PS1, which slightly reduces the observed pair fraction. The lowest success rate of the WISE-only targets is likely due to the lack of optical imaging information to exclude the contamination from extended host galaxies or the complicated selection criteria to remove the low-z extended galaxy that might also reject possible doubles. 

To better demonstrate the host galaxy contamination at low redshift, we show the pair fraction of the {\it Gaia} single-matched targets as a function of redshift in \autoref{fig:pair_fraction}. The pair fraction is the fraction of doubles/multiples regardless of their physical nature among the observed targets. The targets at low redshift ($z<1$) are more contaminated by extended host emission and have a low success rate of \hst-resolved pairs. However, the pair fraction increases dramatically at $z>1$. The high pair fraction at $z>1$ suggests that the {\tt varstrometry} technique is highly effective in searching for sub-arcsec quasar pairs at high redshift. While our follow-up observations only produced a small number of confirmed dual quasars, there is evidence (\autoref{fig:pair_fraction}) that the {\tt varstrometry} selection is efficient in discovering genuine $z>1$ double quasars (either physical pairs or lensed quasars). 


\subsection{Separation predicted from varstrometry}

\begin{figure}
  \centering
    \includegraphics[width=0.95\columnwidth]{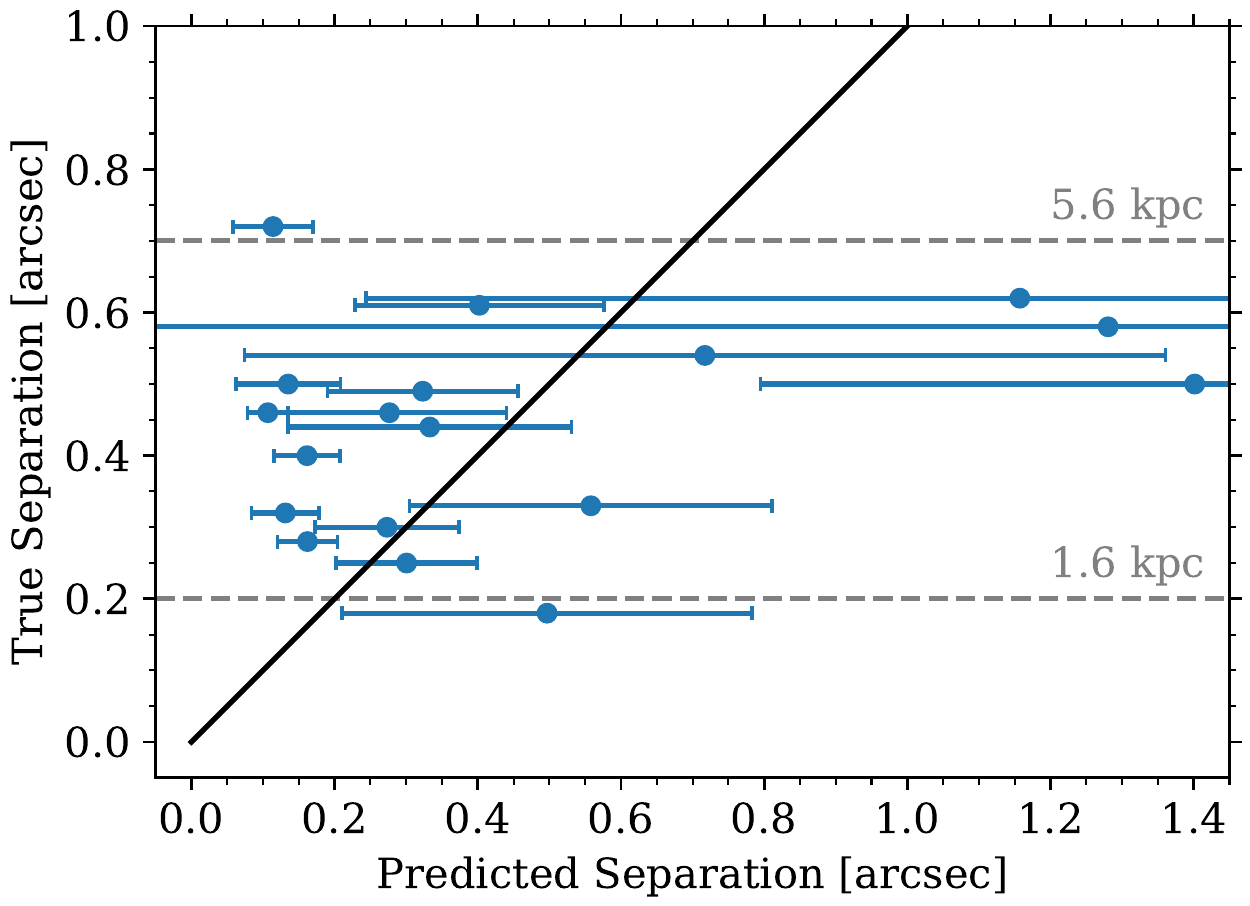}
    \caption{Comparison between the predicted separations from {\tt varstrometry} and the true separations for the 18 pairs with single {\it Gaia} detections. The black line shows the unity relation. The grey lines are the corresponding physical separations for 0.2 arcsec (1.6 kpc)  and 0.7 arcsec (5.6 kpc) at $z=1$. At these scales, both black holes are within the potential of the merged galaxy and follow orbital decay due to dynamical friction. 
    }
    \label{fig:sep_predicted}
\end{figure}

We test if the separations predicted from {\tt varstrometry} are consistent with the true separations measured in the \hst\ images for the 18 pairs with single {\it Gaia} detections. \citet{HwangShen2020} gives the relation
between astrometric jitter, separation and fractional photometric variability if  astrometric jitter is solely due to the asynchronous flux variations of both members in an unresolved pair. For an unresolved double quasar, if both quasars have the similar fractional variability, the predicted separation (using Equation 3 in \citealt{HwangShen2020}) is
\begin{equation}
    D \sim \frac{\sigma_{\rm astro}}{\sqrt{\frac{2q^2}{(1+q)^2(1+q^2)}}\frac{\sqrt{\langle\Delta f^2\rangle}}{\bar{f}}},
\end{equation}
where $D$ is the separation, $\sigma_{\rm astro}$ is the astrometric jitter, $q$ is the flux ratio, $\sqrt{\langle\Delta f^2\rangle}$ is the total flux variability of the system and $\bar{f}$ is the mean flux of the system. For an unresolved star-quasar superposition, in which only one member is variable, the predicted separation (using Equation 5 in \citealt{HwangShen2020}) is 
\begin{equation}
    D \sim \frac{\sigma_{\rm astro}}{\frac{q}{1+q}\frac{\sqrt{\langle\Delta f^2\rangle}}{\bar{f}}}.
\end{equation}
The astrometric jitter $\sigma_{\rm astro}$ is calculated as the root sum of square of {\tt astrometric\_excess\_noise}, parallax and proper motion in the 22-month observing period of {\it Gaia} DR2. The flux variability $\sqrt{\langle\Delta f^2\rangle}$ is estimated as {\tt phot\_g\_mean\_flux\_error} times the square root of {\tt phot\_g\_n\_obs}. We subtract the flux variability by the measurement uncertainty (using all sources with similar magnitudes ($\Delta m<0.1$) within a radius of 0.5 degree) to obtain the quasar-induced intrinsic photometric variability. The flux ratio in {\it Gaia} $G$-band is estimated from the interpolation of the  \hst\ dual-band fluxes. We apply Equation 1 to the targets classified as dual/lensed quasars and Equation 2 to those classified as star-quasar superpositions. \autoref{fig:sep_predicted} shows the comparison between the predicted separations from {\tt varstrometry} and the true separations for the 18 pairs with single {\it Gaia} detections. Two thirds of the targets are within 1-$\sigma$ uncertainties. The predicted separations are broadly consistent with the true separations, supporting {\tt varstrometry} as the origin of astrometric jitters, although the data points are not tight enough to reject other systematics. We found that targets with separations of $\gtrsim 0\farcs5$ have larger offsets from the expected values, which is likely because the large pair separation makes the light profile deviate from a single PSF.  In practice, it is difficult to predict the separations precisely based on the approximate {\tt varstrometry} formula. We suspect that the 0.2-0.7 arcsec range is mostly a selection effect. For pairs with separations of $\gtrsim$0.8 arcsec, Gaia would likely already resolve the source; on the other hand, \hst\ cannot resolve the pair with separations of $\lesssim$0.2 arcsec), but the expected shorter dynamical time at smaller separations might play some role here as well.

\subsection{HST-unresolved Sources}


Besides the sub-arcsec doubles/multiples resolved by \hst, there are 39 \hst\ unresolved targets. Several possibilities could explain those unresolved sources: {\tt varstrometry} may be able to detect doubles that are unresolved by \hst, extended host emission could contribute extra astrometric noise, or they are caused by {\it Gaia} systematics.
13 ($\sim$33\%) of the 39 \hst\ unresolved sources contain significant extended emission from host galaxies (\autoref{fig:N_class}) and are at low redshift ($z\lesssim1$) based on the SDSS spectra or photo-$z$s. Most of the systems classified as extended hosts fall in the categories without spectroscopic redshifts (Categories 2 and 3), which is expected because these categories mainly rely on the astrometric excess noise; extended host emission can contribute to this astrometric excess noise \citep{HwangShen2020}. 
As for the systems from SDSS (Category 1), since low-redshift targets ($z<0.5$) had been removed, the point-like targets could either be small-scale ($<0.1$ arcsec) doubles that are unresolved by \hst\ or could be caused by {\it Gaia} systematics. 

There are some interesting targets that cannot be modeled by a single PSF or a PSF with S${\rm \acute{e}}$rsic components. One such case is J0041$-$7347, which shows an elongated ellipse with a similar color profile across the ellipse. However, \citet{Leisy1996} have identified J0041$-$7347 as a carbon-rich planetary nebula based on optical and ultraviolet spectroscopic observations.
The other cases are J0620$-$0711 and J0258+6905, whose central components deviate from the PSF and show faint offset residuals. The targets with the small offset residuals could be double cores with a separation smaller than the \hst\ resolution ($<0\farcs1$); however the possibility of an imperfect PSF model cannot be excluded. 

\subsection{Comparison to other methods}

There are several other techniques for finding dual quasars at (sub)kpc scales that are unresolved in ground-based observations. One popular method is double-peaked profiles in narrow emission lines in quasar spectra \citep{wang09,Liu2010b,Smith2010}. However, most dual quasars found from the double-peaked narrow-line samples \citep{Liu10,Fu2012,Comerford15} are at low redshifts ($z<0.5$). In addition, in most cases such double-peaked structure is due to the complex kinematics of the narrow-line region such as rotation and outflows \citep{Shen2011a,Fu2012,Zakamska2014,Muller-Sanchez2016}; the dual quasar fraction among the double-peaked narrow-line sample is usually few percent \citep{Shen2011a,Fu2012}. Direct imaging with high-resolution radio interferometers (e.g., VLA and VLBA) is also used to confirm the nature of dual quasar candidates at kpc scales \citep{Fu15,Fu2015b,Shen2021} or even at pc scales \citep{Rodriguez2006} although it requires the source to be radio-bright.

By design, the varstrometry selection technique only applies to (optically) unobscured broad-line (i.e., type 1) quasars, where the AGN variability is detectable by Gaia. It is possible that kpc dual quasars in high-redshift mergers are more likely to be obscured systems, similar to the cases observed in MIR-selected low-redshift dual AGNs \citep{Satyapal17,Pfeifle2019}. We will investigate this possibility in future work by comparing our observational results with cosmological hydrodynamic simulations \citep{Blecha2013a,Blecha2018}.

Other indirect methods such as the radial velocity shift of the broad emission lines \citep{eracleous11,Shen2013,Liu2014,Runnoe2017,Wang2017,Guo2019} and periodicity in the optical light curves \citep{Graham2015,Charisi2016,Liutt2019,ChenYC2020,Liao2021} are sensitive to binary SMBHs at sub-pc scales, where the two SMBHs become gravitationally bound. Both methods require decade-long multi-epoch imaging or spectroscopy. Those candidates are difficult to resolve with radio interferometers even with VLBI; none of them are confirmed by direct imaging so far.

The high pair fraction of $\sim$30\% in the single-{\it Gaia}-matched quasar sample indicates that the {\tt varstrometry} technique is very effective in finding unresolved pairs. Assuming that $\sim$40\% of the pairs are dual/lensed quasars, we expect that $\sim$12\% of the candidates selected from {\tt varstrometry} are dual/lensed quasars. The fraction is even higher in the spectroscopically confirmed quasar sample (Category 1).

\subsection{Refining Targeting and Follow-up Strategy}\label{sec:refine}

Building on the results from this pilot \hst\ program, we can refine the targeting strategy to search for dual quasars with high-spatial-resolution imaging/spectroscopy, especially with {\tt varstrometry}, to improve the success rate:

\begin{enumerate}
    \item The new {\it Gaia} EDR3 \citep{GaiaEDR3} provides better astrometric measurements overall. The position uncertainties of objects at the faint end decreased by nearly half, for example, from 0.7 mas in DR2 to 0.4 mas in EDR3 at $G=20$ \citep{Lindegren2021}. In addition, EDR3 resolves more sub-arcsec pairs than DR2, has better treatments of extended sources, and the reliability of the astrometric excess noise parameter is generally improved over DR2. 
     From the same {\tt varstrometry} selection criteria, the 38 HST-unresolved targets (excluding J0322$-$5515) show {\tt astrometric\_excess\_noise} of 1.66$\pm$0.14 mas, and the 18 HST-resolved pairs show higher values of 2.50$\pm$1.36 mas, a $\sim$5$\sigma$ difference. The higher {\tt astrometric\_excess\_noise} values of the HST-resolved pairs support the hypothesis of {\tt varstrometry} technique.
    For the \hst-unresolved targets, the {\tt astrometric\_excess\_noise} values in EDR3 decrease by $~$40\% on average compared with those in DR2. 
    The improvement on the astrometric excess noise would be particularly useful to reduce false positives selected with {\tt varstrometry} (e.g., those appeared unresolved or with extended host emission in \hst\ imaging). We also expect less spurious {\it Gaia} source detections such as those in J0322$-$5515. 
    
    
    \item Additional IR imaging with \hst\ (or ground-based adaptive optics) is needed to confirm/reject the lensing scenario. The current \hst\ F814W optical imaging data are too shallow to detect the lens galaxies, especially for the high-$z$ systems \citep{Shen2021}. Even with spatially-resolved spectroscopy, it is often difficult to rule out the lensing scenario \citep{Shen2021}. High-resolution IR imaging, even at the \hst\ SNAP depth, is highly efficient to rule out the lensing scenario by the non-detection of lens galaxies.
    
    \item High-resolution radio imaging could be used to reject star-quasar superposition. Detecting two flat-spectrum compact radio cores will confirm the candidate as a dual/lensed quasar. In addition, if two radio cores have significantly different spectral indices, it will suggest that the source is a dual quasar instead of a lensed quasar. \citet{Shen2021} reported preliminary VLBA observations of J0749+2255, supporting the dual quasar scenario. A few of our targets have  detections in publicly available radio surveys such as FIRST \citep{FIRST}, NVSS \citep{NVSS}, and/or VLASS \citep{VLASS}, but the angular resolutions of those public radio data are not high enough to resolve the sub-arcsecond pairs. 
    
    \item It is more efficient to target spectroscopically confirmed quasars at $z>1$ than photometric quasar candidates, given the redshift information and potential identification of star-quasar superposition in the spectrum. Indeed, the SDSS {\tt varstrometry}-selected candidates show a high pair detection rate (53\%), while the WISE-only candidates only have a pair fraction of 15\%, which suggests that the redshift information is crucial to mitigate host galaxy contamination in the astrometric measurements. 
    
    \item Wide-field multi-band imaging surveys such as the Dark Energy Survey \citep{DES} and DECaLS \citep{DECaLS} can provide extra morphology and color information. Because of the excellent image qualities (median $r$-band FWHM of 1\farcs18 in DECaLS and 0\farcs96 in DES), some pairs with wider separations are marginally resolved and identified. The multi-band colors could also help remove star-quasar superpositions. Among the 14 quasar+star pairs that are covered by DECaLS and DES, we find 8 of them have been identified as two sources, whose separations are $\gtrsim0\farcs6$. For the rest 4 quasar+star pairs, 2 of them also show significant red emissions in the residual map.
    
\end{enumerate}

\section{Conclusions}\label{sec:conclusions}

In this paper, we present \hst\ dual-band (F475W and F814W) optical imaging for 56 {\it Gaia}-unresolved dual quasar candidates selected by {\tt varstrometry} and 28 {\it Gaia}-resolved dual quasar candidates. These dual quasar candidates are in the sub-arcsec regime, and represent the long-sought kpc-scale dual quasar population. 

Our \hst\ imaging of the 84 targets reveals 45 resolved pairs (or multiples), among which 17 resolved pairs are from the {\tt varstrometry} selection. The fraction of resolved pairs among the {\tt varstrometry}-selected targets is $\sim 30-50\%$, increasing toward high redshift ($\sim 60-80\%$ at $z>1.5$), with the highest success rate ($\gtrsim$80\%) in spectroscopically confirmed quasar targets. 

We discuss the nature of the 45 \hst-resolved quasar targets based on \hst\ and supplementary data. Given that star-quasar superpositions are a significant contaminant for our sample, we develop color criteria that can successfully reject most of them ($\sim$80\%). A substantial fraction (19/45 $\sim40\%$) of the \hst-resolved pairs are likely physical quasar pairs or gravitationally lensed quasars. It is more probable that most of them are dual quasars instead of lenses, which require very massive lenses at high redshifts given the non-detection of the lens galaxy in the \hst\ image. These systems fill in an important redshift--separation regime of dual SMBHs that has been poorly explored in earlier searches (\autoref{fig:z_rp}). Besides the candidates selected by colors, we also discover a quadruply lensed quasar, which shows a foreground lens in the \hst\ images.

This program with \hst\ optical imaging demonstrates the potential of using {\tt varstrometry} and {\it Gaia} data to systematically discover genuine $\sim$kpc-scale dual quasars, especially at $z>1$. Our results provide important guidelines to significantly refine the targeting and follow-up strategy and to facilitate the classifications (\S\ref{sec:refine}), with improved astrometric measurements from {\it Gaia} EDR3 and future data releases. 


\facilities{{\it Gaia}, {\it HST} (WFC3), Gemini (GMOS)}
\software{dustmaps \citep{Green2018}, astropy \citep{Astropy}, pysynphot \citep{pysynphot}
}

\acknowledgments

We thank J. Blakeslee for granting us Gemini DDT, and K. Chiboucas, H. Kim, and A. Nitta for their help with conducting the Gemini observations, and the referee for useful comments that improved the manuscript. This work is supported by the Heising-Simons Foundation and Research Corporation for Science Advancement, and NSF grant AST-2108162 (YCC, YS, XL). YCC and XL acknowledge support from the University of Illinois Campus Research Board. YCC acknowledges support by the government scholarship to study aboard from the ministry of education of Taiwan and support by the Illinois Survey Science Graduate Student Fellowship. YS acknowledges partial support from NSF grant AST-2009947. NLZ and HCH acknowledge support by the HST-SNAP-15900 grant administered by the STScI. 

Based on observations made with the NASA/ESA Hubble Space Telescope, obtained at the Space Telescope Science Institute, which is operated by the Association of Universities for Research in Astronomy, Inc., under NASA contract NAS 5-26555. These observations are associated with program number SNAP-15900.

Based on observations obtained at the international Gemini Observatory (Program
IDs GN-2020A-DD-106, GS-2020A-DD-106, and GN-2020A-Q-232), a program of NSF’s NOIRLab, which is managed by the Association of Universities for Research in Astronomy (AURA) under a cooperative agreement with the National Science Foundation. on behalf of the Gemini Observatory partnership: the National Science Foundation (United States), National Research Council (Canada), Agencia Nacional de Investigaci\'{o}n y Desarrollo (Chile), Ministerio de Ciencia, Tecnolog\'{i}a e Innovaci\'{o}n (Argentina), Minist\'{e}rio da Ci\^{e}ncia, Tecnologia, Inova\c{c}\~{o}es e Comunica\c{c}\~{o}es (Brazil), and Korea Astronomy and Space Science Institute (Republic of Korea). This work was enabled by observations made from the Gemini North telescope, located within the Maunakea Science Reserve and adjacent to the summit of Maunakea. We are grateful for the privilege of observing the Universe from a place that is unique in both its astronomical quality and its cultural significance.


\bibliography{ref}

\begin{thebibliography}{}
\expandafter\ifx\csname natexlab\endcsname\relax\def\natexlab#1{#1}\fi
\providecommand{\url}[1]{\href{#1}{#1}}

\bibitem[{{Abbott} {et~al.}(2018){Abbott}, {Abdalla}, {Allam}, {Amara},
  {Annis}, {Asorey}, {Avila}, {Ballester}, {Banerji}, {Barkhouse}, {Baruah},
  {Baumer}, {Bechtol}, {Becker}, {Benoit-L{\'e}vy}, {Bernstein}, {Bertin},
  {Blazek}, {Bocquet}, {Brooks}, {Brout}, {Buckley-Geer}, {Burke}, {Busti},
  {Campisano}, {Cardiel-Sas}, {Carnero Rosell}, {Carrasco Kind}, {Carretero},
  {Castander}, {Cawthon}, {Chang}, {Chen}, {Conselice}, {Costa}, {Crocce},
  {Cunha}, {D'Andrea}, {da Costa}, {Das}, {Daues}, {Davis}, {Davis}, {De
  Vicente}, {DePoy}, {DeRose}, {Desai}, {Diehl}, {Dietrich}, {Dodelson},
  {Doel}, {Drlica-Wagner}, {Eifler}, {Elliott}, {Evrard}, {Farahi}, {Fausti
  Neto}, {Fernandez}, {Finley}, {Flaugher}, {Foley}, {Fosalba}, {Friedel},
  {Frieman}, {Garc{\'\i}a-Bellido}, {Gaztanaga}, {Gerdes}, {Giannantonio},
  {Gill}, {Glazebrook}, {Goldstein}, {Gower}, {Gruen}, {Gruendl}, {Gschwend},
  {Gupta}, {Gutierrez}, {Hamilton}, {Hartley}, {Hinton}, {Hislop}, {Hollowood},
  {Honscheid}, {Hoyle}, {Huterer}, {Jain}, {James}, {Jeltema}, {Johnson},
  {Johnson}, {Kacprzak}, {Kent}, {Khullar}, {Klein}, {Kovacs}, {Koziol},
  {Krause}, {Kremin}, {Kron}, {Kuehn}, {Kuhlmann}, {Kuropatkin}, {Lahav},
  {Lasker}, {Li}, {Li}, {Liddle}, {Lima}, {Lin}, {L{\'o}pez-Reyes}, {MacCrann},
  {Maia}, {Maloney}, {Manera}, {March}, {Marriner}, {Marshall}, {Martini},
  {McClintock}, {McKay}, {McMahon}, {Melchior}, {Menanteau}, {Miller},
  {Miquel}, {Mohr}, {Morganson}, {Mould}, {Neilsen}, {Nichol}, {Nogueira},
  {Nord}, {Nugent}, {Nunes}, {Ogando}, {Old}, {Pace}, {Palmese},
  {Paz-Chinch{\'o}n}, {Peiris}, {Percival}, {Petravick}, {Plazas}, {Poh},
  {Pond}, {Porredon}, {Pujol}, {Refregier}, {Reil}, {Ricker}, {Rollins},
  {Romer}, {Roodman}, {Rooney}, {Ross}, {Rykoff}, {Sako}, {Sanchez}, {Sanchez},
  {Santiago}, {Saro}, {Scarpine}, {Scolnic}, {Serrano}, {Sevilla-Noarbe},
  {Sheldon}, {Shipp}, {Silveira}, {Smith}, {Smith}, {Smith}, {Soares-Santos},
  {Sobreira}, {Song}, {Stebbins}, {Suchyta}, {Sullivan}, {Swanson}, {Tarle},
  {Thaler}, {Thomas}, {Thomas}, {Troxel}, {Tucker}, {Vikram}, {Vivas},
  {Walker}, {Wechsler}, {Weller}, {Wester}, {Wolf}, {Wu}, {Yanny}, {Zenteno},
  {Zhang}, {Zuntz}, {DES Collaboration}, {Juneau}, {Fitzpatrick}, {Nikutta},
  {Nidever}, {Olsen}, {Scott}, \& {NOAO Data Lab}}]{DES}
{Abbott}, T.~M.~C., {Abdalla}, F.~B., {Allam}, S., {et~al.} 2018, \apjs, 239,
  18

\bibitem[{{Anguita} {et~al.}(2018){Anguita}, {Schechter}, {Kuropatkin},
  {Morgan}, {Ostrovski}, {Abramson}, {Agnello}, {Apostolovski}, {Fassnacht},
  {Hsueh}, {Motta}, {Rojas}, {Rusu}, {Treu}, {Williams}, {Auger},
  {Buckley-Geer}, {Lin}, {McMahon}, {Abbott}, {Allam}, {Annis}, {Bernstein},
  {Bertin}, {Brooks}, {Burke}, {Carnero Rosell}, {Carrasco-Kind}, {Carretero},
  {Cunha}, {D'Andrea}, {De Vicente}, {DePoy}, {Desai}, {Diehl}, {Doel},
  {Flaugher}, {Garc{\'\i}a-Bellido}, {Gerdes}, {Gruen}, {Gruendl}, {Gschwend},
  {Hartley}, {Hollowood}, {Honscheid}, {James}, {Kuehn}, {Lima}, {Maia},
  {Miquel}, {Plazas}, {Sanchez}, {Scarpine}, {Smith}, {Soares-Santos},
  {Sobreira}, {Suchyta}, {Tarle}, \& {Walker}}]{Anguita18}
{Anguita}, T., {Schechter}, P.~L., {Kuropatkin}, N., {et~al.} 2018, \mnras,
  480, 5017

\bibitem[{{Arenou} {et~al.}(2018){Arenou}, {Luri}, {Babusiaux}, {Fabricius},
  {Helmi}, {Muraveva}, {Robin}, {Spoto}, {Vallenari}, {Antoja},
  {Cantat-Gaudin}, {Jordi}, {Leclerc}, {Reyl{\'e}}, {Romero-G{\'o}mez}, {Shih},
  {Soria}, {Barache}, {Bossini}, {Bragaglia}, {Breddels}, {Fabrizio},
  {Lambert}, {Marrese}, {Massari}, {Moitinho}, {Robichon}, {Ruiz-Dern},
  {Sordo}, {Veljanoski}, {Eyer}, {Jasniewicz}, {Pancino}, {Soubiran}, {Spagna},
  {Tanga}, {Turon}, \& {Zurbach}}]{Arenou2018}
{Arenou}, F., {Luri}, X., {Babusiaux}, C., {et~al.} 2018, \aap, 616, A17

\bibitem[{{Arzoumanian} {et~al.}(2018){Arzoumanian}, {Baker}, {Brazier},
  {Burke-Spolaor}, {Chamberlin}, {Chatterjee}, {Christy}, {Cordes}, {Cornish},
  {Crawford}, {Thankful Cromartie}, {Crowter}, {DeCesar}, {Demorest}, {Dolch},
  {Ellis}, {Ferdman}, {Ferrara}, {Folkner}, {Fonseca}, {Garver-Daniels},
  {Gentile}, {Haas}, {Hazboun}, {Huerta}, {Islo}, {Jones}, {Jones}, {Kaplan},
  {Kaspi}, {Lam}, {Lazio}, {Levin}, {Lommen}, {Lorimer}, {Luo}, {Lynch},
  {Madison}, {McLaughlin}, {McWilliams}, {Mingarelli}, {Ng}, {Nice}, {Park},
  {Pennucci}, {Pol}, {Ransom}, {Ray}, {Rasskazov}, {Siemens}, {Simon},
  {Spiewak}, {Stairs}, {Stinebring}, {Stovall}, {Swiggum}, {Taylor},
  {Vallisneri}, {van Haasteren}, {Vigeland}, {Zhu}, \& {NANOGrav
  Collaboration}}]{Arzoumanian2018a}
{Arzoumanian}, Z., {Baker}, P.~T., {Brazier}, A., {et~al.} 2018, \apj, 859, 47

\bibitem[{{Astropy Collaboration} {et~al.}(2018){Astropy Collaboration},
  {Price-Whelan}, {Sip{\H{o}}cz}, {G{\"u}nther}, {Lim}, {Crawford}, {Conseil},
  {Shupe}, {Craig}, {Dencheva}, {Ginsburg}, {VanderPlas}, {Bradley},
  {P{\'e}rez-Su{\'a}rez}, {de Val-Borro}, {Aldcroft}, {Cruz}, {Robitaille},
  {Tollerud}, {Ardelean}, {Babej}, {Bach}, {Bachetti}, {Bakanov}, {Bamford},
  {Barentsen}, {Barmby}, {Baumbach}, {Berry}, {Biscani}, {Boquien}, {Bostroem},
  {Bouma}, {Brammer}, {Bray}, {Breytenbach}, {Buddelmeijer}, {Burke},
  {Calderone}, {Cano Rodr{\'\i}guez}, {Cara}, {Cardoso}, {Cheedella}, {Copin},
  {Corrales}, {Crichton}, {D'Avella}, {Deil}, {Depagne}, {Dietrich}, {Donath},
  {Droettboom}, {Earl}, {Erben}, {Fabbro}, {Ferreira}, {Finethy}, {Fox},
  {Garrison}, {Gibbons}, {Goldstein}, {Gommers}, {Greco}, {Greenfield},
  {Groener}, {Grollier}, {Hagen}, {Hirst}, {Homeier}, {Horton}, {Hosseinzadeh},
  {Hu}, {Hunkeler}, {Ivezi{\'c}}, {Jain}, {Jenness}, {Kanarek}, {Kendrew},
  {Kern}, {Kerzendorf}, {Khvalko}, {King}, {Kirkby}, {Kulkarni}, {Kumar},
  {Lee}, {Lenz}, {Littlefair}, {Ma}, {Macleod}, {Mastropietro}, {McCully},
  {Montagnac}, {Morris}, {Mueller}, {Mumford}, {Muna}, {Murphy}, {Nelson},
  {Nguyen}, {Ninan}, {N{\"o}the}, {Ogaz}, {Oh}, {Parejko}, {Parley}, {Pascual},
  {Patil}, {Patil}, {Plunkett}, {Prochaska}, {Rastogi}, {Reddy Janga},
  {Sabater}, {Sakurikar}, {Seifert}, {Sherbert}, {Sherwood-Taylor}, {Shih},
  {Sick}, {Silbiger}, {Singanamalla}, {Singer}, {Sladen}, {Sooley},
  {Sornarajah}, {Streicher}, {Teuben}, {Thomas}, {Tremblay}, {Turner},
  {Terr{\'o}n}, {van Kerkwijk}, {de la Vega}, {Watkins}, {Weaver}, {Whitmore},
  {Woillez}, {Zabalza}, \& {Astropy Contributors}}]{Astropy}
{Astropy Collaboration}, {Price-Whelan}, A.~M., {Sip{\H{o}}cz}, B.~M., {et~al.}
  2018, \aj, 156, 123

\bibitem[{{Becker} {et~al.}(1995){Becker}, {White}, \& {Helfand}}]{FIRST}
{Becker}, R.~H., {White}, R.~L., \& {Helfand}, D.~J. 1995, \apj, 450, 559

\bibitem[{{Begelman} {et~al.}(1980){Begelman}, {Blandford}, \&
  {Rees}}]{begelman80}
{Begelman}, M.~C., {Blandford}, R.~D., \& {Rees}, M.~J. 1980, \nat, 287, 307

\bibitem[{{Bianchi} {et~al.}(2008){Bianchi}, {Chiaberge}, {Piconcelli},
  {Guainazzi}, \& {Matt}}]{Bianchi08}
{Bianchi}, S., {Chiaberge}, M., {Piconcelli}, E., {Guainazzi}, M., \& {Matt},
  G. 2008, \mnras, 386, 105

\bibitem[{{Blackburne} {et~al.}(2008){Blackburne}, {Wisotzki}, \&
  {Schechter}}]{Blackburne2008}
{Blackburne}, J.~A., {Wisotzki}, L., \& {Schechter}, P.~L. 2008, \aj, 135, 374

\bibitem[{{Blecha} {et~al.}(2013){Blecha}, {Loeb}, \& {Narayan}}]{Blecha2013a}
{Blecha}, L., {Loeb}, A., \& {Narayan}, R. 2013, \mnras, 429, 2594

\bibitem[{{Blecha} {et~al.}(2018){Blecha}, {Snyder}, {Satyapal}, \&
  {Ellison}}]{Blecha2018}
{Blecha}, L., {Snyder}, G.~F., {Satyapal}, S., \& {Ellison}, S.~L. 2018,
  \mnras, 478, 3056

\bibitem[{Bogdanovic {et~al.}(2021)Bogdanovic, Miller, \&
  Blecha}]{Bogdanovic2021}
Bogdanovic, T., Miller, M.~C., \& Blecha, L. 2021, Living Reviews in
  Relativity, arXiv:2109.03262

\bibitem[{{Bothun} {et~al.}(1989){Bothun}, {Halpern}, {Lonsdale}, {Impey}, \&
  {Schmitz}}]{Bothun89}
{Bothun}, G.~D., {Halpern}, J.~P., {Lonsdale}, C.~J., {Impey}, C., \&
  {Schmitz}, M. 1989, \apjs, 70, 271

\bibitem[{{Bovy}(2017)}]{Bovy2017}
{Bovy}, J. 2017, \mnras, 470, 1360

\bibitem[{{Brotherton} {et~al.}(1999){Brotherton}, {Gregg}, {Becker},
  {Laurent-Muehleisen}, {White}, \& {Stanford}}]{Brotherton99}
{Brotherton}, M.~S., {Gregg}, M.~D., {Becker}, R.~H., {et~al.} 1999, \apjl,
  514, L61

\bibitem[{{Burke-Spolaor}(2011)}]{burke11}
{Burke-Spolaor}, S. 2011, \mnras, 410, 2113

\bibitem[{{Centrella} {et~al.}(2010){Centrella}, {Baker}, {Kelly}, \& {van
  Meter}}]{Centrella2010}
{Centrella}, J., {Baker}, J.~G., {Kelly}, B.~J., \& {van Meter}, J.~R. 2010,
  Reviews of Modern Physics, 82, 3069

\bibitem[{{Chambers} {et~al.}(2016){Chambers}, {Magnier}, {Metcalfe},
  {Flewelling}, {Huber}, {Waters}, {Denneau}, {Draper}, {Farrow}, {Finkbeiner},
  {Holmberg}, {Koppenhoefer}, {Price}, {Rest}, {Saglia}, {Schlafly}, {Smartt},
  {Sweeney}, {Wainscoat}, {Burgett}, {Chastel}, {Grav}, {Heasley}, {Hodapp},
  {Jedicke}, {Kaiser}, {Kudritzki}, {Luppino}, {Lupton}, {Monet}, {Morgan},
  {Onaka}, {Shiao}, {Stubbs}, {Tonry}, {White}, {Ba{\~n}ados}, {Bell},
  {Bender}, {Bernard}, {Boegner}, {Boffi}, {Botticella}, {Calamida},
  {Casertano}, {Chen}, {Chen}, {Cole}, {Deacon}, {Frenk}, {Fitzsimmons},
  {Gezari}, {Gibbs}, {Goessl}, {Goggia}, {Gourgue}, {Goldman}, {Grant},
  {Grebel}, {Hambly}, {Hasinger}, {Heavens}, {Heckman}, {Henderson}, {Henning},
  {Holman}, {Hopp}, {Ip}, {Isani}, {Jackson}, {Keyes}, {Koekemoer}, {Kotak},
  {Le}, {Liska}, {Long}, {Lucey}, {Liu}, {Martin}, {Masci}, {McLean}, {Mindel},
  {Misra}, {Morganson}, {Murphy}, {Obaika}, {Narayan}, {Nieto-Santisteban},
  {Norberg}, {Peacock}, {Pier}, {Postman}, {Primak}, {Rae}, {Rai}, {Riess},
  {Riffeser}, {Rix}, {R{\"o}ser}, {Russel}, {Rutz}, {Schilbach}, {Schultz},
  {Scolnic}, {Strolger}, {Szalay}, {Seitz}, {Small}, {Smith}, {Soderblom},
  {Taylor}, {Thomson}, {Taylor}, {Thakar}, {Thiel}, {Thilker}, {Unger},
  {Urata}, {Valenti}, {Wagner}, {Walder}, {Walter}, {Watters}, {Werner},
  {Wood-Vasey}, \& {Wyse}}]{Chambers2016}
{Chambers}, K.~C., {Magnier}, E.~A., {Metcalfe}, N., {et~al.} 2016, arXiv
  e-prints, arXiv:1612.05560

\bibitem[{{Charisi} {et~al.}(2016){Charisi}, {Bartos}, {Haiman},
  {Price-Whelan}, {Graham}, {Bellm}, {Laher}, \& {M{\'a}rka}}]{Charisi2016}
{Charisi}, M., {Bartos}, I., {Haiman}, Z., {et~al.} 2016, \mnras, 463, 2145

\bibitem[{{Chen} {et~al.}(2020{\natexlab{a}}){Chen}, {Yu}, \&
  {Lu}}]{ChenYF2020}
{Chen}, Y., {Yu}, Q., \& {Lu}, Y. 2020{\natexlab{a}}, \apj, 897, 86

\bibitem[{{Chen} {et~al.}(2020{\natexlab{b}}){Chen}, {Liu}, {Liao}, {Holgado},
  {Guo}, {Gruendl}, {Morganson}, {Shen}, {Zhang}, {Abbott}, {Aguena}, {Allam},
  {Avila}, {Bertin}, {Bhargava}, {Brooks}, {Burke}, {Carnero Rosell},
  {Carollo}, {Carrasco Kind}, {Carretero}, {Costanzi}, {da Costa}, {Davis}, {De
  Vicente}, {Desai}, {Diehl}, {Doel}, {Everett}, {Flaugher}, {Friedel},
  {Frieman}, {Garc{\'\i}a-Bellido}, {Gaztanaga}, {Glazebrook}, {Gruen},
  {Gutierrez}, {Hinton}, {Hollowood}, {James}, {Kim}, {Kuehn}, {Kuropatkin},
  {Lewis}, {Lidman}, {Lima}, {Maia}, {March}, {Marshall}, {Menanteau},
  {Miquel}, {Palmese}, {Paz-Chinch{\'o}n}, {Plazas}, {Sanchez}, {Schubnell},
  {Serrano}, {Sevilla-Noarbe}, {Smith}, {Suchyta}, {Swanson}, {Tarle},
  {Tucker}, {Norbert Varga}, \& {Walker}}]{ChenYC2020}
{Chen}, Y.-C., {Liu}, X., {Liao}, W.-T., {et~al.} 2020{\natexlab{b}}, \mnras,
  499, 2245

\bibitem[{{Civano} {et~al.}(2010){Civano}, {Elvis}, {Lanzuisi}, {Jahnke},
  {Zamorani}, {Blecha}, {Bongiorno}, {Brusa}, {Comastri}, {Hao}, {Leauthaud},
  {Loeb}, {Mainieri}, {Piconcelli}, {Salvato}, {Scoville}, {Trump}, {Vignali},
  {Aldcroft}, {Bolzonella}, {Bressert}, {Finoguenov}, {Fruscione}, {Koekemoer},
  {Cappelluti}, {Fiore}, {Giodini}, {Gilli}, {Impey}, {Lilly}, {Lusso},
  {Puccetti}, {Silverman}, {Aussel}, {Capak}, {Frayer}, {Le Floch},
  {McCracken}, {Sanders}, {Schiminovich}, \& {Taniguchi}}]{Civano10}
{Civano}, F., {Elvis}, M., {Lanzuisi}, G., {et~al.} 2010, \apj, 717, 209

\bibitem[{{Comerford} {et~al.}(2015){Comerford}, {Pooley}, {Barrows}, {Greene},
  {Zakamska}, {Madejski}, \& {Cooper}}]{Comerford15}
{Comerford}, J.~M., {Pooley}, D., {Barrows}, R.~S., {et~al.} 2015, \apj, 806,
  219

\bibitem[{{Comerford} {et~al.}(2011){Comerford}, {Pooley}, {Gerke}, \&
  {Madejski}}]{Comerford11}
{Comerford}, J.~M., {Pooley}, D., {Gerke}, B.~F., \& {Madejski}, G.~M. 2011,
  \apjl, 737, L19

\bibitem[{{Condon} {et~al.}(1998){Condon}, {Cotton}, {Greisen}, {Yin},
  {Perley}, {Taylor}, \& {Broderick}}]{NVSS}
{Condon}, J.~J., {Cotton}, W.~D., {Greisen}, E.~W., {et~al.} 1998, \aj, 115,
  1693

\bibitem[{{Crampton} {et~al.}(1988){Crampton}, {Cowley}, {Hickson}, {Kindl},
  {Wagner}, {Tyson}, \& {Gullixson}}]{Crampton88}
{Crampton}, D., {Cowley}, A.~P., {Hickson}, P., {et~al.} 1988, \apj, 330, 184

\bibitem[{{Crotts} {et~al.}(1994){Crotts}, {Bechtold}, {Fang}, \&
  {Duncan}}]{Crotts94}
{Crotts}, A. P.~S., {Bechtold}, J., {Fang}, Y., \& {Duncan}, R.~C. 1994, \apjl,
  437, L79

\bibitem[{{De Rosa} {et~al.}(2020){De Rosa}, {Vignali}, {Bogdanovi{\'c}},
  {Capelo}, {Charisi}, {Dotti}, {Husemann}, {Lusso}, {Mayer}, {Paragi},
  {Runnoe}, {Sesana}, {Steinborn}, {Bianchi}, {Colpi}, {Del Valle}, {Frey},
  {Gab{\'a}nyi}, {Giustini}, {Guainazzi}, {Haiman}, {Herrera Ruiz},
  {Herrero-Illana}, {Iwasawa}, {Komossa}, {Lena}, {Loiseau}, {Perez-Torres},
  {Piconcelli}, \& {Volonteri}}]{DeRosa2020}
{De Rosa}, A., {Vignali}, C., {Bogdanovi{\'c}}, T., {et~al.} 2020, arXiv
  e-prints, arXiv:2001.06293

\bibitem[{{Dey} {et~al.}(2019){Dey}, {Schlegel}, {Lang}, {Blum}, {Burleigh},
  {Fan}, {Findlay}, {Finkbeiner}, {Herrera}, {Juneau}, {Landriau}, {Levi},
  {McGreer}, {Meisner}, {Myers}, {Moustakas}, {Nugent}, {Patej}, {Schlafly},
  {Walker}, {Valdes}, {Weaver}, {Y{\`e}che}, {Zou}, {Zhou}, {Abareshi},
  {Abbott}, {Abolfathi}, {Aguilera}, {Alam}, {Allen}, {Alvarez}, {Annis},
  {Ansarinejad}, {Aubert}, {Beechert}, {Bell}, {BenZvi}, {Beutler}, {Bielby},
  {Bolton}, {Brice{\~n}o}, {Buckley-Geer}, {Butler}, {Calamida}, {Carlberg},
  {Carter}, {Casas}, {Castander}, {Choi}, {Comparat}, {Cukanovaite}, {Delubac},
  {DeVries}, {Dey}, {Dhungana}, {Dickinson}, {Ding}, {Donaldson}, {Duan},
  {Duckworth}, {Eftekharzadeh}, {Eisenstein}, {Etourneau}, {Fagrelius},
  {Farihi}, {Fitzpatrick}, {Font-Ribera}, {Fulmer}, {G{\"a}nsicke},
  {Gaztanaga}, {George}, {Gerdes}, {Gontcho}, {Gorgoni}, {Green}, {Guy},
  {Harmer}, {Hernandez}, {Honscheid}, {Huang}, {James}, {Jannuzi}, {Jiang},
  {Joyce}, {Karcher}, {Karkar}, {Kehoe}, {Kneib}, {Kueter-Young}, {Lan},
  {Lauer}, {Le Guillou}, {Le Van Suu}, {Lee}, {Lesser}, {Perreault Levasseur},
  {Li}, {Mann}, {Marshall}, {Mart{\'\i}nez-V{\'a}zquez}, {Martini}, {du Mas des
  Bourboux}, {McManus}, {Meier}, {M{\'e}nard}, {Metcalfe},
  {Mu{\~n}oz-Guti{\'e}rrez}, {Najita}, {Napier}, {Narayan}, {Newman}, {Nie},
  {Nord}, {Norman}, {Olsen}, {Paat}, {Palanque-Delabrouille}, {Peng},
  {Poppett}, {Poremba}, {Prakash}, {Rabinowitz}, {Raichoor}, {Rezaie},
  {Robertson}, {Roe}, {Ross}, {Ross}, {Rudnick}, {Safonova}, {Saha},
  {S{\'a}nchez}, {Savary}, {Schweiker}, {Scott}, {Seo}, {Shan}, {Silva},
  {Slepian}, {Soto}, {Sprayberry}, {Staten}, {Stillman}, {Stupak}, {Summers},
  {Sien Tie}, {Tirado}, {Vargas-Maga{\~n}a}, {Vivas}, {Wechsler}, {Williams},
  {Yang}, {Yang}, {Yapici}, {Zaritsky}, {Zenteno}, {Zhang}, {Zhang}, {Zhou}, \&
  {Zhou}}]{DECaLS}
{Dey}, A., {Schlegel}, D.~J., {Lang}, D., {et~al.} 2019, \aj, 157, 168

\bibitem[{{Eftekharzadeh} {et~al.}(2017){Eftekharzadeh}, {Myers}, {Hennawi},
  {Djorgovski}, {Richards}, {Mahabal}, \& {Graham}}]{Eftekharzadeh17}
{Eftekharzadeh}, S., {Myers}, A.~D., {Hennawi}, J.~F., {et~al.} 2017, \mnras,
  468, 77

\bibitem[{{Ellison} {et~al.}(2017){Ellison}, {Secrest}, {Mendel}, {Satyapal},
  \& {Simard}}]{Ellison17}
{Ellison}, S.~L., {Secrest}, N.~J., {Mendel}, J.~T., {Satyapal}, S., \&
  {Simard}, L. 2017, \mnras, 470, L49

\bibitem[{{Eracleous} {et~al.}(2012){Eracleous}, {Boroson}, {Halpern}, \&
  {Liu}}]{eracleous11}
{Eracleous}, M., {Boroson}, T.~A., {Halpern}, J.~P., \& {Liu}, J. 2012, \apjs,
  201, 23

\bibitem[{{Foord} {et~al.}(2021){Foord}, {G{\"u}ltekin}, {Runnoe}, \&
  {Koss}}]{Foord2021}
{Foord}, A., {G{\"u}ltekin}, K., {Runnoe}, J.~C., \& {Koss}, M.~J. 2021, \apj,
  907, 71

\bibitem[{{Frey} {et~al.}(2012){Frey}, {Paragi}, {An}, \&
  {Gab{\'a}nyi}}]{Frey12}
{Frey}, S., {Paragi}, Z., {An}, T., \& {Gab{\'a}nyi}, K.~{\'E}. 2012, \mnras,
  425, 1185

\bibitem[{{Fu} {et~al.}(2015{\natexlab{a}}){Fu}, {Myers}, {Djorgovski}, {Yan},
  {Wrobel}, \& {Stockton}}]{Fu15}
{Fu}, H., {Myers}, A.~D., {Djorgovski}, S.~G., {et~al.} 2015{\natexlab{a}},
  \apj, 799, 72

\bibitem[{{Fu} {et~al.}(2015{\natexlab{b}}){Fu}, {Wrobel}, {Myers},
  {Djorgovski}, \& {Yan}}]{Fu2015b}
{Fu}, H., {Wrobel}, J.~M., {Myers}, A.~D., {Djorgovski}, S.~G., \& {Yan}, L.
  2015{\natexlab{b}}, \apjl, 815, L6

\bibitem[{{Fu} {et~al.}(2012){Fu}, {Yan}, {Myers}, {Stockton}, {Djorgovski},
  {Aldering}, \& {Rich}}]{Fu2012}
{Fu}, H., {Yan}, L., {Myers}, A.~D., {et~al.} 2012, \apj, 745, 67

\bibitem[{{Fu} {et~al.}(2011){Fu}, {Zhang}, {Assef}, {Stockton}, {Myers},
  {Yan}, {Djorgovski}, {Wrobel}, \& {Riechers}}]{Fu11}
{Fu}, H., {Zhang}, Z.-Y., {Assef}, R.~J., {et~al.} 2011, \apjl, 740, L44

\bibitem[{{Gaia Collaboration} {et~al.}(2018){Gaia Collaboration}, {Brown},
  {Vallenari}, {Prusti}, {de Bruijne}, {Babusiaux}, {Bailer-Jones}, {Biermann},
  {Evans}, {Eyer}, {Jansen}, {Jordi}, {Klioner}, {Lammers}, {Lindegren},
  {Luri}, {Mignard}, {Panem}, {Pourbaix}, {Randich}, {Sartoretti}, {Siddiqui},
  {Soubiran}, {van Leeuwen}, {Walton}, {Arenou}, {Bastian}, {Cropper},
  {Drimmel}, {Katz}, {Lattanzi}, {Bakker}, {Cacciari}, {Casta{\~n}eda},
  {Chaoul}, {Cheek}, {De Angeli}, {Fabricius}, {Guerra}, {Holl}, {Masana},
  {Messineo}, {Mowlavi}, {Nienartowicz}, {Panuzzo}, {Portell}, {Riello},
  {Seabroke}, {Tanga}, {Th{\'e}venin}, {Gracia-Abril}, {Comoretto},
  {Garcia-Reinaldos}, {Teyssier}, {Altmann}, {Andrae}, {Audard},
  {Bellas-Velidis}, {Benson}, {Berthier}, {Blomme}, {Burgess}, {Busso},
  {Carry}, {Cellino}, {Clementini}, {Clotet}, {Creevey}, {Davidson}, {De
  Ridder}, {Delchambre}, {Dell'Oro}, {Ducourant},
  {Fern{\'a}ndez-Hern{\'a}ndez}, {Fouesneau}, {Fr{\'e}mat}, {Galluccio},
  {Garc{\'\i}a-Torres}, {Gonz{\'a}lez-N{\'u}{\~n}ez}, {Gonz{\'a}lez-Vidal},
  {Gosset}, {Guy}, {Halbwachs}, {Hambly}, {Harrison}, {Hern{\'a}ndez},
  {Hestroffer}, {Hodgkin}, {Hutton}, {Jasniewicz}, {Jean-Antoine-Piccolo},
  {Jordan}, {Korn}, {Krone-Martins}, {Lanzafame}, {Lebzelter}, {L{\"o}ffler},
  {Manteiga}, {Marrese}, {Mart{\'\i}n-Fleitas}, {Moitinho}, {Mora}, {Muinonen},
  {Osinde}, {Pancino}, {Pauwels}, {Petit}, {Recio-Blanco}, {Richards},
  {Rimoldini}, {Robin}, {Sarro}, {Siopis}, {Smith}, {Sozzetti}, {S{\"u}veges},
  {Torra}, {van Reeven}, {Abbas}, {Abreu Aramburu}, {Accart}, {Aerts},
  {Altavilla}, {{\'A}lvarez}, {Alvarez}, {Alves}, {Anderson}, {Andrei},
  {Anglada Varela}, {Antiche}, {Antoja}, {Arcay}, {Astraatmadja}, {Bach},
  {Baker}, {Balaguer-N{\'u}{\~n}ez}, {Balm}, {Barache}, {Barata}, {Barbato},
  {Barblan}, {Barklem}, {Barrado}, {Barros}, {Barstow}, {Bartholom{\'e}
  Mu{\~n}oz}, {Bassilana}, {Becciani}, {Bellazzini}, {Berihuete}, {Bertone},
  {Bianchi}, {Bienaym{\'e}}, {Blanco-Cuaresma}, {Boch}, {Boeche}, {Bombrun},
  {Borrachero}, {Bossini}, {Bouquillon}, {Bourda}, {Bragaglia}, {Bramante},
  {Breddels}, {Bressan}, {Brouillet}, {Br{\"u}semeister}, {Brugaletta},
  {Bucciarelli}, {Burlacu}, {Busonero}, {Butkevich}, {Buzzi}, {Caffau},
  {Cancelliere}, {Cannizzaro}, {Cantat-Gaudin}, {Carballo}, {Carlucci},
  {Carrasco}, {Casamiquela}, {Castellani}, {Castro-Ginard}, {Charlot},
  {Chemin}, {Chiavassa}, {Cocozza}, {Costigan}, {Cowell}, {Crifo}, {Crosta},
  {Crowley}, {Cuypers}, {Dafonte}, {Damerdji}, {Dapergolas}, {David}, {David},
  {de Laverny}, {De Luise}, {De March}, {de Martino}, {de Souza}, {de Torres},
  {Debosscher}, {del Pozo}, {Delbo}, {Delgado}, {Delgado}, {Di Matteo},
  {Diakite}, {Diener}, {Distefano}, {Dolding}, {Drazinos}, {Dur{\'a}n},
  {Edvardsson}, {Enke}, {Eriksson}, {Esquej}, {Eynard Bontemps}, {Fabre},
  {Fabrizio}, {Faigler}, {Falc{\~a}o}, {Farr{\`a}s Casas}, {Federici},
  {Fedorets}, {Fernique}, {Figueras}, {Filippi}, {Findeisen}, {Fonti},
  {Fraile}, {Fraser}, {Fr{\'e}zouls}, {Gai}, {Galleti}, {Garabato},
  {Garc{\'\i}a-Sedano}, {Garofalo}, {Garralda}, {Gavel}, {Gavras}, {Gerssen},
  {Geyer}, {Giacobbe}, {Gilmore}, {Girona}, {Giuffrida}, {Glass}, {Gomes},
  {Granvik}, {Gueguen}, {Guerrier}, {Guiraud}, {Guti{\'e}rrez-S{\'a}nchez},
  {Haigron}, {Hatzidimitriou}, {Hauser}, {Haywood}, {Heiter}, {Helmi}, {Heu},
  {Hilger}, {Hobbs}, {Hofmann}, {Holland}, {Huckle}, {Hypki}, {Icardi},
  {Jan{\ss}en}, {Jevardat de Fombelle}, {Jonker}, {Juh{\'a}sz}, {Julbe},
  {Karampelas}, {Kewley}, {Klar}, {Kochoska}, {Kohley}, {Kolenberg},
  {Kontizas}, {Kontizas}, {Koposov}, {Kordopatis}, {Kostrzewa-Rutkowska},
  {Koubsky}, {Lambert}, {Lanza}, {Lasne}, {Lavigne}, {Le Fustec}, {Le
  Poncin-Lafitte}, {Lebreton}, {Leccia}, {Leclerc}, {Lecoeur-Taibi},
  {Lenhardt}, {Leroux}, {Liao}, {Licata}, {Lindstr{\o}m}, {Lister}, {Livanou},
  {Lobel}, {L{\'o}pez}, {Managau}, {Mann}, {Mantelet}, {Marchal}, {Marchant},
  {Marconi}, {Marinoni}, {Marschalk{\'o}}, {Marshall}, {Martino}, {Marton},
  {Mary}, {Massari}, {Matijevi{\v{c}}}, {Mazeh}, {McMillan}, {Messina},
  {Michalik}, {Millar}, {Molina}, {Molinaro}, {Moln{\'a}r}, {Montegriffo},
  {Mor}, {Morbidelli}, {Morel}, {Morris}, {Mulone}, {Muraveva}, {Musella},
  {Nelemans}, {Nicastro}, {Noval}, {O'Mullane}, {Ord{\'e}novic},
  {Ord{\'o}{\~n}ez-Blanco}, {Osborne}, {Pagani}, {Pagano}, {Pailler},
  {Palacin}, {Palaversa}, {Panahi}, {Pawlak}, {Piersimoni}, {Pineau}, {Plachy},
  {Plum}, {Poggio}, {Poujoulet}, {Pr{\v{s}}a}, {Pulone}, {Racero}, {Ragaini},
  {Rambaux}, {Ramos-Lerate}, {Regibo}, {Reyl{\'e}}, {Riclet}, {Ripepi}, {Riva},
  {Rivard}, {Rixon}, {Roegiers}, {Roelens}, {Romero-G{\'o}mez}, {Rowell},
  {Royer}, {Ruiz-Dern}, {Sadowski}, {Sagrist{\`a} Sell{\'e}s}, {Sahlmann},
  {Salgado}, {Salguero}, {Sanna}, {Santana-Ros}, {Sarasso}, {Savietto},
  {Schultheis}, {Sciacca}, {Segol}, {Segovia}, {S{\'e}gransan}, {Shih},
  {Siltala}, {Silva}, {Smart}, {Smith}, {Solano}, {Solitro}, {Sordo}, {Soria
  Nieto}, {Souchay}, {Spagna}, {Spoto}, {Stampa}, {Steele},
  {Steidelm{\"u}ller}, {Stephenson}, {Stoev}, {Suess}, {Surdej}, {Szabados},
  {Szegedi-Elek}, {Tapiador}, {Taris}, {Tauran}, {Taylor}, {Teixeira},
  {Terrett}, {Teyssandier}, {Thuillot}, {Titarenko}, {Torra Clotet}, {Turon},
  {Ulla}, {Utrilla}, {Uzzi}, {Vaillant}, {Valentini}, {Valette}, {van Elteren},
  {Van Hemelryck}, {van Leeuwen}, {Vaschetto}, {Vecchiato}, {Veljanoski},
  {Viala}, {Vicente}, {Vogt}, {von Essen}, {Voss}, {Votruba}, {Voutsinas},
  {Walmsley}, {Weiler}, {Wertz}, {Wevers}, {Wyrzykowski}, {Yoldas},
  {{\v{Z}}erjal}, {Ziaeepour}, {Zorec}, {Zschocke}, {Zucker}, {Zurbach}, \&
  {Zwitter}}]{GaiaDR2}
{Gaia Collaboration}, {Brown}, A.~G.~A., {Vallenari}, A., {et~al.} 2018, \aap,
  616, A1

\bibitem[{{Gaia Collaboration} {et~al.}(2021){Gaia Collaboration}, {Brown},
  {Vallenari}, {Prusti}, {de Bruijne}, {Babusiaux}, {Biermann}, {Creevey},
  {Evans}, {Eyer}, {Hutton}, {Jansen}, {Jordi}, {Klioner}, {Lammers},
  {Lindegren}, {Luri}, {Mignard}, {Panem}, {Pourbaix}, {Randich}, {Sartoretti},
  {Soubiran}, {Walton}, {Arenou}, {Bailer-Jones}, {Bastian}, {Cropper},
  {Drimmel}, {Katz}, {Lattanzi}, {van Leeuwen}, {Bakker}, {Cacciari},
  {Casta{\~n}eda}, {De Angeli}, {Ducourant}, {Fabricius}, {Fouesneau},
  {Fr{\'e}mat}, {Guerra}, {Guerrier}, {Guiraud}, {Jean-Antoine Piccolo},
  {Masana}, {Messineo}, {Mowlavi}, {Nicolas}, {Nienartowicz}, {Pailler},
  {Panuzzo}, {Riclet}, {Roux}, {Seabroke}, {Sordo}, {Tanga}, {Th{\'e}venin},
  {Gracia-Abril}, {Portell}, {Teyssier}, {Altmann}, {Andrae}, {Bellas-Velidis},
  {Benson}, {Berthier}, {Blomme}, {Brugaletta}, {Burgess}, {Busso}, {Carry},
  {Cellino}, {Cheek}, {Clementini}, {Damerdji}, {Davidson}, {Delchambre},
  {Dell'Oro}, {Fern{\'a}ndez-Hern{\'a}ndez}, {Galluccio}, {Garc{\'\i}a-Lario},
  {Garcia-Reinaldos}, {Gonz{\'a}lez-N{\'u}{\~n}ez}, {Gosset}, {Haigron},
  {Halbwachs}, {Hambly}, {Harrison}, {Hatzidimitriou}, {Heiter},
  {Hern{\'a}ndez}, {Hestroffer}, {Hodgkin}, {Holl}, {Jan{\ss}en}, {Jevardat de
  Fombelle}, {Jordan}, {Krone-Martins}, {Lanzafame}, {L{\"o}ffler}, {Lorca},
  {Manteiga}, {Marchal}, {Marrese}, {Moitinho}, {Mora}, {Muinonen}, {Osborne},
  {Pancino}, {Pauwels}, {Petit}, {Recio-Blanco}, {Richards}, {Riello},
  {Rimoldini}, {Robin}, {Roegiers}, {Rybizki}, {Sarro}, {Siopis}, {Smith},
  {Sozzetti}, {Ulla}, {Utrilla}, {van Leeuwen}, {van Reeven}, {Abbas}, {Abreu
  Aramburu}, {Accart}, {Aerts}, {Aguado}, {Ajaj}, {Altavilla}, {{\'A}lvarez},
  {{\'A}lvarez Cid-Fuentes}, {Alves}, {Anderson}, {Anglada Varela}, {Antoja},
  {Audard}, {Baines}, {Baker}, {Balaguer-N{\'u}{\~n}ez}, {Balbinot}, {Balog},
  {Barache}, {Barbato}, {Barros}, {Barstow}, {Bartolom{\'e}}, {Bassilana},
  {Bauchet}, {Baudesson-Stella}, {Becciani}, {Bellazzini}, {Bernet}, {Bertone},
  {Bianchi}, {Blanco-Cuaresma}, {Boch}, {Bombrun}, {Bossini}, {Bouquillon},
  {Bragaglia}, {Bramante}, {Breedt}, {Bressan}, {Brouillet}, {Bucciarelli},
  {Burlacu}, {Busonero}, {Butkevich}, {Buzzi}, {Caffau}, {Cancelliere},
  {C{\'a}novas}, {Cantat-Gaudin}, {Carballo}, {Carlucci}, {Carnerero},
  {Carrasco}, {Casamiquela}, {Castellani}, {Castro-Ginard}, {Castro Sampol},
  {Chaoul}, {Charlot}, {Chemin}, {Chiavassa}, {Cioni}, {Comoretto}, {Cooper},
  {Cornez}, {Cowell}, {Crifo}, {Crosta}, {Crowley}, {Dafonte}, {Dapergolas},
  {David}, {David}, {de Laverny}, {De Luise}, {De March}, {De Ridder}, {de
  Souza}, {de Teodoro}, {de Torres}, {del Peloso}, {del Pozo}, {Delbo},
  {Delgado}, {Delgado}, {Delisle}, {Di Matteo}, {Diakite}, {Diener},
  {Distefano}, {Dolding}, {Eappachen}, {Edvardsson}, {Enke}, {Esquej}, {Fabre},
  {Fabrizio}, {Faigler}, {Fedorets}, {Fernique}, {Fienga}, {Figueras},
  {Fouron}, {Fragkoudi}, {Fraile}, {Franke}, {Gai}, {Garabato},
  {Garcia-Gutierrez}, {Garc{\'\i}a-Torres}, {Garofalo}, {Gavras}, {Gerlach},
  {Geyer}, {Giacobbe}, {Gilmore}, {Girona}, {Giuffrida}, {Gomel}, {Gomez},
  {Gonzalez-Santamaria}, {Gonz{\'a}lez-Vidal}, {Granvik},
  {Guti{\'e}rrez-S{\'a}nchez}, {Guy}, {Hauser}, {Haywood}, {Helmi}, {Hidalgo},
  {Hilger}, {H{\l}adczuk}, {Hobbs}, {Holland}, {Huckle}, {Jasniewicz},
  {Jonker}, {Juaristi Campillo}, {Julbe}, {Karbevska}, {Kervella}, {Khanna},
  {Kochoska}, {Kontizas}, {Kordopatis}, {Korn}, {Kostrzewa-Rutkowska},
  {Kruszy{\'n}ska}, {Lambert}, {Lanza}, {Lasne}, {Le Campion}, {Le Fustec},
  {Lebreton}, {Lebzelter}, {Leccia}, {Leclerc}, {Lecoeur-Taibi}, {Liao},
  {Licata}, {Lindstr{\o}m}, {Lister}, {Livanou}, {Lobel}, {Madrero Pardo},
  {Managau}, {Mann}, {Marchant}, {Marconi}, {Marcos Santos}, {Marinoni},
  {Marocco}, {Marshall}, {Martin Polo}, {Mart{\'\i}n-Fleitas}, {Masip},
  {Massari}, {Mastrobuono-Battisti}, {Mazeh}, {McMillan}, {Messina},
  {Michalik}, {Millar}, {Mints}, {Molina}, {Molinaro}, {Moln{\'a}r},
  {Montegriffo}, {Mor}, {Morbidelli}, {Morel}, {Morris}, {Mulone}, {Munoz},
  {Muraveva}, {Murphy}, {Musella}, {Noval}, {Ord{\'e}novic}, {Orr{\`u}},
  {Osinde}, {Pagani}, {Pagano}, {Palaversa}, {Palicio}, {Panahi}, {Pawlak},
  {Pe{\~n}alosa Esteller}, {Penttil{\"a}}, {Piersimoni}, {Pineau}, {Plachy},
  {Plum}, {Poggio}, {Poretti}, {Poujoulet}, {Pr{\v{s}}a}, {Pulone}, {Racero},
  {Ragaini}, {Rainer}, {Raiteri}, {Rambaux}, {Ramos}, {Ramos-Lerate}, {Re
  Fiorentin}, {Regibo}, {Reyl{\'e}}, {Ripepi}, {Riva}, {Rixon}, {Robichon},
  {Robin}, {Roelens}, {Rohrbasser}, {Romero-G{\'o}mez}, {Rowell}, {Royer},
  {Rybicki}, {Sadowski}, {Sagrist{\`a} Sell{\'e}s}, {Sahlmann}, {Salgado},
  {Salguero}, {Samaras}, {Sanchez Gimenez}, {Sanna}, {Santove{\~n}a},
  {Sarasso}, {Schultheis}, {Sciacca}, {Segol}, {Segovia}, {S{\'e}gransan},
  {Semeux}, {Shahaf}, {Siddiqui}, {Siebert}, {Siltala}, {Slezak}, {Smart},
  {Solano}, {Solitro}, {Souami}, {Souchay}, {Spagna}, {Spoto}, {Steele},
  {Steidelm{\"u}ller}, {Stephenson}, {S{\"u}veges}, {Szabados}, {Szegedi-Elek},
  {Taris}, {Tauran}, {Taylor}, {Teixeira}, {Thuillot}, {Tonello}, {Torra},
  {Torra}, {Turon}, {Unger}, {Vaillant}, {van Dillen}, {Vanel}, {Vecchiato},
  {Viala}, {Vicente}, {Voutsinas}, {Weiler}, {Wevers}, {Wyrzykowski}, {Yoldas},
  {Yvard}, {Zhao}, {Zorec}, {Zucker}, {Zurbach}, \& {Zwitter}}]{GaiaEDR3}
---. 2021, \aap, 649, A1

\bibitem[{{Glikman} {et~al.}(2006){Glikman}, {Helfand}, \&
  {White}}]{Glikman2006}
{Glikman}, E., {Helfand}, D.~J., \& {White}, R.~L. 2006, \apj, 640, 579

\bibitem[{{Goulding} {et~al.}(2019){Goulding}, {Pardo}, {Greene}, {Mingarelli},
  {Nyland}, \& {Strauss}}]{Goulding19}
{Goulding}, A.~D., {Pardo}, K., {Greene}, J.~E., {et~al.} 2019, \apjl, 879, L21

\bibitem[{{Graham} {et~al.}(2015){Graham}, {Djorgovski}, {Stern}, {Drake},
  {Mahabal}, {Donalek}, {Glikman}, {Larson}, \& {Christensen}}]{Graham2015}
{Graham}, M.~J., {Djorgovski}, S.~G., {Stern}, D., {et~al.} 2015, \mnras, 453,
  1562

\bibitem[{{Green}(2018)}]{Green2018}
{Green}, G. 2018, The Journal of Open Source Software, 3, 695

\bibitem[{{Green} {et~al.}(2011){Green}, {Myers}, {Barkhouse}, {Aldcroft},
  {Trichas}, {Richards}, {Ruiz}, \& {Hopkins}}]{Green11}
{Green}, P.~J., {Myers}, A.~D., {Barkhouse}, W.~A., {et~al.} 2011, \apj, 743,
  81

\bibitem[{{Gregg} {et~al.}(2002){Gregg}, {Becker}, {White}, {Richards},
  {Chaffee}, \& {Fan}}]{Gregg02}
{Gregg}, M.~D., {Becker}, R.~H., {White}, R.~L., {et~al.} 2002, \apjl, 573, L85

\bibitem[{{Guo} {et~al.}(2019){Guo}, {Liu}, {Shen}, {Loeb}, {Monroe}, \&
  {Prochaska}}]{Guo2019}
{Guo}, H., {Liu}, X., {Shen}, Y., {et~al.} 2019, \mnras, 482, 3288

\bibitem[{{Hagen} {et~al.}(1996){Hagen}, {Hopp}, {Engels}, \&
  {Reimers}}]{Hagen96}
{Hagen}, H.~J., {Hopp}, U., {Engels}, D., \& {Reimers}, D. 1996, \aap, 308, L25

\bibitem[{{Hennawi} {et~al.}(2006){Hennawi}, {Strauss}, {Oguri}, {Inada},
  {Richards}, {Pindor}, {Schneider}, {Becker}, {Gregg}, {Hall}, {Johnston},
  {Fan}, {Burles}, {Schlegel}, {Gunn}, {Lupton}, {Bahcall}, {Brunner}, \&
  {Brinkmann}}]{Hennawi06}
{Hennawi}, J.~F., {Strauss}, M.~A., {Oguri}, M., {et~al.} 2006, \aj, 131, 1

\bibitem[{{Hennawi} {et~al.}(2010){Hennawi}, {Myers}, {Shen}, {Strauss},
  {Djorgovski}, {Fan}, {Glikman}, {Mahabal}, {Martin}, {Richards}, {Schneider},
  \& {Shankar}}]{Hennawi10}
{Hennawi}, J.~F., {Myers}, A.~D., {Shen}, Y., {et~al.} 2010, \apj, 719, 1672

\bibitem[{{Hewett} {et~al.}(1998){Hewett}, {Foltz}, {Harding}, \&
  {Lewis}}]{Hewett98}
{Hewett}, P.~C., {Foltz}, C.~B., {Harding}, M.~E., \& {Lewis}, G.~F. 1998, \aj,
  115, 383

\bibitem[{{Hopkins} {et~al.}(2008){Hopkins}, {Hernquist}, {Cox}, \& {Kere{\v
  s}}}]{hopkins08}
{Hopkins}, P.~F., {Hernquist}, L., {Cox}, T.~J., \& {Kere{\v s}}, D. 2008,
  \apjs, 175, 356

\bibitem[{{Huang} {et~al.}(2014){Huang}, {Liu}, {Yuan}, {Xiang}, {Huo}, {Hou},
  {Jin}, {Zhang}, \& {Zhou}}]{Huang14}
{Huang}, Y., {Liu}, X.~W., {Yuan}, H.~B., {et~al.} 2014, \mnras, 439, 2927

\bibitem[{{Hudson} {et~al.}(2006){Hudson}, {Reiprich}, {Clarke}, \&
  {Sarazin}}]{Hudson06}
{Hudson}, D.~S., {Reiprich}, T.~H., {Clarke}, T.~E., \& {Sarazin}, C.~L. 2006,
  \aap, 453, 433

\bibitem[{{Hughes}(2009)}]{hughes09}
{Hughes}, S.~A. 2009, \araa, 47, 107

\bibitem[{{Hwang} {et~al.}(2020){Hwang}, {Shen}, {Zakamska}, \&
  {Liu}}]{HwangShen2020}
{Hwang}, H.-C., {Shen}, Y., {Zakamska}, N., \& {Liu}, X. 2020, \apj, 888, 73

\bibitem[{{Inada} {et~al.}(2008){Inada}, {Oguri}, {Becker}, {Shin}, {Richards},
  {Hennawi}, {White}, {Pindor}, {Strauss}, {Kochanek}, {Johnston}, {Gregg},
  {Kayo}, {Eisenstein}, {Hall}, {Castander}, {Clocchiatti}, {Anderson},
  {Schneider}, {York}, {Lupton}, {Chiu}, {Kawano}, {Scranton}, {Frieman},
  {Keeton}, {Morokuma}, {Rix}, {Turner}, {Burles}, {Brunner}, {Sheldon},
  {Bahcall}, \& {Masataka}}]{Inada08}
{Inada}, N., {Oguri}, M., {Becker}, R.~H., {et~al.} 2008, \aj, 135, 496

\bibitem[{{Inada} {et~al.}(2012){Inada}, {Oguri}, {Shin}, {Kayo}, {Strauss},
  {Morokuma}, {Rusu}, {Fukugita}, {Kochanek}, {Richards}, {Schneider}, {York},
  {Bahcall}, {Frieman}, {Hall}, \& {White}}]{Inada12}
{Inada}, N., {Oguri}, M., {Shin}, M.-S., {et~al.} 2012, \aj, 143, 119

\bibitem[{{Junkkarinen} {et~al.}(2001){Junkkarinen}, {Shields}, {Beaver},
  {Burbidge}, {Cohen}, {Hamann}, \& {Lyons}}]{junkkarinen01}
{Junkkarinen}, V., {Shields}, G.~A., {Beaver}, E.~A., {et~al.} 2001, \apjl,
  549, L155

\bibitem[{{Kharb} {et~al.}(2017){Kharb}, {Lal}, \& {Merritt}}]{Kharb2017}
{Kharb}, P., {Lal}, D.~V., \& {Merritt}, D. 2017, Nature Astronomy, 1, 727

\bibitem[{{Kirkpatrick} {et~al.}(2012){Kirkpatrick}, {Gelino}, {Cushing},
  {Mace}, {Griffith}, {Skrutskie}, {Marsh}, {Wright}, {Eisenhardt}, {McLean},
  {Mainzer}, {Burgasser}, {Tinney}, {Parker}, \& {Salter}}]{Kirkpatrick2012}
{Kirkpatrick}, J.~D., {Gelino}, C.~R., {Cushing}, M.~C., {et~al.} 2012, \apj,
  753, 156

\bibitem[{{Komossa} {et~al.}(2003){Komossa}, {Burwitz}, {Hasinger}, {Predehl},
  {Kaastra}, \& {Ikebe}}]{Komossa03}
{Komossa}, S., {Burwitz}, V., {Hasinger}, G., {et~al.} 2003, \apjl, 582, L15

\bibitem[{{Kormendy} \& {Ho}(2013)}]{KormendyHo2013}
{Kormendy}, J., \& {Ho}, L.~C. 2013, \araa, 51, 511

\bibitem[{{Kormendy} \& {Richstone}(1995)}]{kormendy95}
{Kormendy}, J., \& {Richstone}, D. 1995, \araa, 33, 581

\bibitem[{{Koss} {et~al.}(2012){Koss}, {Mushotzky}, {Treister}, {Veilleux},
  {Vasudevan}, \& {Trippe}}]{Koss12}
{Koss}, M., {Mushotzky}, R., {Treister}, E., {et~al.} 2012, \apjl, 746, L22

\bibitem[{{Koss} {et~al.}(2011){Koss}, {Mushotzky}, {Treister}, {Veilleux},
  {Vasudevan}, {Miller}, {Sand ers}, {Schawinski}, \& {Trippe}}]{Koss11}
---. 2011, \apjl, 735, L42

\bibitem[{{Kroupa} {et~al.}(1993){Kroupa}, {Tout}, \& {Gilmore}}]{Kroupa1993}
{Kroupa}, P., {Tout}, C.~A., \& {Gilmore}, G. 1993, \mnras, 262, 545

\bibitem[{{Lacy} {et~al.}(2020){Lacy}, {Baum}, {Chandler}, {Chatterjee},
  {Clarke}, {Deustua}, {English}, {Farnes}, {Gaensler}, {Gugliucci},
  {Hallinan}, {Kent}, {Kimball}, {Law}, {Lazio}, {Marvil}, {Mao}, {Medlin},
  {Mooley}, {Murphy}, {Myers}, {Osten}, {Richards}, {Rosolowsky}, {Rudnick},
  {Schinzel}, {Sivakoff}, {Sjouwerman}, {Taylor}, {White}, {Wrobel},
  {Andernach}, {Beasley}, {Berger}, {Bhatnager}, {Birkinshaw}, {Bower},
  {Brandt}, {Brown}, {Burke-Spolaor}, {Butler}, {Comerford}, {Demorest}, {Fu},
  {Giacintucci}, {Golap}, {G{\"u}th}, {Hales}, {Hiriart}, {Hodge}, {Horesh},
  {Ivezi{\'c}}, {Jarvis}, {Kamble}, {Kassim}, {Liu}, {Loinard}, {Lyons},
  {Masters}, {Mezcua}, {Moellenbrock}, {Mroczkowski}, {Nyland},
  {O{\textquoteright}Dea}, {O{\textquoteright}Sullivan}, {Peters}, {Radford},
  {Rao}, {Robnett}, {Salcido}, {Shen}, {Sobotka}, {Witz}, {Vaccari}, {van
  Weeren}, {Vargas}, {Williams}, \& {Yoon}}]{VLASS}
{Lacy}, M., {Baum}, S.~A., {Chandler}, C.~J., {et~al.} 2020, \pasp, 132, 035001

\bibitem[{{Leisy} \& {Dennefeld}(1996)}]{Leisy1996}
{Leisy}, P., \& {Dennefeld}, M. 1996, \aaps, 116, 95

\bibitem[{{Lemon} {et~al.}(2020){Lemon}, {Auger}, {McMahon}, {Anguita},
  {Apostolovski}, {Chen}, {Fassnacht}, {Melo}, {Motta}, {Shajib}, {Treu},
  {Agnello}, {Buckley-Geer}, {Schechter}, {Birrer}, {Collett}, {Courbin},
  {Rusu}, {Abbott}, {Allam}, {Annis}, {Avila}, {Bertin}, {Brooks}, {Burke},
  {Carnero Rosell}, {Carrasco Kind}, {Carretero}, {Costanzi}, {da Costa}, {De
  Vicente}, {Desai}, {Eifler}, {Flaugher}, {Frieman}, {Garc{\'\i}a-Bellido},
  {Gaztanaga}, {Gerdes}, {Gruen}, {Gruendl}, {Gschwend}, {Gutierrez},
  {Honscheid}, {James}, {Kim}, {Krause}, {Kuehn}, {Kuropatkin}, {Lahav},
  {Lima}, {Lin}, {Maia}, {March}, {Marshall}, {Menanteau}, {Miquel}, {Palmese},
  {Paz-Chinch{\'o}n}, {Plazas}, {Roodman}, {Sanchez}, {Schubnell}, {Serrano},
  {Smith}, {Soares-Santos}, {Suchyta}, {Tarle}, \& {Walker}}]{Lemon20}
{Lemon}, C., {Auger}, M.~W., {McMahon}, R., {et~al.} 2020, \mnras, 494, 3491

\bibitem[{{Lemon} {et~al.}(2019){Lemon}, {Auger}, \& {McMahon}}]{Lemon19}
{Lemon}, C.~A., {Auger}, M.~W., \& {McMahon}, R.~G. 2019, \mnras, 483, 4242

\bibitem[{{Lemon} {et~al.}(2018){Lemon}, {Auger}, {McMahon}, \&
  {Ostrovski}}]{Lemon18}
{Lemon}, C.~A., {Auger}, M.~W., {McMahon}, R.~G., \& {Ostrovski}, F. 2018,
  \mnras, 479, 5060

\bibitem[{{Liao} {et~al.}(2021){Liao}, {Chen}, {Liu}, {Holgado}, {Guo},
  {Gruendl}, {Morganson}, {Shen}, {Davis}, {Kessler}, {Martini}, {McMahon},
  {Allam}, {Annis}, {Avila}, {Banerji}, {Bechtol}, {Bertin}, {Brooks},
  {Buckley-Geer}, {Carnero Rosell}, {Carrasco Kind}, {Carretero}, {Javier
  Castander}, {Cunha}, {D'Andrea}, {da Costa}, {Davis}, {De Vicente}, {Desai},
  {Thomas Diehl}, {Doel}, {Eifler}, {Evrard}, {Flaugher}, {Fosalba}, {Frieman},
  {Garcia-Bellido}, {Gaztanaga}, {Glazebrook}, {Gruen}, {Gschwend},
  {Gutierrez}, {Hartley}, {Hollowood}, {Honscheid}, {Hoyle}, {James}, {Krause},
  {Kuehn}, {Lima}, {Maia}, {Marshall}, {Menanteau}, {Miquel}, {Plazas
  Malag{\'o}n}, {Roodman}, {Sanchez}, {Scarpine}, {Schubnell}, {Serrano},
  {Smith}, {Smith}, {Soares-Santos}, {Sobreira}, {Suchyta}, {Swanson}, {Tarle},
  {Vikram}, \& {Walker}}]{Liao2021}
{Liao}, W.-T., {Chen}, Y.-C., {Liu}, X., {et~al.} 2021, \mnras, 500, 4025

\bibitem[{{Lindegren} {et~al.}(2012){Lindegren}, {Lammers}, {Hobbs},
  {O'Mullane}, {Bastian}, \& {Hern{\'a}ndez}}]{Lindegren2012}
{Lindegren}, L., {Lammers}, U., {Hobbs}, D., {et~al.} 2012, \aap, 538, A78

\bibitem[{{Lindegren} {et~al.}(2018){Lindegren}, {Hern{\'a}ndez}, {Bombrun},
  {Klioner}, {Bastian}, {Ramos-Lerate}, {de Torres}, {Steidelm{\"u}ller},
  {Stephenson}, {Hobbs}, {Lammers}, {Biermann}, {Geyer}, {Hilger}, {Michalik},
  {Stampa}, {McMillan}, {Casta{\~n}eda}, {Clotet}, {Comoretto}, {Davidson},
  {Fabricius}, {Gracia}, {Hambly}, {Hutton}, {Mora}, {Portell}, {van Leeuwen},
  {Abbas}, {Abreu}, {Altmann}, {Andrei}, {Anglada}, {Balaguer-N{\'u}{\~n}ez},
  {Barache}, {Becciani}, {Bertone}, {Bianchi}, {Bouquillon}, {Bourda},
  {Br{\"u}semeister}, {Bucciarelli}, {Busonero}, {Buzzi}, {Cancelliere},
  {Carlucci}, {Charlot}, {Cheek}, {Crosta}, {Crowley}, {de Bruijne}, {de
  Felice}, {Drimmel}, {Esquej}, {Fienga}, {Fraile}, {Gai}, {Garralda},
  {Gonz{\'a}lez-Vidal}, {Guerra}, {Hauser}, {Hofmann}, {Holl}, {Jordan},
  {Lattanzi}, {Lenhardt}, {Liao}, {Licata}, {Lister}, {L{\"o}ffler},
  {Marchant}, {Martin-Fleitas}, {Messineo}, {Mignard}, {Morbidelli}, {Poggio},
  {Riva}, {Rowell}, {Salguero}, {Sarasso}, {Sciacca}, {Siddiqui}, {Smart},
  {Spagna}, {Steele}, {Taris}, {Torra}, {van Elteren}, {van Reeven}, \&
  {Vecchiato}}]{Lindegren2018}
{Lindegren}, L., {Hern{\'a}ndez}, J., {Bombrun}, A., {et~al.} 2018, \aap, 616,
  A2

\bibitem[{{Lindegren} {et~al.}(2021){Lindegren}, {Klioner}, {Hern{\'a}ndez},
  {Bombrun}, {Ramos-Lerate}, {Steidelm{\"u}ller}, {Bastian}, {Biermann}, {de
  Torres}, {Gerlach}, {Geyer}, {Hilger}, {Hobbs}, {Lammers}, {McMillan},
  {Stephenson}, {Casta{\~n}eda}, {Davidson}, {Fabricius}, {Gracia-Abril},
  {Portell}, {Rowell}, {Teyssier}, {Torra}, {Bartolom{\'e}}, {Clotet},
  {Garralda}, {Gonz{\'a}lez-Vidal}, {Torra}, {Abbas}, {Altmann}, {Anglada
  Varela}, {Balaguer-N{\'u}{\~n}ez}, {Balog}, {Barache}, {Becciani}, {Bernet},
  {Bertone}, {Bianchi}, {Bouquillon}, {Brown}, {Bucciarelli}, {Busonero},
  {Butkevich}, {Buzzi}, {Cancelliere}, {Carlucci}, {Charlot}, {Cioni},
  {Crosta}, {Crowley}, {del Peloso}, {del Pozo}, {Drimmel}, {Esquej}, {Fienga},
  {Fraile}, {Gai}, {Garcia-Reinaldos}, {Guerra}, {Hambly}, {Hauser},
  {Jan{\ss}en}, {Jordan}, {Kostrzewa-Rutkowska}, {Lattanzi}, {Liao}, {Licata},
  {Lister}, {L{\"o}ffler}, {Marchant}, {Masip}, {Mignard}, {Mints}, {Molina},
  {Mora}, {Morbidelli}, {Murphy}, {Pagani}, {Panuzzo}, {Pe{\~n}alosa Esteller},
  {Poggio}, {Re Fiorentin}, {Riva}, {Sagrist{\`a} Sell{\'e}s}, {Sanchez
  Gimenez}, {Sarasso}, {Sciacca}, {Siddiqui}, {Smart}, {Souami}, {Spagna},
  {Steele}, {Taris}, {Utrilla}, {van Reeven}, \& {Vecchiato}}]{Lindegren2021}
{Lindegren}, L., {Klioner}, S.~A., {Hern{\'a}ndez}, J., {et~al.} 2021, \aap,
  649, A2

\bibitem[{{Liu} {et~al.}(2019){Liu}, {Gezari}, {Ayers}, {Burgett}, {Chambers},
  {Hodapp}, {Huber}, {Kudritzki}, {Metcalfe}, {Tonry}, {Wainscoat}, \&
  {Waters}}]{Liutt2019}
{Liu}, T., {Gezari}, S., {Ayers}, M., {et~al.} 2019, \apj, 884, 36

\bibitem[{{Liu} {et~al.}(2013){Liu}, {Civano}, {Shen}, {Green}, {Greene}, \&
  {Strauss}}]{Liu13}
{Liu}, X., {Civano}, F., {Shen}, Y., {et~al.} 2013, \apj, 762, 110

\bibitem[{{Liu} {et~al.}(2010{\natexlab{a}}){Liu}, {Greene}, {Shen}, \&
  {Strauss}}]{Liu10}
{Liu}, X., {Greene}, J.~E., {Shen}, Y., \& {Strauss}, M.~A. 2010{\natexlab{a}},
  \apjl, 715, L30

\bibitem[{{Liu} {et~al.}(2014){Liu}, {Shen}, {Bian}, {Loeb}, \&
  {Tremaine}}]{Liu2014}
{Liu}, X., {Shen}, Y., {Bian}, F., {Loeb}, A., \& {Tremaine}, S. 2014, \apj,
  789, 140

\bibitem[{{Liu} {et~al.}(2010{\natexlab{b}}){Liu}, {Shen}, {Strauss}, \&
  {Greene}}]{Liu2010b}
{Liu}, X., {Shen}, Y., {Strauss}, M.~A., \& {Greene}, J.~E. 2010{\natexlab{b}},
  \apj, 708, 427

\bibitem[{{Liu} {et~al.}(2011){Liu}, {Shen}, {Strauss}, \& {Hao}}]{Liu11}
{Liu}, X., {Shen}, Y., {Strauss}, M.~A., \& {Hao}, L. 2011, \apj, 737, 101

\bibitem[{{Liu}(2015)}]{LiuY_2015}
{Liu}, Y. 2015, \aap, 580, A133

\bibitem[{{Madau} \& {Dickinson}(2014)}]{Madau2014}
{Madau}, P., \& {Dickinson}, M. 2014, \araa, 52, 415

\bibitem[{{Marrese} {et~al.}(2019){Marrese}, {Marinoni}, {Fabrizio}, \&
  {Altavilla}}]{Marrese2019}
{Marrese}, P.~M., {Marinoni}, S., {Fabrizio}, M., \& {Altavilla}, G. 2019,
  \aap, 621, A144

\bibitem[{{Mateos} {et~al.}(2012){Mateos}, {Alonso-Herrero}, {Carrera},
  {Blain}, {Watson}, {Barcons}, {Braito}, {Severgnini}, {Donley}, \&
  {Stern}}]{Mateos2012}
{Mateos}, S., {Alonso-Herrero}, A., {Carrera}, F.~J., {et~al.} 2012, \mnras,
  426, 3271

\bibitem[{{Mazzarella} {et~al.}(2012){Mazzarella}, {Iwasawa}, {Vavilkin},
  {Armus}, {Kim}, {Bothun}, {Evans}, {Spoon}, {Haan}, {Howell}, {Lord},
  {Marshall}, {Ishida}, {Xu}, {Petric}, {Sanders}, {Surace}, {Appleton},
  {Chan}, {Frayer}, {Inami}, {Khachikian}, {Madore}, {Privon}, {Sturm}, {U}, \&
  {Veilleux}}]{Mazzarella12}
{Mazzarella}, J.~M., {Iwasawa}, K., {Vavilkin}, T., {et~al.} 2012, \aj, 144,
  125

\bibitem[{{McGurk} {et~al.}(2011){McGurk}, {Max}, {Rosario}, {Shields},
  {Smith}, \& {Wright}}]{McGurk11}
{McGurk}, R.~C., {Max}, C.~E., {Rosario}, D.~J., {et~al.} 2011, \apjl, 738, L2

\bibitem[{{Merritt}(2013)}]{Merritt2013}
{Merritt}, D. 2013, {Dynamics and Evolution of Galactic Nuclei}

\bibitem[{{More} {et~al.}(2016){More}, {Oguri}, {Kayo}, {Zinn}, {Strauss},
  {Santiago}, {Mosquera}, {Inada}, {Kochanek}, {Rusu}, {Brownstein}, {da
  Costa}, {Kneib}, {Maia}, {Quimby}, {Schneider}, {Streblyanska}, \&
  {York}}]{More16}
{More}, A., {Oguri}, M., {Kayo}, I., {et~al.} 2016, \mnras, 456, 1595

\bibitem[{{Morgan} {et~al.}(2000){Morgan}, {Burley}, {Costa}, {Maza},
  {Persson}, {Ruiz}, {Schechter}, {Thompson}, \& {Winn}}]{Morgan00}
{Morgan}, N.~D., {Burley}, G., {Costa}, E., {et~al.} 2000, \aj, 119, 1083

\bibitem[{{M{\"u}ller-S{\'a}nchez} {et~al.}(2015){M{\"u}ller-S{\'a}nchez},
  {Comerford}, {Nevin}, {Barrows}, {Cooper}, \& {Greene}}]{Muller-Sanchez15}
{M{\"u}ller-S{\'a}nchez}, F., {Comerford}, J.~M., {Nevin}, R., {et~al.} 2015,
  \apj, 813, 103

\bibitem[{{M{\"u}ller-S{\'a}nchez} {et~al.}(2016){M{\"u}ller-S{\'a}nchez},
  {Comerford}, {Stern}, \& {Harrison}}]{Muller-Sanchez2016}
{M{\"u}ller-S{\'a}nchez}, F., {Comerford}, J.~M., {Stern}, D., \& {Harrison},
  F.~A. 2016, ArXiv e-prints 1606.07446, arXiv:1606.07446

\bibitem[{{Murga} {et~al.}(2015){Murga}, {Zhu}, {M{\'e}nard}, \&
  {Lan}}]{Murga2015}
{Murga}, M., {Zhu}, G., {M{\'e}nard}, B., \& {Lan}, T.-W. 2015, \mnras, 452,
  511

\bibitem[{{Murgia} {et~al.}(2001){Murgia}, {Parma}, {de Ruiter}, {Bondi},
  {Ekers}, {Fanti}, \& {Fomalont}}]{Murgia01}
{Murgia}, M., {Parma}, P., {de Ruiter}, H.~R., {et~al.} 2001, \aap, 380, 102

\bibitem[{{Owen} {et~al.}(1985){Owen}, {O'Dea}, {Inoue}, \& {Eilek}}]{Owen85}
{Owen}, F.~N., {O'Dea}, C.~P., {Inoue}, M., \& {Eilek}, J.~A. 1985, \apjl, 294,
  L85

\bibitem[{{P{\^a}ris} {et~al.}(2018){P{\^a}ris}, {Petitjean}, {Aubourg},
  {Myers}, {Streblyanska}, {Lyke}, {Anderson}, {Armengaud}, {Bautista},
  {Blanton}, {Blomqvist}, {Brinkmann}, {Brownstein}, {Brandt}, {Burtin},
  {Dawson}, {de la Torre}, {Georgakakis}, {Gil-Mar{\'{\i}}n}, {Green}, {Hall},
  {Kneib}, {LaMassa}, {Le Goff}, {MacLeod}, {Mariappan}, {McGreer}, {Merloni},
  {Noterdaeme}, {Palanque-Delabrouille}, {Percival}, {Ross}, {Rossi},
  {Schneider}, {Seo}, {Tojeiro}, {Weaver}, {Weijmans}, {Y{\`e}che}, {Zarrouk},
  \& {Zhao}}]{Paris2018}
{P{\^a}ris}, I., {Petitjean}, P., {Aubourg}, {\'E}., {et~al.} 2018, \aap, 613,
  A51

\bibitem[{{Peng} {et~al.}(2010){Peng}, {Ho}, {Impey}, \& {Rix}}]{Peng2010}
{Peng}, C.~Y., {Ho}, L.~C., {Impey}, C.~D., \& {Rix}, H.-W. 2010, \aj, 139,
  2097

\bibitem[{{Pfeifle} {et~al.}(2019){Pfeifle}, {Satyapal}, {Secrest}, {Gliozzi},
  {Ricci}, {Ellison}, {Rothberg}, {Cann}, {Blecha}, {Williams}, \&
  {Constantin}}]{Pfeifle2019}
{Pfeifle}, R.~W., {Satyapal}, S., {Secrest}, N.~J., {et~al.} 2019, \apj, 875,
  117

\bibitem[{{Pickles}(1998)}]{Pickles1998}
{Pickles}, A.~J. 1998, \pasp, 110, 863

\bibitem[{{Pindor} {et~al.}(2006){Pindor}, {Eisenstein}, {Gregg}, {Becker},
  {Inada}, {Oguri}, {Hall}, {Johnston}, {Richards}, {Schneider}, {Turner},
  {Brasi}, {Hinz}, {Kenworthy}, {Miller}, {Barentine}, {Brewington},
  {Brinkmann}, {Harvanek}, {Kleinman}, {Krzesinski}, {Long}, {Neilsen},
  {Newman}, {Nitta}, {Snedden}, \& {York}}]{Pindor06}
{Pindor}, B., {Eisenstein}, D.~J., {Gregg}, M.~D., {et~al.} 2006, \aj, 131, 41

\bibitem[{{Richards} {et~al.}(2006){Richards}, {Strauss}, {Fan}, {Hall},
  {Jester}, {Schneider}, {Vanden Berk}, {Stoughton}, {Anderson}, {Brunner},
  {Gray}, {Gunn}, {Ivezi{\'c}}, {Kirkland}, {Knapp}, {Loveday}, {Meiksin},
  {Pope}, {Szalay}, {Thakar}, {Yanny}, {York}, {Barentine}, {Brewington},
  {Brinkmann}, {Fukugita}, {Harvanek}, {Kent}, {Kleinman}, {Krzesi{\'n}ski},
  {Long}, {Lupton}, {Nash}, {Neilsen}, {Nitta}, {Schlegel}, \&
  {Snedden}}]{richards06}
{Richards}, G.~T., {Strauss}, M.~A., {Fan}, X., {et~al.} 2006, \aj, 131, 2766

\bibitem[{{Riello} {et~al.}(2018){Riello}, {De Angeli}, {Evans}, {Busso},
  {Hambly}, {Davidson}, {Burgess}, {Montegriffo}, {Osborne}, {Kewley},
  {Carrasco}, {Fabricius}, {Jordi}, {Cacciari}, {van Leeuwen}, \&
  {Holland}}]{Riello2018}
{Riello}, M., {De Angeli}, F., {Evans}, D.~W., {et~al.} 2018, \aap, 616, A3

\bibitem[{{Rodriguez} {et~al.}(2006){Rodriguez}, {Taylor}, {Zavala}, {Peck},
  {Pollack}, \& {Romani}}]{Rodriguez2006}
{Rodriguez}, C., {Taylor}, G.~B., {Zavala}, R.~T., {et~al.} 2006, \apj, 646, 49

\bibitem[{{Runnoe} {et~al.}(2017){Runnoe}, {Eracleous}, {Pennell}, {Mathes},
  {Boroson}, {Sigurethsson}, {Bogdanovi{\'c}}, {Halpern}, {Liu}, \&
  {Brown}}]{Runnoe2017}
{Runnoe}, J.~C., {Eracleous}, M., {Pennell}, A., {et~al.} 2017, \mnras, 468,
  1683

\bibitem[{{Satyapal} {et~al.}(2017){Satyapal}, {Secrest}, {Ricci}, {Ellison},
  {Rothberg}, {Blecha}, {Constantin}, {Gliozzi}, {McNulty}, \&
  {Ferguson}}]{Satyapal17}
{Satyapal}, S., {Secrest}, N.~J., {Ricci}, C., {et~al.} 2017, \apj, 848, 126

\bibitem[{{Schechter} {et~al.}(2017){Schechter}, {Morgan}, {Chehade},
  {Metcalfe}, {Shanks}, \& {McDonald}}]{Schechter17}
{Schechter}, P.~L., {Morgan}, N.~D., {Chehade}, B., {et~al.} 2017, \aj, 153,
  219

\bibitem[{{Schlafly} \& {Finkbeiner}(2011)}]{Schlafly2011}
{Schlafly}, E.~F., \& {Finkbeiner}, D.~P. 2011, \apj, 737, 103

\bibitem[{{Schlegel} {et~al.}(1998){Schlegel}, {Finkbeiner}, \&
  {Davis}}]{Schlegel1998}
{Schlegel}, D.~J., {Finkbeiner}, D.~P., \& {Davis}, M. 1998, \apj, 500, 525

\bibitem[{{Secrest} {et~al.}(2015){Secrest}, {Dudik}, {Dorland}, {Zacharias},
  {Makarov}, {Fey}, {Frouard}, \& {Finch}}]{Secrest2015}
{Secrest}, N.~J., {Dudik}, R.~P., {Dorland}, B.~N., {et~al.} 2015, \apjs, 221,
  12

\bibitem[{{Shen} {et~al.}(2019){Shen}, {Hwang}, {Zakamska}, \&
  {Liu}}]{ShenHwang2019}
{Shen}, Y., {Hwang}, H.-C., {Zakamska}, N., \& {Liu}, X. 2019, \apjl, 885, L4

\bibitem[{{Shen} {et~al.}(2011{\natexlab{a}}){Shen}, {Liu}, {Greene}, \&
  {Strauss}}]{Shen2011a}
{Shen}, Y., {Liu}, X., {Greene}, J.~E., \& {Strauss}, M.~A. 2011{\natexlab{a}},
  \apj, 735, 48

\bibitem[{{Shen} {et~al.}(2013){Shen}, {Liu}, {Loeb}, \& {Tremaine}}]{Shen2013}
{Shen}, Y., {Liu}, X., {Loeb}, A., \& {Tremaine}, S. 2013, \apj, 775, 49

\bibitem[{{Shen} {et~al.}(2011{\natexlab{b}}){Shen}, {Richards}, {Strauss},
  {Hall}, {Schneider}, {Snedden}, {Bizyaev}, {Brewington}, {Malanushenko},
  {Malanushenko}, {Oravetz}, {Pan}, \& {Simmons}}]{Shen2011b}
{Shen}, Y., {Richards}, G.~T., {Strauss}, M.~A., {et~al.} 2011{\natexlab{b}},
  \apjs, 194, 45

\bibitem[{{Shen} {et~al.}(2021){Shen}, {Chen}, {Hwang}, {Liu}, {Zakamska},
  {Oguri}, {Li}, {Lazio}, \& {Breiding}}]{Shen2021}
{Shen}, Y., {Chen}, Y.-C., {Hwang}, H.-C., {et~al.} 2021, Nature Astronomy, 5,
  569

\bibitem[{{Shields} {et~al.}(2012){Shields}, {Rosario}, {Junkkarinen},
  {Chapman}, {Bonning}, \& {Chiba}}]{Shields12}
{Shields}, G.~A., {Rosario}, D.~J., {Junkkarinen}, V., {et~al.} 2012, \apj,
  744, 151

\bibitem[{{Silverman} {et~al.}(2020){Silverman}, {Tang}, {Lee}, {Hartwig},
  {Goulding}, {Strauss}, {Schramm}, {Ding}, {Riffel}, {Fujimoto}, {Hikage},
  {Imanishi}, {Iwasawa}, {Jahnke}, {Kayo}, {Kashikawa}, {Kawaguchi}, {Kohno},
  {Luo}, {Matsuoka}, {Matsuda}, {Nagao}, {Oguri}, {Ono}, {Onoue}, {Ouchi},
  {Shimasaku}, {Suh}, {Suzuki}, {Taniguchi}, {Toba}, {Ueda}, \&
  {Yasuda}}]{Silverman20}
{Silverman}, J.~D., {Tang}, S., {Lee}, K.-G., {et~al.} 2020, arXiv e-prints,
  arXiv:2007.05581

\bibitem[{{Smith} {et~al.}(2010){Smith}, {Shields}, {Bonning}, {McMullen},
  {Rosario}, \& {Salviander}}]{Smith2010}
{Smith}, K.~L., {Shields}, G.~A., {Bonning}, E.~W., {et~al.} 2010, \apj, 716,
  866

\bibitem[{{Springer} \& {Ofek}(2021)}]{Springer2021}
{Springer}, O.~M., \& {Ofek}, E.~O. 2021, arXiv e-prints, arXiv:2101.11024

\bibitem[{{STScI Development Team}(2013)}]{pysynphot}
{STScI Development Team}. 2013, {pysynphot: Synthetic photometry software
  package}, , , ascl:1303.023

\bibitem[{{Tang} {et~al.}(2021){Tang}, {Silverman}, {Ding}, {Li}, {Lee},
  {Strauss}, {Goulding}, {Schramm}, {Kawinwanichakij}, {Prochaska}, {Hennawi},
  {Imanishi}, {Iwasawa}, {Toba}, {Kayo}, {Oguri}, {Matsuoka}, {Ichikawa},
  {Hartwig}, {Kashikawa}, {Kawaguchi}, {Kohno}, {Matsuda}, {Nagao}, {Ono},
  {Onoue}, {Ouchi}, {Shimasaku}, {Suh}, {Suzuki}, {Taniguchi}, {Ueda}, \&
  {Yasuda}}]{Tang21}
{Tang}, S., {Silverman}, J.~D., {Ding}, X., {et~al.} 2021, arXiv e-prints,
  arXiv:2105.10163

\bibitem[{{Teng} {et~al.}(2012){Teng}, {Schawinski}, {Urry}, {Darg}, {Kaviraj},
  {Oh}, {Bonning}, {Cardamone}, {Keel}, {Lintott}, {Simmons}, \&
  {Treister}}]{Teng12}
{Teng}, S.~H., {Schawinski}, K., {Urry}, C.~M., {et~al.} 2012, \apj, 753, 165

\bibitem[{{Vanden Berk} {et~al.}(2001){Vanden Berk}, {Richards}, {Bauer},
  {Strauss}, {Schneider}, {Heckman}, {York}, {Hall}, {Fan}, {Knapp},
  {Anderson}, {Annis}, {Bahcall}, {Bernardi}, {Briggs}, {Brinkmann}, {Brunner},
  {Burles}, {Carey}, {Castander}, {Connolly}, {Crocker}, {Csabai}, {Doi},
  {Finkbeiner}, {Friedman}, {Frieman}, {Fukugita}, {Gunn}, {Hennessy},
  {Ivezi{\'c}}, {Kent}, {Kunszt}, {Lamb}, {Leger}, {Long}, {Loveday}, {Lupton},
  {Meiksin}, {Merelli}, {Munn}, {Newberg}, {Newcomb}, {Nichol}, {Owen}, {Pier},
  {Pope}, {Rockosi}, {Schlegel}, {Siegmund}, {Smee}, {Snir}, {Stoughton},
  {Stubbs}, {SubbaRao}, {Szalay}, {Szokoly}, {Tremonti}, {Uomoto}, {Waddell},
  {Yanny}, \& {Zheng}}]{VandenBerk2001}
{Vanden Berk}, D.~E., {Richards}, G.~T., {Bauer}, A., {et~al.} 2001, \aj, 122,
  549

\bibitem[{{Wang} {et~al.}(2009){Wang}, {Chen}, {Hu}, {Mao}, {Zhang}, \&
  {Bian}}]{wang09}
{Wang}, J., {Chen}, Y., {Hu}, C., {et~al.} 2009, \apjl, 705, L76

\bibitem[{{Wang} {et~al.}(2017){Wang}, {Greene}, {Ju}, {Rafikov}, {Ruan}, \&
  {Schneider}}]{Wang2017}
{Wang}, L., {Greene}, J.~E., {Ju}, W., {et~al.} 2017, \apj, 834, 129

\bibitem[{{Williams} \& {Saha}(1995)}]{williams1995}
{Williams}, L.~L.~R., \& {Saha}, P. 1995, \aj, 110, 1471

\bibitem[{{Woo} {et~al.}(2014){Woo}, {Cho}, {Husemann}, {Komossa}, {Park}, \&
  {Bennert}}]{Woo14}
{Woo}, J.-H., {Cho}, H., {Husemann}, B., {et~al.} 2014, \mnras, 437, 32

\bibitem[{{Wright} {et~al.}(2010){Wright}, {Eisenhardt}, {Mainzer}, {Ressler},
  {Cutri}, {Jarrett}, {Kirkpatrick}, {Padgett}, {McMillan}, {Skrutskie},
  {Stanford}, {Cohen}, {Walker}, {Mather}, {Leisawitz}, {Gautier}, {McLean},
  {Benford}, {Lonsdale}, {Blain}, {Mendez}, {Irace}, {Duval}, {Liu}, {Royer},
  {Heinrichsen}, {Howard}, {Shannon}, {Kendall}, {Walsh}, {Larsen}, {Cardon},
  {Schick}, {Schwalm}, {Abid}, {Fabinsky}, {Naes}, \& {Tsai}}]{WISE}
{Wright}, E.~L., {Eisenhardt}, P. R.~M., {Mainzer}, A.~K., {et~al.} 2010, \aj,
  140, 1868

\bibitem[{{Yang} {et~al.}(2017){Yang}, {Wu}, {Fan}, {Jiang}, {McGreer},
  {Green}, {Yang}, {Schindler}, {Wang}, {Zuo}, \& {Fu}}]{Yang2017}
{Yang}, Q., {Wu}, X.-B., {Fan}, X., {et~al.} 2017, \aj, 154, 269

\bibitem[{{Yu}(2002)}]{Yu2002}
{Yu}, Q. 2002, \mnras, 331, 935

\bibitem[{{Zakamska} \& {Greene}(2014)}]{Zakamska2014}
{Zakamska}, N.~L., \& {Greene}, J.~E. 2014, \mnras, 442, 784

\end{thebibliography}
\end{document}